\documentclass[accepted=2018-10-09,a4paper,pra,longbibliography]{quantumarticle}
\pdfoutput=1

	\usepackage{bbm}
	\usepackage{amsmath,amssymb}
	\usepackage{dcolumn} 
	\usepackage{bm} 
	\usepackage{graphicx}
	\usepackage{color}
	\usepackage{hyperref}
	\usepackage{tikz}
	\usetikzlibrary{decorations.pathreplacing}
	\usetikzlibrary{automata,positioning}
	\usepackage{physics}
	\usepackage{hhline} 
	\usepackage{stmaryrd} 
	\usepackage{tocloft}

	\renewcommand{\phi}{\varphi}
	\renewcommand{\epsilon}{\varepsilon}

	\newcommand{\defeq}{\mathrel{\mathop:}=}

	\newcommand{\cC}{\mathcal{C}}
	\newcommand{\cM}{\mathcal{M}}
	\newcommand{\cS}{\mathcal{S}}
	\newcommand{\bZ}{\mathbbm{Z}}
	\newcommand{\one}{\mathbbm{1}}

	\newcommand*\braid[2]{%
		\raisebox{-7 pt}{%
		\begin{tikzpicture}
			\draw [thick] (0.0,0.0) to[in=-90,out=90] (0.5,0.5);
			\draw [line width=3pt,white] (0.5,0.0) to[in=-90,out=90] (0.0,0.5);
			\draw [thick] (0.5,0.0) to[in=-90,out=90] (0.0,0.5);
			\node[anchor = base] (a) at (0.0,0.55) {$#1$};
			\node[anchor = base] (a) at (0.5,0.55) {$#2$};
		\end{tikzpicture}}
	}
	\newcommand*\antibraid[2]{%
		\raisebox{-7 pt}{%
		\begin{tikzpicture}
			\draw [thick] (0.5,0.0) to[in=-90,out=90] (0.0,0.5);
			\draw [line width=3pt,white] (0.0,0.0) to[in=-90,out=90] (0.5,0.5);
			\draw [thick] (0.0,0.0) to[in=-90,out=90] (0.5,0.5);
			\node[anchor = base] (a) at (0.0,0.55) {$#1$};
			\node[anchor = base] (a) at (0.5,0.55) {$#2$};
		\end{tikzpicture}}
	}
	\newcommand*\trivialbraid[2]{%
		\raisebox{-7 pt}{%
		\begin{tikzpicture}
			\draw [thick] (0.0,0.0) to[in=-90,out=90] (0.0,0.5);
			\draw [thick] (0.5,0.0) to[in=-90,out=90] (0.5,0.5);
			\node[anchor = base] (a) at (0.0,0.55) {$#1$};
			\node[anchor = base] (a) at (0.5,0.55) {$#2$};
		\end{tikzpicture}}
	}
	\newcommand*\fullbraid[2]{%
		\raisebox{-17 pt}{%
		\begin{tikzpicture}
			\node (b) at (0,0) {\braid{\textcolor{white}{#1}}{\textcolor{white}{#2}}};
			\node (t) at (0,.5) {\braid{#1}{#2}};
		\end{tikzpicture}}
	}
	\newcommand*\fulltrivialbraid[2]{%
		\raisebox{-17 pt}{%
		\begin{tikzpicture}
			\node (b) at (0,0) {\trivialbraid{\textcolor{white}{#1}}{\textcolor{white}{#2}}};
			\node (t) at (0,.5) {\trivialbraid{#1}{#2}};
		\end{tikzpicture}}
	}

\begin{document}

	\title{The boundaries and twist defects of the color code and their applications to topological quantum computation}
	\author{Markus S. Kesselring}
	\email{markus.kesselring@fu-berlin.de}
	\affiliation{Dahlem Center for Complex Quantum Systems, Freie Universit\"at Berlin, 14195 Berlin, Germany}
	\author{Fernando Pastawski}
	\affiliation{Dahlem Center for Complex Quantum Systems, Freie Universit\"at Berlin, 14195 Berlin, Germany}
	\author{Jens Eisert}
	\affiliation{Dahlem Center for Complex Quantum Systems, Freie Universit\"at Berlin, 14195 Berlin, Germany}
	\author{Benjamin J. Brown}
	\affiliation{Centre for Engineered Quantum Systems, School of Physics, University of Sydney, Sydney, New South Wales 2006, Australia}
	\affiliation{Niels Bohr International Academy, Niels Bohr Institute, Blegdamsvej 17, 2100 Copenhagen, Denmark}
	\date{\today}

	\begin{abstract}
		The color code is both an interesting example of an exactly solvable topologically ordered phase of matter and also among the most promising candidate models to realize fault-tolerant quantum computation with minimal resource overhead.
		The contributions of this work are threefold.
		First of all, we build upon the abstract theory of boundaries and domain walls of topological phases of matter to comprehensively catalog the objects as they are realizable in color codes.
		Together with our classification we also provide lattice representations of these objects which include three new types of boundaries as well as a generating set for all $72$ color code twist defects.
		Our work thus provides an explicit toy model that will help to better understand the abstract theory of domain walls.
		Secondly, building upon the established framework, we discover a number of interesting new applications of the cataloged objects for devising quantum information protocols.
		These include improved methods for performing quantum computations by code deformation, a new four-qubit error-detecting code, as well as families of new quantum error-correcting codes we call stellated color codes, which encode logical qubits at the same distance as the next best color code, but using approximately half the number of physical qubits.
		To the best of our knowledge, our new topological codes have the highest encoding rate of local stabilizer codes with bounded-weight stabilizers in two dimensions.
		Finally, we show how the boundaries and twist defects of the color code are represented by multiple copies of other phases.
		Indeed, in addition to the well studied comparison between the color code and two copies of the surface code, we also compare the color code to two copies of the three-fermion model.
		In particular, we find that this analogy offers a very clear lens through which we can view the symmetries of the color code which gives rise to its multitude of domain walls.
	\end{abstract}

	\maketitle

	\setlength\cftbeforesecskip{2pt}
	\setlength\cftaftertoctitleskip{5pt}
	\tableofcontents


\section{Introduction}

	Topological phases of matter are of significant interest both from a condensed matter perspective~\cite{Wen04} as well as for their potential application in fault-tolerant quantum computation~\cite{Kitaev03, Dennis02, Preskill17lectures, Nayak08, Pachos12, Terhal15, Brown16, Campbell17}.
	Indeed, among the leading architectures that have been proposed for scalable quantum-information processing are topological quantum error-correcting codes.
	The physics of these phases is enriched further when their low-energy excitations are connected by symmetries, in which case they give rise to non-trivial domain walls and point-like topological defects~\cite{Kitaev06, Barkeshli10, Bombin10, Beigi11, Kitaev12, Barkeshli14, Barter18} that have been utilized in a number of instances~\cite{Bombin10, Bombin11, Barkeshli13, Brown14, Hastings15, Wootton15a, Yoder16, Brown17, Yoshida17, Roberts17, Zhu17, Lavasani18} to improve protocols for topological quantum computation and to reduce overhead requirements.
	Advancing the study of topological defects and their corresponding symmetries gives us new tools to improve architectures for topological quantum computation and, moreover, extends our understanding of the underlying topological phases that support them.

	The color code~\cite{Bombin06} is a particularly interesting model due to both its symmetries and its ability to realize fault-tolerant quantum computational operations.
	The importance of the use of different boundaries to implement the Clifford gates fault-tolerantly using transversal rotations has been recognized since its inception~\cite{Bombin06}.
	Further work on the computational power of color codes and their higher-dimensional counterparts have also been studied directly with various boundary configurations in Refs.~\cite{Bombin15, Kubica15a, Watson15, CampbellBlog} and in the guise of the unfolded surface code model in Ref.~\cite{Kubica15, Vasmer18}.
	Remarkably, these models are known to saturate bounds on the computational power of local models using local gates~\cite{Bravyi13, Pastawski15, OConnor17, Webster17}.
	It has also been shown that the boundaries of the color code can be exploited to perform Clifford gates using code deformation~\cite{Raussendorf06, Bombin09}
	either by the braiding of punctures~\cite{Fowler11} or by use of lattice surgery~\cite{Horsman12} in Refs.~\cite{Landahl14, Lavasani18}.
	Further to this, certain color code twist defects~\cite{Bombin11, Teo14}, and their application to quantum computation~\cite{Bombin11, Lavasani18} have also been studied.

	At a more fundamental level, a study in Ref.~\cite{Yoshida15} examines the symmetries of the color code model to identify continuous domain walls;
	the fundamental objects that are used to generate twist defects.
	Beyond this, the symmetries and twist defects with the color code model have been expressed using a tensor network formalism in Ref.~\cite{Bridgeman17}.
	This is complemented by Ref.~\cite{Williamson17} where a general study of twist defects and 
	domain walls using the framework of tensor networks is given.

	A recent surge in activity~\cite{Reed12, Barends14, Nigg14, Corcoles15, Albrecht16, Takita16, Linke17} has shown that we may be on the verge of experimental demonstrations of the fault-tolerant components that will make up a scalable quantum computer.
	Moreover, with its aforementioned versatility for performing logical gates, together with its high threshold error rates~\cite{Bombin12a} and its relatively economical resource demands~\cite{Landahl11}, the color code is regarded among the leading candidate architectures for the realization of topological quantum computation.
	As such, we have recently seen experimental proposals to realize color code models with a number of physical implementations and platforms.
	In particular Majorana devices~\cite{Terhal12, Aasen16, Plugge16, Landau16, Litinski17, Litinski17a} may be suitable as they allow for the measurement of relatively high-weight stabilizers.

	The color code model evidently has a very rich structure~\cite{Yoshida15, Bridgeman17} that we may be able to exploit further to discover better and more resource efficient protocols to perform quantum-computational operations in a fault-tolerant manner.
	It is the primary goal of the present manuscript to systematically explore the different objects that can be realized with modifications to the color code lattice through a study of its underlying topological excitations.
	We discover three new types of boundary, together with lattice realizations of all $72$ different twist defects that can be realized with the color code model.
	We catalog all of these objects and quantify their properties for use with topological quantum computation.

	We go on to find a number of new applications of the boundaries and twist defects of the color code for quantum information processing.
	For example, we argue that the new boundaries we discover, which we refer to as Pauli boundaries, may be of practical benefit when performing topological quantum computation.
	Specifically, we find that, like with the surface code model~\cite{Bombin09}, we can move punctures with Pauli boundaries in the color code model using single-qubit measurements.
	In contrast, to the best of our knowledge, we know of no way that we can move punctures with the known color code boundaries without the use of two-qubit Bell-type measurements~\cite{Fowler11}.
	We additionally find a new four-qubit error detecting code as a small instance of a new variant of the color code that has Pauli boundaries, one that could well serve as a blueprint for an experimental realization of quantum error correcting protocols, reminiscent of the ones presented in Refs.~\cite{Nigg14, CampbellBlog}.

	Beyond these examples, we find that our study of twist defects reveals new local codes with high encoding rates.
	It is known that two-dimensional commuting projector codes obey the bound $kd^2 \le c n$~\cite{Bravyi10}, where $k$ is the number of encoded qubits, $d$ is the code distance, $n$ is the number of physical qubits in the code and $c$ is a constant to maximize.
	By considering color codes where we place a twist defect in the center of the lattice, we find a code where $c$ converges towards $4$ as the rotational symmetry of the code increases.
	In particular we present two codes with bounded stabilizer weights $w$, for $w \le 8$ we find a code with $c = 4$, and for $w \le 6$ a code with $c = 8/3 \approx 2.7$.
	In contrast, the twisted surface code~\cite{Yoder16} has a value $c = 4/3$ and the standard color code has a value $c = 2$ ($c = 4/3$) when $w \le 8$ ($w \le 6$).
	The standard surface code has a value $c = 1$ for a lattice that is appropriately oriented~\cite{Wen03, Hastings15}, or when punctures are arranged in a suitable pattern~\cite{Delfosse16b}.
	With our simple example we see that there may be many other benefits to be gleaned by exploring all of the twists available with the color code.
	We present our new family of stellated color codes, as well as our new four-qubit error-detecting code, in self-contained sections that can be understood with only prerequisite knowledge of the stabilizer formalism for the reader interested only in the quantum information applications of our work.

	Finally, we study how the new objects we introduce to the color code lattice can be represented on multiple copies of other topological phases through unfolding~\cite{Bombin12, Kubica15, Bhagoji15, Criger16, Bridgeman17, Zhu17}.
	In particular, in addition to representing the new objects we discover on two copies of the surface code, we also explore a lesser known~\cite{Wang17} connection between the color code and two copies of the three-fermion model~\cite{Rowell09}.
	We find this connection to give a very natural presentation of the underlying structure of the color code model.
	Developing the connections between the color code model and copies of other, simpler phases may help to discover new and improved schemes for topological quantum computation.
	Examples of which have been demonstrated, for instance in Refs.~\cite{Moussa16, Zhu17}.
	Similar mappings have also provided new ways of writing decoding algorithms for the color code~\cite{Bombin12}.

	The remainder of the manuscript is structured as follows.
	In Sec.~\ref{sec:ReviewChapter} we review the general theory of anyon models and in Sec.~\ref{sec:Review2DColorCode} we review the stabilizer formalism, the color code and the low-energy excitations of this model.
	In Sec.~\ref{sec:Boundaries} we examine the theory of boundaries in abelian topological phases and discover new boundaries with the color code.
	Using these new boundaries we discuss improved methods of deforming color code punctures and a new four-qubit error-detecting code in Sec.~\ref{sec:ApplicationsPauliBdry}.
	In Sec.~\ref{sec:Domainwalls} we present the symmetries of the color code model and introduce domain walls to the lattice that map excitations non-trivially while preserving the basic data of the particle model.
	Then, in Sec.~\ref{sec:CCTwists} we extend the discussion of domain walls by terminating lattice dislocations to introduce the twists of the color code.
	Through the study of twists we find new two-dimensional codes with high encoding rates, discussed in Sec.~\ref{sec:TwistsApplications}.
	Lastly, in Sec.~\ref{sec:2xTC} we study the connection between the color code and two copies of the surface code model to show how color code defects are represented in this simpler model.
	We finally conclude by proposing new directions of research that make use of the toolbox we have elucidated in Sec.~\ref{sec:SummaryAndOutlook}.

	We also give supplemental information on the color code in the Appendices.
	In App.~\ref{app:CCFermions} we discuss the six fermionic excitations of the color code, and in App.~\ref{app:ThreeFermionModel} we discuss the connection between the color code and two copies of the three-fermion model.
	In App.~\ref{app:PachnerMoves} we discuss the dual color-code lattice, and consider systematically introducing certain twists to the model with Pachner moves.
	Finally, in App.~\ref{app:StellatedSurfaceCodes} we present a new family of surface codes, namely, the stellated surface codes, which are related to the new color codes we present.

\section{Anyon models}
	\label{sec:ReviewChapter}

	We require notation to describe the anyonic low-energy excitations of the model of interest without explicitly referring to the underlying system.
	Here we review topological quantum field theory which describes anyonic particles while remaining agnostic of microscopic details.
	For a more detailed and complete discussion of the theory of anyons see Refs.~\cite{Preskill17lectures, Nayak08, Pachos12, Barkeshli14} or the Appendix E of Ref.~\cite{Kitaev06}.

		\label{sec:ReviewAnyonModels}

		The low-energy excitations of a topological phase of matter are described by an anyon model.
		An anyon model is a set of excitations together with additional data that describes interactions between these excitations.
		We write a set of labels $a,b,c,\dots \in \cC$ to denote the distinct charges of the system.
		Included in all anyon models is the vacuum charge, $1 \in \cC$.
		The vacuum is the label that indicates `no charge'.
		
		Fusion describes the combination of two particles $a,\,b \in \cC$.
		The fusion product of two particles $a$ and $b$ is denoted with the binary operation $a \times b$.
		The fusion rules of an anyon model are described by the fusion multiplicities $N^c_{a,b}$, such that
		\begin{align}
			a \times b = \sum_c N^c_{a,b} ~ c,
			\label{eq:FusionGeneral}
		\end{align}
		where $c \in \cC$.
		Each anyon $a$ has a unique antiparticle labeled $\overline{a}$ which can fuse with it to the vacuum charge, $N_{a,\overline{a}}^1 =1$.

		Anyons have a Hilbert space associated with them known as a fusion space.
		It is possible for the dimension of the fusion space to increase as anyons are added to the system.
		The factor with which the fusion space increases for an anyon $a$ is called its quantum dimension, denoted $d_a \ge 1$, which is determined by the equation
		\begin{align}
			 d_a^2 ={\sum_{c \in \cC} d_c N_{a,c}^{a}},
			\label{eq:GeneralQuantumDimensionAnyons}
		\end{align}
		where $d_1=1$.
		Anyon $a$ is abelian if $d_a = 1$ and is non-abelian if $d_a > 1$.
		An anyon model is abelian if it contains only abelian anyons.
		In fact, for all abelian anyon models, fusion outcomes are deterministic such that for each $a, b$ there exists a unique label $c$ such that $N_{a,b}^c = 1$, and $N_{a,b}^d = 0$ for all $d \neq c$.

		Lastly, we address the exchange and braid statistics of anyons.
		Exchanging two anyons of charge $a$ results in a phase $\theta_{a}$, called the topological spin.
		Diagrammatically we express this with the equation
		\begin{align}
			\theta_a ~ \trivialbraid{a}{a} = \braid{a}{a},
			\label{eq:SelfExchangeGeneral}
		\end{align}
		where the lines indicate the trajectories of particles.
		Note that $\theta_{a} = \theta_{\overline{a}}$.
		We call $a$ bosonic if it exchanges trivially with itself, i.e. $\theta_{a} = +1$, and fermionic if $\theta_{a} = -1$.

		Braiding two abelian anyons $a$ and $b$ clockwise results in a phase $M_{a,b}$, called the monodromy.
		We can similarly express this operation with a diagrammatic equation
		\begin{align}
			M_{a,b} ~ \fulltrivialbraid{a}{b} = \fullbraid{a}{b}.
			\label{eq:BraidingGeneral}
		\end{align}
		This equation can be transformed by adding a braid to the two sides of Eq.~\eqref{eq:BraidingGeneral} to obtain
		\begin{align}
			M_{a,b} \antibraid{a}{b} = \braid{a}{b},
			\label{eq:BraidingChangeOrder}
		\end{align}
		an expression which will provide useful to calculate the monodromy of anyons in the considered model.
		If $M_{a,b} = +1$, we say $a$ and $b$ braid trivially.

		Lastly, we can derive the data involving anyons if we know enough about the particles into which we can decompose it.
		If an anyon $c$ is the only fusion product of $a$ and $b$, as is the case in abelian anyon models, then its self-exchange statistic is
		\begin{align}
			\theta_c = \theta_a \theta_b M_{a,b},
			\label{eq:CompositeSelfExchange}
		\end{align}
		and the monodromy with another anyon $d \in \cC$ is given by
		\begin{align}
			M_{c,d} = M_{a,d} M_{b,d}.
			\label{eq:CompositeBraiding}
		\end{align}

\section{The two-dimensional color code}
	\label{sec:Review2DColorCode}

	We next review the two-dimensional color code~\cite{Bombin06}, and summarize how the microscopic description gives rise to its corresponding anyon model.
	Refs.~\cite{Landahl11, Bombin15, Kubica15a, Watson15} provide alternative reviews of the model.

	\subsection{The stabilizer formalism}
		\label{sec:ReviewStaibilizerCodes}

		A useful tool to describe a large class of quantum error-correcting codes and many-body quantum states is the stabilizer formalism~\cite{Gottesman97}.
		Stabilizer codes are specified by their stabilizer group, $\cS$.
		This is an abelian subgroup of the Pauli group acting on $n$ qubits.
		We denote the generators of the Pauli group $X_j$ and $Z_j$, where indices $1 \le j \le n$ denote individual spin-half particles, or qubits, of the system.

		The stabilizer group defines a code subspace spanned by basis  vectors $\ket{\psi}$.
		The code subspace, or code-space for short, is the common $+1$ eigenspace of all the stabilizers of the stabilizer group.
		Specifically, we have
		\begin{align}
			s \ket{\psi} = \ket{\psi}, 
		\end{align}
		for all $s \in \cS$.
		The additional requirement that $-\one\not\in \cS$ guarantees that the stabilized subspace is non-empty.
		The dimension of the code-space is $2^k$, where $k = n - m$, is the difference between the number of physical qubits $n$ and the number of independent generators of the stabilizer group  $m$.
		When interpreting the stabilized subspace as a logical code-space, $k$ is the number of logical qubits it supports.

		Pauli operators which commute with the stabilizer group but are not included in it are logical operators.
		More formally, the set of logical operators is such that $\mathcal{L} \defeq \mathcal{N}(\cS) \backslash \cS$, where $\mathcal{N}(\cS)$ is the normalizer of $\cS$.
		Operators in $\mathcal{L}$ preserve the code-space yet act non-trivially on it.
		The weight $w(P)$ of a Pauli operator is the number of qubits upon which $P$ acts non-trivially.
		It allows defining the code distance, $d \defeq \min_{P\in\mathcal{L}}w(P)$, which is commonly used as a first order approximation to quantify the degree of protection provided by the code-space.

	\subsection{The color code lattice and stabilizers}
		\label{sec:ReviewCCStabilizers}

\begin{figure}
			\centering
			\includegraphics[width=.9\linewidth]{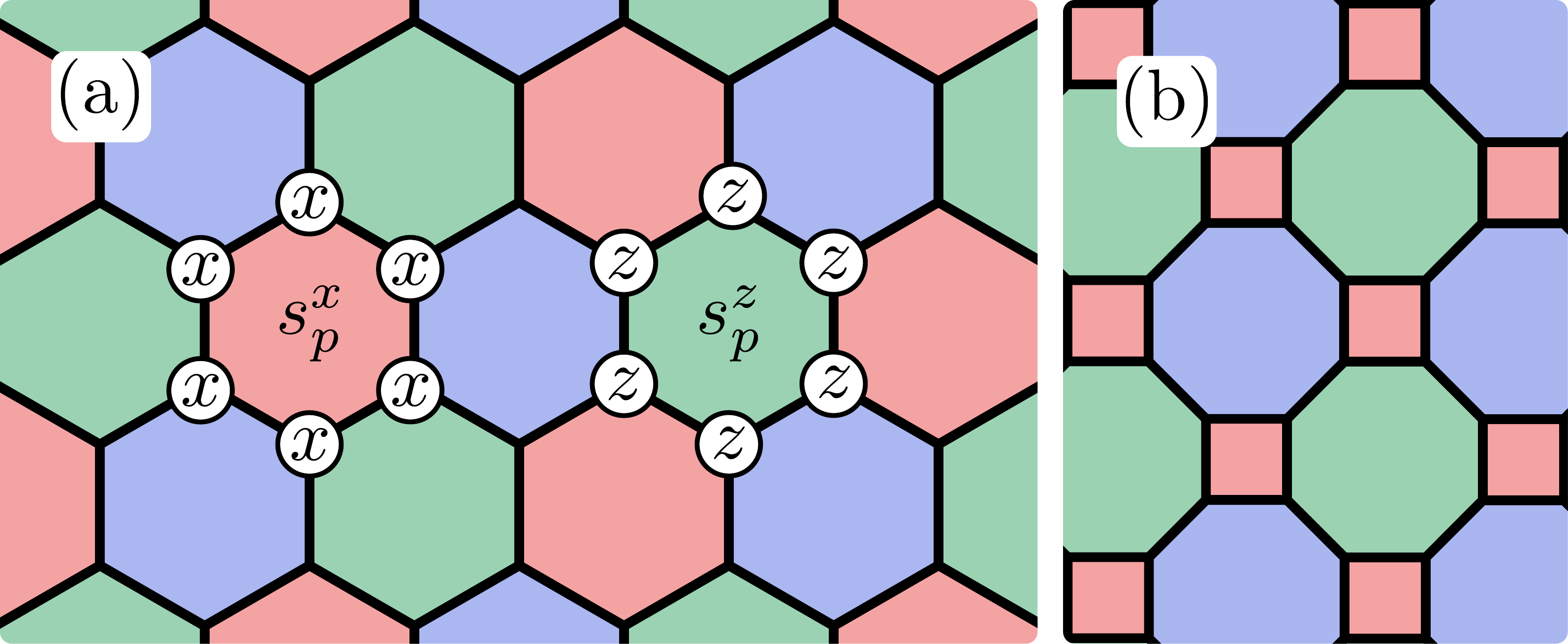}
			\caption{
				\textbf{(a)} the 6.6.6 and \textbf{(b)} the 4.8.8 color code lattice.
				Physical qubits live on the vertices.
				Each plaquette $p$ hosts two stabilizers, one measuring the $Z$ parity of the qubits on the boundary of $p$, and the other measures the $X$ parity.
				Examples of stabilizers are shown in panel \textbf{(a)}.
			}
			\label{fig:CCLatticesAndStabilizers}
		\end{figure}

		The color code is defined on a trivalent three-colorable lattice, regular examples of which are shown in Fig.~\ref{fig:CCLatticesAndStabilizers}.
		Specifically, each vertex of the lattice has exactly three incident edges and each plaquette, indexed $p$, can be assigned one of three colors, red, $r$; green, $g$; or blue, $b$, such that no two plaquettes of the same color touch.

		With the lattice defined we consider a system where a qubit is placed on each vertex of the lattice.
		The stabilizer group of the color code is then generated by operators associated to the plaquettes of the lattice.
		Specifically we have two stabilizer generators,
		\begin{align}
			s^x_p = \prod_{j \in \partial p} X_j
			\qquad \text{and} \qquad
			s^z_p = \prod_{j \in \partial p} Z_j,
			\label{eq:CCStabilizers}
		\end{align}
		for each plaquette $p$, where the product is taken over all the qubits that lie on the boundary of $p$.
		We show examples of stabilizers on the lattice in Fig.~\ref{fig:CCLatticesAndStabilizers}.

	\subsection{Color code anyons}
		\label{sec:ReviewCCAnyons}

		We have claimed that the color code model gives rise to anyonic quasiparticle excitations.
		However, we have thus far only introduced the color code as a code-space of some larger physical Hilbert space using the stabilizer formalism.
		In fact, we have also implicitly defined a Hamiltonian system whose low-energy excitations are anyons~\cite{Kitaev03, Bombin06}.
		The code-space of the model is also the ground state subspace of the exactly solvable Hamiltonian $H$, whose interaction terms are stabilizer generators, which for the color code is
		\begin{align}
			H = - \sum_p s^z_p - \sum_p s^x_p.
			\label{eq:CCHamiltonain}
		\end{align}
		Applying Pauli matrices, $P$, to states in the code-space yields excited eigenstates of $H$.
		Plaquettes where at least one stabilizer generator is violated by, i.e., anti-commutes with, $P$ are then interpreted as quasi-particle excitations, namely, anyons.
		The type of anyonic quasi-particle is determined by which stabilizer(s) are violated on a given plaquette, and the color of the plaquette as we will describe in more detail below.
		For a detailed discussion on color code excitations, see Ref.~\cite{Bombin06} and for an in depth discussion on the connection between the stabilizer code and commuting Pauli Hamiltonians see Ref.~\cite{Brown16}.
		Given the connection between the Hamiltonian presented in Eq.~\eqref{eq:CCHamiltonain}, we will continue the discussion in terms of stabilizer states as we defined them previously.
 
		We can characterize the sixteen anyonic charges of the color code by the operators that create them.
		The operators that create excitations are Pauli operators with string-like support.
		Importantly, string operators commute with all the plaquette operators except for those at their endpoints.
		As such, anyonic excitations are created at the endpoints of string operators where plaquette operators are violated.

		To study the sixteen color code anyons, we first focus on a subset of nine charges.
		This subset is an over-complete generating set, in that it can generate any charge from the color code anyon model through fusion.
		Even though four generators would also suffice, we find this generating set particularly convenient to characterize the the model.
		The nine charges in this set can be denoted by two labels which are a Pauli label, $ P =x,\,y,\,z$, and a color label, $C = r,\,g,\,b$.
		Given that such excitation is specified by one of three colors and one of three Pauli labels we arrive at nine anyons which are labeled as $C P$.
		Shortly, it will become convenient to arrange the generating charges in a $3 {\times} 3$ grid as in Tab.~\ref{tab:CC3x3anyons} where color labels are arranged in columns, and Pauli labels in rows.
		\begin{table}
			\centering
			\begin{tikzpicture}
				\draw (-.8,.2) -- (.8,.2);
				\draw (-.8,-.2) -- (.8,-.2);
				\draw (-.25,-.5) -- (-.25,.5);
				\draw (.25,-.5) -- (.25,.5);

				\node (anyons) at (0,0) {
					$\begin{matrix}
						rx & gx & bx \\
						ry & gy & by \\
						rz & gz & bz
					\end{matrix}$
				};
			\end{tikzpicture}
			\caption{
				The nine generating anyons in the color code can be labeled using a color label and a Pauli label.
				The three colors, $r$, $g$, and $b$ are give in the columns, the Pauli labels $x$, $y$, and $z$ are associated with rows.
				This arrangement of anyons will prove useful in what follows, as it succinctly conveys braid statistics and fusion rules of the color code anyon model.
			}
			\label{tab:CC3x3anyons}
		\end{table}

		The excitations we create on the lattice are determined by the string-operator used to create the excitation.
		The color label is determined by the color of the violated stabilizer at the endpoint of a string and the Pauli label corresponds to the Pauli operator which was applied to create the excitation.
		For instance, in Fig.~\ref{fig:CCAnyonsFusion}~(a) we show a string operator that creates two $gx$ excitations at its endpoints.
		This is because the string is composed of Pauli-X operators, and the operator shown in the figure violates the two green $s^Z_p$ plaquettes at its endpoints.
		One can readily check that the other stabilizers on the lattice commute with this operator.

	\subsection{Fusion}
		\label{sec:ReviewCCFusion}

\begin{figure}[b!]
			\centering
			\includegraphics[width=0.9\linewidth]{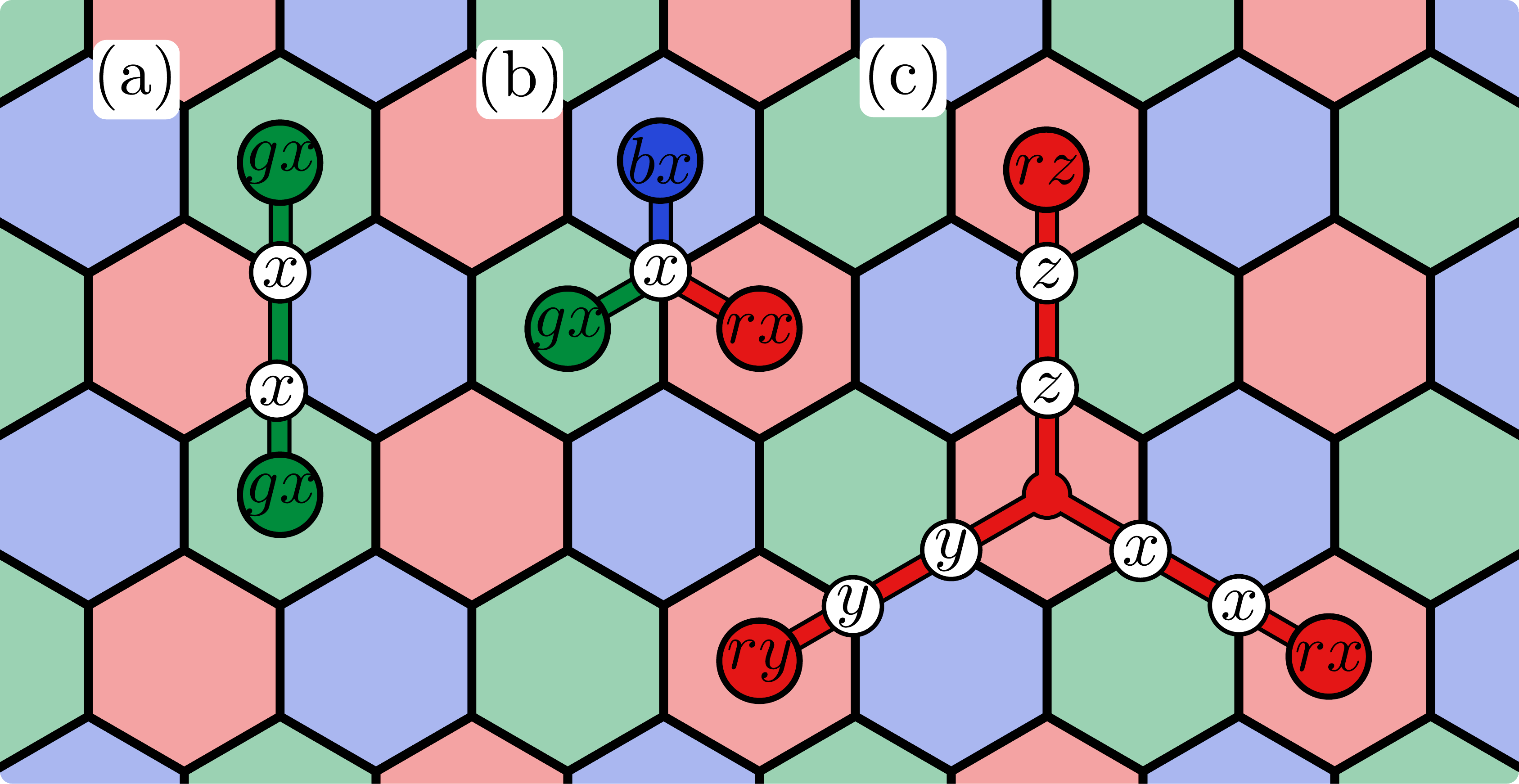}
			\caption{
				The fusion rules for color code anyons.
				\textbf{(a)} Each color code anyon is its own antiparticle.
				Here, as an example, an operator is shown which annihilates (or creates) a pair of $gx$ anyons.
				Corresponding operators can be found for pairs of all color code anyons.
				\textbf{(b)} Three anyons sharing the Pauli label but each with a different color label annihilate.
				This can also be read as a fusion (or decay) process:
				From bottom to top for example, it corresponds to the fusion of a red and a green anyon to form a blue anyon.
				Notice, the Pauli label $x$ was chosen arbitrarily; the same holds for $y$ or $z$.
				\textbf{(c)} Three anyons sharing the same color label but each with one of the three Pauli labels can also be annihilated.
				Again, and alternative reading is the fusion of an $x$ anyon and a $y$ anyon to form a $z$ anyon.
				The red color label can be replaced by either green or blue.
			}
			\label{fig:CCAnyonsFusion}
		\end{figure}

		We now offer a brief intuition on how the fusion rules arise from the Hamiltonian of the color code, Eq.~\eqref{eq:CCHamiltonain}.
		A more detailed discussion is given in Ref.~\cite{Bombin06}.
		Following the example shown in Fig.~\ref{fig:CCAnyonsFusion}~(a) we have already seen that two anyons of the same type are created at the endpoints of the string.
		In fact, we can extend the string by applying Pauli-X rotations on the pairs of qubits on the green edges adjacent to the plaquettes that support the excitations.
		Green edges are those that connect two green plaquettes.
	
		In general, it holds that pairs of excitations of a given color are created by applying strings of the appropriate Pauli rotations to edges of its corresponding color~\cite{Bombin06}.
		Given that we can separate pairs of excitations of the same type $CP$ by extending a string of Pauli operators $P$ type via edges of color $C$, we arrive at the fusion rule
		\begin{align}
			C P \times C P  = 1.
			\label{eq:CCFusionSameCandP}
		\end{align}
		Physically, this fusion rule represents $\bZ_2$ charge conservation as it shows that anyons of a given type must be created in pairs.
		This rule can also be interpreted to say that color code anyons are their own anti-particles.

		We can similarly study string operators acting on the color code lattice to determine fusion rules between anyons with different color and Pauli labels.
		In Fig.~\ref{fig:CCAnyonsFusion}~(b) a Pauli-X operator is applied to a single site.
		This operator violates stabilizers on its three adjacent plaquettes, each of which has a different color due to the three-colorability of the lattice.
		The excitations created by this operator can be separated arbitrarily far from the other excitations with the application of another appropriately chosen string.
		We thus arrive at the fusion rule
		\begin{align}
			C_1 P \times C_2 P = C_3 P,
			\label{eq:CCFusionSameP}
		\end{align}
		for any Pauli label $P$ and pairwise distinct color labels $C_1, C_2, C_3$.
		We thus observe a conservation law between the color labels of the color code excitations:
		Two charges of a common Pauli-type and distinct colors fuse to give the charge of the same Pauli type and the third color.

		Finally, we can also observe a fusion rule between different Pauli labels.
		In Fig.~\ref{fig:CCAnyonsFusion}~(c) we show a string operator composed of Pauli-X, Pauli-Y and Pauli-Z operators with three red endpoints where stabilizers are violated.
		At two of the endpoints we have a single violated $s^x_p$ ($s^z_p$) stabilizer which indicates an $rz$ ($rx$) excitation.
		At the third endpoint both stabilizer generators are violated, indicating an $ry$ excitation at this location.
		Generalizing, we obtain the fusion rule that connects Pauli-labels of the excitations
		\begin{align}
			C P_1 \times C P_2 = C P_3,
			\label{eq:CCFusionSameC}
		\end{align}
		for arbitrary color $C$ and pairwise distinct Pauli labels $P_1, P_2, P_3$.
		The reader may recognize a duality between the color label and the Pauli label in Eq.~\eqref{eq:CCFusionSameP} and Eq.~\eqref{eq:CCFusionSameC}.
		Indeed, this duality is a key feature of the color code anyon model as will become clear throughout the manuscript.

		With the fusion rules we have developed so far we begin to see the utility of Tab.~\ref{tab:CC3x3anyons}.
		Indeed, the table echos Eq.~\eqref{eq:CCFusionSameP} as the fusion product of two distinct excitations from the same row give the third excitation of the given row.
		Similarly, the fusion product of two excitations of different Pauli label from a common column fuse to give the third excitation from the given column.
		This reflects the fusion rule shown in Eq.~\eqref{eq:CCFusionSameC}.
		More trivially, we have that the fusion product of two excitations from the same element of the table fuse to give the vacuum excitation as in Eq.~\eqref{eq:CCFusionSameCandP}.

		With the rows and columns of Tab.~\ref{tab:CC3x3anyons} now separately understood, it remains to consider the fusion product of pairs of charges that share neither a row nor a column.
		As we will show in the following subsection, the nine excitations in Tab.~\ref{tab:CC3x3anyons} are in fact bosonic excitations.
		Aside from the vacuum particle, the excitations of the color code that are not shown in Tab.~\ref{tab:CC3x3anyons} are fermionic excitations.
		All of the remaining excitations can be expressed as the fusion product of two of elements.
		In fact, the color code gives rise to six distinct fermions.
		For more details on the fermionic excitations of the color code we refer the reader to App.~\ref{app:CCFermions}.
		Moreover, in App.~\ref{app:ThreeFermionModel} we show that we can describe the excitations of the color code using two non-interacting copies of a model of three fermions.

	\subsection{Self-exchange and braiding}
		\label{sec:ReviewCCBraiding}

		\begin{figure}[b!]
			\centering
			\includegraphics[width=0.9\linewidth]{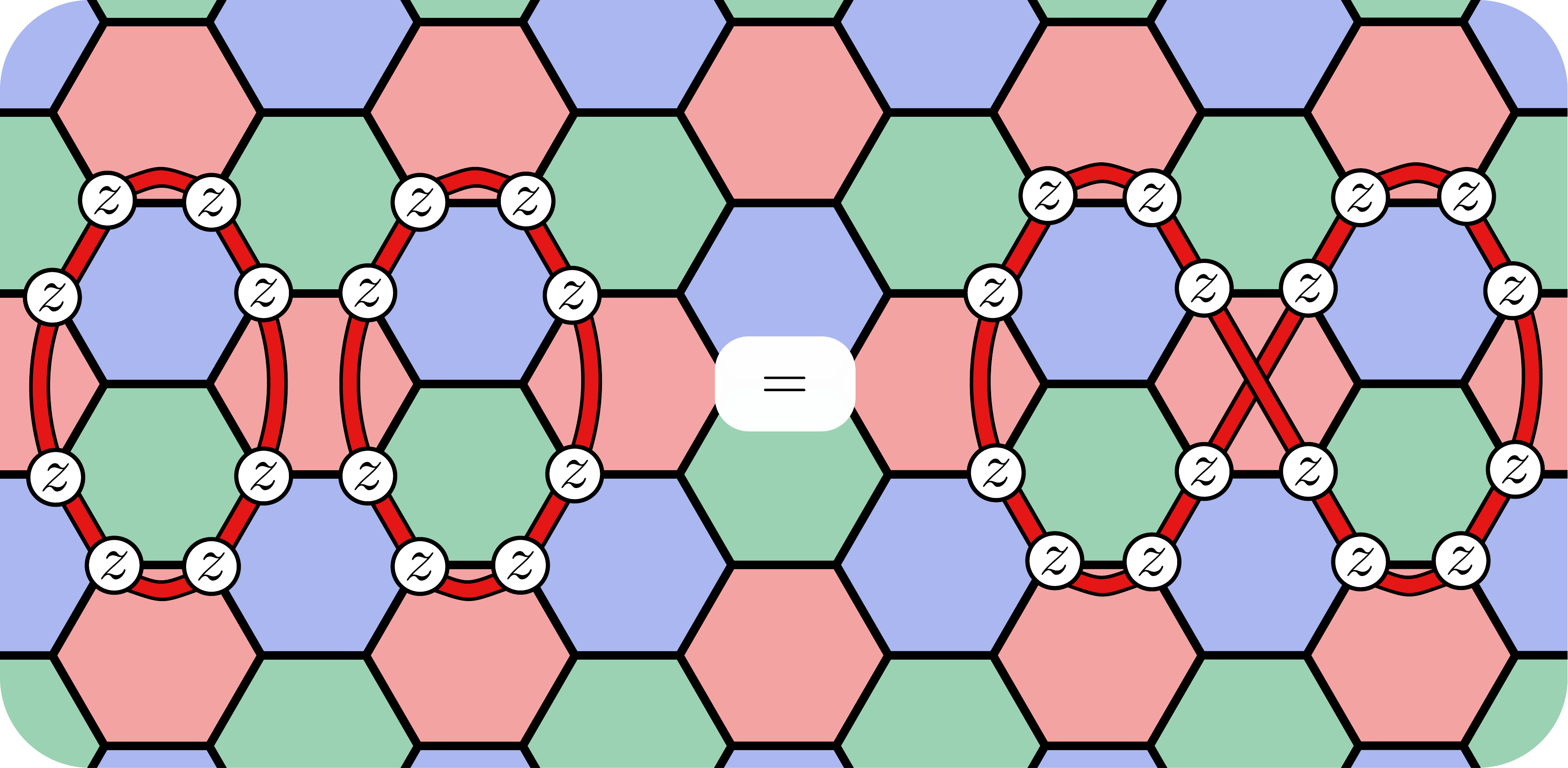}
			\caption{
				Bosons in the color code have trivial spin $\theta_a = +1$.
				As an example compare string operators which move an $rz$ anyon along a closed loop.
				On the left, two circular paths looping back to their initial positions are shown.
				Each of these operators is the product of the $z$ type stabilizers on the blue and green plaquettes they enclose and hence itself again a stabilizer.
				On the right however, the middle section is changed, such that path crosses itself once in the middle to form a single larger loop.
				However, this redrawing of the path did not change how the operators act, and the left-hand side is exactly equal to the right-hand side.
				Further more, according to Eq.~\eqref{eq:SelfExchangeGeneral}, the difference between these operators gives us the spin of the moved anyon, here $\theta_{rz} = +1$.
				The same holds for all nine anyons shown in Tab.~\ref{tab:CC3x3anyons}.
			}
			\label{fig:CCAnyonsSelfExchange}
		\end{figure}
		We referred to color code anyons as either bosonic or fermionic.
		To justify these names we now derive their self-exchange statistics.
		Consider Fig.~\ref{fig:CCAnyonsSelfExchange}, where we compare two sets of string operators which move $rz$ anyons along closed loops.
		On the left, two separate loops are shown.
		This is the left-hand side of Eq.~\eqref{eq:CompositeSelfExchange}, only that the bottom-left(right) and top-left(right) ends of the trajectories are here connected.
		On the right-hand side of Fig.~\ref{fig:CCAnyonsSelfExchange}, a single closed loop of an $rz$ operator which crosses itself in the middle is depicted.
		This is the right-hand side of Eq.~\eqref{eq:CompositeSelfExchange}, again with the bottom-left(right) end of the trajectories connected to the top left(right).
		Hence, the difference between the left and right half of the figure corresponds to the self-exchange statistic of $rz$.
		But clearly, the effect of the shown operators is exactly the same and hence $\theta_{rz} = +1$, $rz$ is a boson.
		This is true for all nine non-trivial bosons, as can be seen by changing the Pauli basis in which the considered operators act, or by moving the string operators such that they move anyons of different colors.
		The self-exchange of the color code fermions can be derived using Eq.~\eqref{eq:CompositeSelfExchange} and Eq.~\eqref{eq:CCBraiding} given below together with any decomposition of fermions into two bosons shown in Eq.~\eqref{eq:CCFermions} and Tab.~\ref{tab:3x3CCfermions} in App.~\ref{app:CCFermions}.

		Finally, we look at the phase obtained by braiding pairs of color code excitations.
		Braiding a bosonic anyon $a$ with color label $C_a$ and Pauli label $P_a$ with a second boson $b$ with labels $C_b$ and $P_b$ results in the phase
		\begin{align}
			M_{a,b} = (-1)^{\delta(C_a \not= C_b) \delta(P_a \not= P_b)}
			\label{eq:CCBraiding}
		\end{align}
		where $\delta(\mathtt{statement}) $ is an indicator function that returns $1$ if $\mathtt{statement}$ is true and $0$ otherwise.

		\begin{figure}
			\centering
			\includegraphics[width=0.85\linewidth]{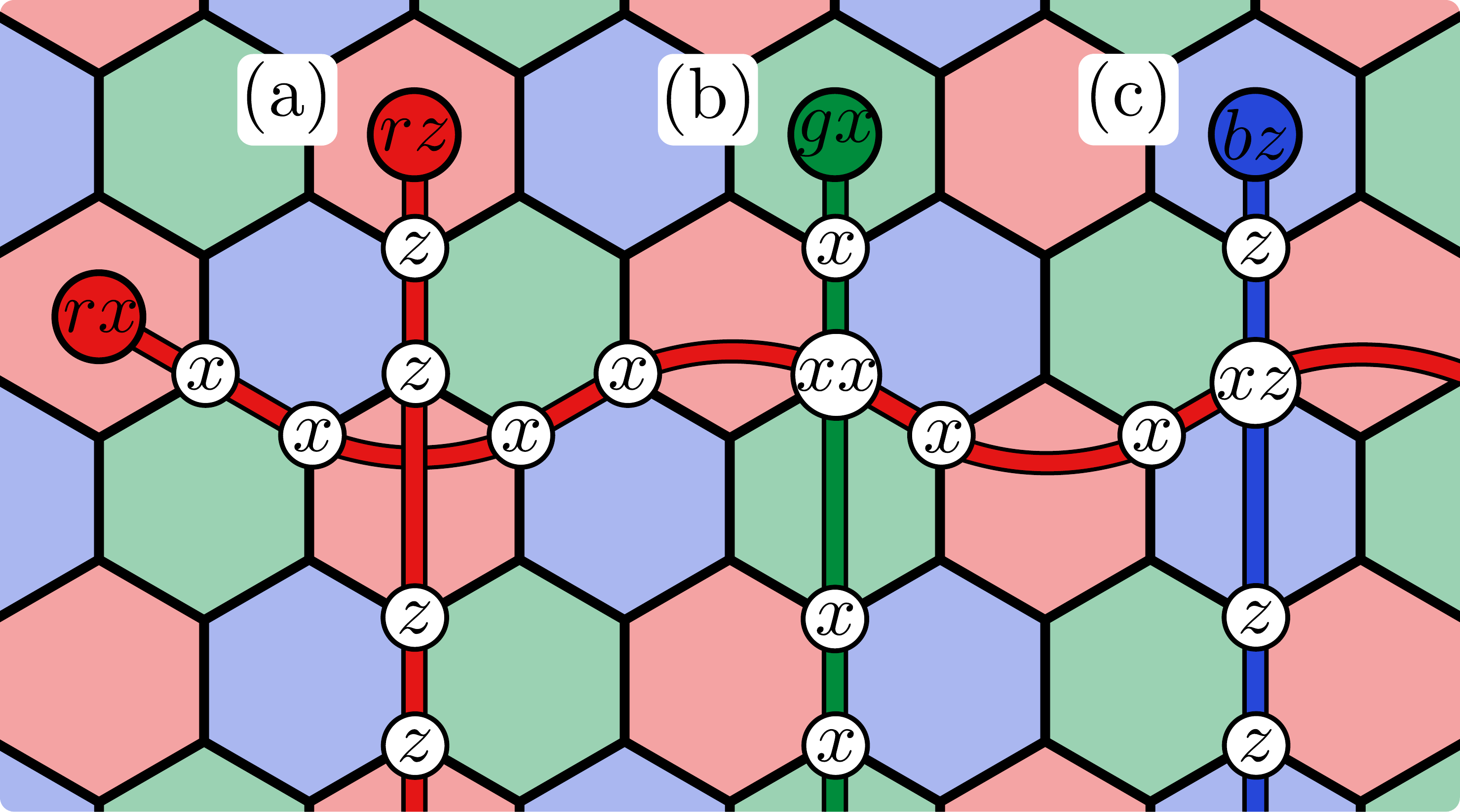}
			\caption{
				To obtain the braid statistics of the color code anyons, we study the commutators of string operators.
				As an illustrative example, an $rx$ anyon is moved from the left to the right out of frame.
				Then, the line along which said anyon was moved is crossed top to bottom by three other anyons.
				Comparing the two different orders in which these operators are applied reveals the phase obtained by braiding the anyons, see Eq.~\eqref{eq:BraidingChangeOrder}.
				In \textbf{(a)} we see that strings moving anyons sharing the color label (red in our example) exchange trivially.
				Here, not a single qubit is acted upon by both strings.
				In general the number of qubits supporting two crossing strings of the same color is even.
				\textbf{(b)} Crossing strings moving anyons with the same Pauli label also exchange trivially, as strings moving them act in the same Pauli basis.
				\textbf{(c)} shows the third case, neither the Pauli label nor the color label of the two anyons associated with the crossing string operators match.
				This leads to an odd number of qubits on which the strings act in anti-commuting bases.
				In our example one qubit is acted on first by an $X$ and then by $Z$.
				This leads to a phase of $-1$ if the anyons are braided.
			}
			\label{fig:CCAnyonsBraiding}
		\end{figure}

		Like for the fusion rules, we find Tab.~\ref{tab:CC3x3anyons} proves a useful tool to visualize the braid statistics between bosonic excitations.
		This is to say, anyons $a$ and $b$ which share a row or a column braid trivially, i.e. $M_{a,b} = +1$.
		If $a$ and $b$ share neither a row nor a column however, we have that $M_{a,b} = -1$.
		To prove the above, we compare the order in which the crossing operators moving anyons $a$ and $b$ are applied.
		As in Eq.~\eqref{eq:BraidingChangeOrder}, the difference reveals the phase $M_{a,b}$.
		As can be seen in Fig.~\ref{fig:CCAnyonsBraiding}, pairs of string operators associated to anyons with the same color label or Pauli label exchange trivially, and thus we find $M_{a,b}=+1$.
		When both, the color label and the Pauli label is different however, the operators anticommute and we get $M_{a,b}=-1$.
		The braiding rules involving fermions can be derived from the above, using Eq.~\eqref{eq:CompositeBraiding} and any decomposition of fermions into two bosons shown in Eq.~\eqref{eq:CCFermions} and Tab.~\ref{tab:3x3CCfermions} in App.~\ref{app:CCFermions}.

\section{Boundaries}
	\label{sec:Boundaries}

	Unless we are dealing with a closed 2D manifold, the spatial extent of a topological phase must end at some boundary, such as a gapped domain wall with the trivial phase.
	It is important to gain a better grasp on the physics at the boundary of a topological phase.
	Not only is this interesting from a condensed matter perspective~\cite{Beigi11, Kitaev12, Levin13, Barkeshli13a, Cong17a, Burnell17}, but can be exploited in quantum computation~\cite{Raussendorf06, Bombin09, Fowler11, Cong16, Cong17, Brown17}.
	In this section, after reviewing concepts of the boundaries of abelian topological codes, we introduce lattice realizations of three new color code boundaries and show that this completes the boundaries available to the model.
	We finally discuss the prospective applications of these boundaries for quantum computational tasks as well as small scale experimental implementations of quantum error-correcting codes.
	
	\subsection{Lagrangian subgroups}
		\label{sec:GeneralLagrSubgroups}

		Gapped boundaries are characterized by Lagrangian subgroups.
		These mathematical objects are subsets of anyons in the anyon model $\cC$ which can condense at the boundary of interest.
		Anyon condensation is the process wherein an anyon meets and is absorbed by the boundary.
		The reverse process is also possible where a single excitation is created or `emitted' by a boundary.
		Anyons which do not condense are said to be confined by the boundary.
		Boundaries that can condense excitations are important for quantum information applications as these are topologically protected degrees of freedom whose state is determined by the charges they have absorbed.

		The subset of anyons that can condense at a boundary is constrained to obey certain conditions that we will now briefly review.
		Indeed, it is shown in Refs.~\cite{Levin13, Barkeshli13a} that the subsets of excitations in $\cC$ that are absorbed at gapped boundaries correspond to Lagrangian subgroups $\cM$.
		A Lagrangian subgroup fulfills three conditions:
		Firstly, as one might expect, a Lagrangian subgroup is a group, and thus closed under fusion.
		Further, if a particle can condense at a given boundary, so can its antiparticle.
		Likewise, the vacuum particle can be condensed at any gapped boundary.
		Secondly, all of the anyons within the subgroup are bosons and braid trivially with all other anyons in the subgroup.
		The third condition is that $\cM$ is maximal in the sense that no additional anyon can be added to the subgroup such that the other conditions are fulfilled.
		In the following, we revisit the argument given in Sec.~IV of Ref.~\cite{Levin13}.

		\begin{figure}[!bt]
			\centering
			\includegraphics[width=0.55\linewidth]{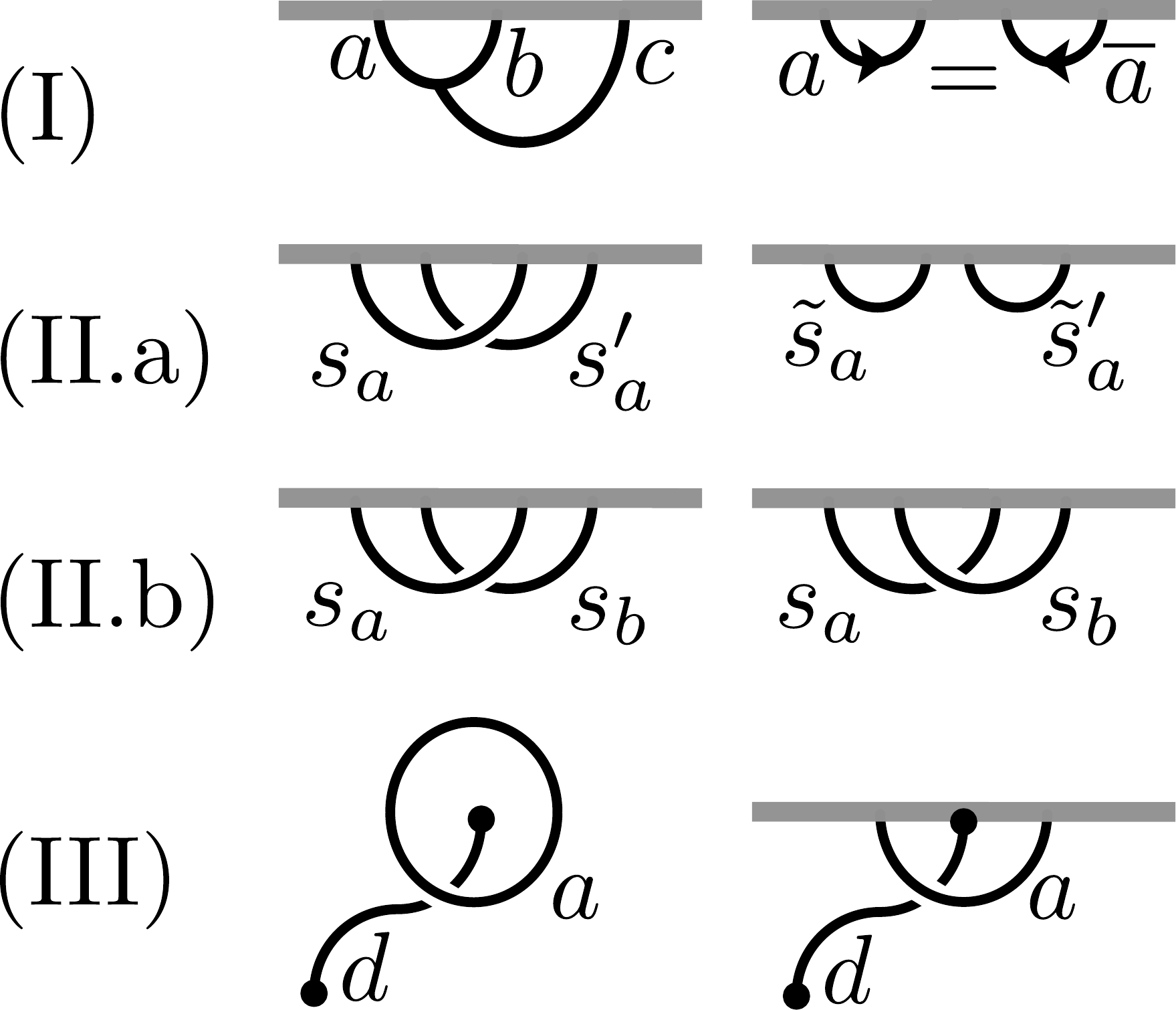}
			\caption{
				The conditions on Lagrangian subgroups pictorially summarized.
				The gray line represents the boundary between the topological phase at the bottom and the trivial phase above.
				\textbf{(I)} If two anyons $a$ and $b$ can condense, so can the outcome of their fusion $c = a \times b$.
				And the emission of an anyon $a$ can be read as the condensation of its antiparticle $\overline{a}$.
				\textbf{(II.a)} Anyons which can condense must have bosonic self-exchange statistics.
				The shown paths are stabilizers, hence the left and the right situation are equal, which implies trivial self-exchange for condensible anyons.
				\textbf{(II.b)} Two anyons which can condense must braid trivially.
				Again, the depicted paths are stabilizers and must hence commute, which implies the trivial braiding of the anyons.
				\textbf{(III)} An anyon $d$ which can not condense must exchange non-trivially with at least one anyon which can condense.
				This means the set of condensible anyons is maximal.
				Sets of anyons which fulfill these conditions are called Lagrangian subgroups of the anyon model.
			}
			\label{fig:ConditionsLagrSubgroup}
		\end{figure}

		First, we start by showing that $\cM$ is closed under fusion, $a \in \cM ~ \text{and} ~ b \in \cM \implies a \times b \in \cM$.
		$a \times b$ can condense by decaying into $a$ and $b$, which can then condense separately.
		These is pictorially shown on the left of Fig.~\ref{fig:ConditionsLagrSubgroup}~(I).
		Before showing that $\cM$ does actually form a group, we first turn out attention to the other claims.

		The second requirement of the boundary can be understood by considering string operators $s_a$.
		These operators act such that they pull an anyon $a \in \cM$ from the boundary, transport it along a trivial trajectory through the bulk and condense it back at the same boundary such that $s_a$ leaves the charge at the boundary unchanged.
		Given that the string operators follow a trivial trajectory and do not change the net charge of the boundary, these operators do not affect the logical state of the system.
		Furthermore, given that such operators do not create excitations, we can regard these operators as stabilizers such that $s_a \ket{\psi} = \ket{\psi}$ for all ground state vectors $\ket{\psi}$.

		Using such operators, we can argue that all anyons which can condense must be bosons as follows:
		Consider the four operators shown in Fig.~\ref{fig:ConditionsLagrSubgroup}~(II.a).
		Comparing $s_a, \, s'_a$ to $\tilde{s}_a, \, \tilde{s}'_a$, we see the endpoints have been exchanged.
		Hence, the left and the right in Fig.~\ref{fig:ConditionsLagrSubgroup}~(II.a) differ, according to Eq.~\eqref{eq:SelfExchangeGeneral}, by a phase $\theta_a$ given by the spin of charge $a$.
		But since these four operators are all stabilizers, they act trivially and hence $\theta_a = +1$.
		This holds for all $a \in \cM$, and thus all condensible anyons are bosons.

		To show the trivial braiding between condensible bosons, consider operators $s_a$ and $s_b$, which move different anyons $a,b \in \cM$.
		Their trajectories intersect in a single point, as depicted in Fig.~\ref{fig:ConditionsLagrSubgroup}~(II.b).
		Note how the order in which they are applied is changed on the right compared to the left.
		Again, since $s_a$ and $s_b$ are stabilizers, the left and the right situation are exactly equivalent.
		And according to Eq.~\eqref{eq:BraidingChangeOrder}, comparing the two cases reveals the phase $M_{a,b}$ obtained when braiding $a$ and $b$.
		We conclude that anyons in a Lagrangian subgroup braid trivially with each other such that $M_{a,b} = 1 ~ \forall ~ a,b \in \cM$.

		The third claim about the maximality of these subgroups can be intuitively understood by requiring braiding non-degeneracy.
		This means every anyonic excitation can be measured by at least one other anyon through braiding, the only exception being charges in $\cM$.
		See the left half of Fig.~\ref{fig:ConditionsLagrSubgroup}~(III).
		This braiding non-degeneracy also holds close to a boundary, if for example anyon $d$ is condensed in or confined at the boundary.
		It can be stated as, $\forall ~ d \not\in \cM ~ \exists ~ a\in\cM ~ \text{s.t.} ~ M_{a,d} \neq +1$.
		In this case, the only anyons which can perform a full braid around $d$ are anyons which are able to condense, $a \in \cM$.
		This can be seen on the right half of Fig.~\ref{fig:ConditionsLagrSubgroup}~(III).
		Thus, an anyon close to the boundary can either be seen by at least one anyon $a \in \cM$, or is itself in $\cM$.

		Now, we can address that $\cM$ indeed forms a group.
		We already discussed that it is closed under fusion.
		From the maximality we can conclude that the trivial charge $1$ is also in $\cM$, as it braids trivially with all anyons.
		To see that for each charge $a \in \cM$ its antiparticle $\overline{a} \in \cM$ holds, we can for example use Eq.~\eqref{eq:CompositeSelfExchange}.
		Choosing $b = \overline{a}$ and $c = 1 = a \times \overline{a}$ and using $\theta_{a} = \theta_{\overline{a}}$, we find $M_{a,\overline{a}} = 1$.

	\subsection{The boundaries of the color code}
		\label{sec:CCBoundaries}

		To find all boundaries of the color code, we explore the Lagrangian subgroups of the excitations of the model.
		Given that Lagrangian subgroups consist only of bosonic excitations, we concentrate only on the bosonic excitations of the color code.
		We find Tab.~\ref{tab:CC3x3anyons} offers a particularly convenient representation of the bosons of the color code model for identifying Lagrangian subgroups.
		Precisely, subsets consisting of the trivial anyon and three bosons from a row or column constitute the Lagrangian subgroups of the color code.

		The fact that these sets of anyons are closed under fusion follows readily from the discussion in Sec.~\ref{sec:ReviewCCFusion}.
		The braiding requirement is met as well, as discussed in Sec.~\ref{sec:ReviewCCBraiding}.
		There we argued that bosons in the table braid trivially with each other if and only if they share a row or a column.
		It is readily checked that the subgroups of the rows and columns of Tab.~\ref{tab:CC3x3anyons} are maximal.
		We consider trying to add additional anyons to the Lagrangian subgroups of a row or column.
		The addition of a fermion would obviously violate the self-exchange criterion of the subgroup.
		And trying to add another boson only leaves bosons from a different column or row.
		But they braid non-trivially with two of the elements of the Lagrangian subgroup.
		We conclude that, together with the trivial charge $1$, the three rows and three columns of Tab.~\ref{tab:CC3x3anyons} correspond exactly to the six possible Lagrangian subgroups of the color code anyon model.

		Alternatively, if the anyon model in question can be decomposed into two layers of a simpler anyon model, boundaries can be found that way.
		Namely, boundaries correspond to folds between the two layers of the simpler model with its domain walls, see Sec.~\ref{sec:Domainwalls} and Ref.~\cite{Barkeshli13a}.
		The famous equivalence of the color code and two copies of the toric code is explored in Sec.~\ref{sec:2xTC}.
		A second equivalence between the color code and two copies of the three-fermion model is discussed in App.~\ref{app:ThreeFermionModel}.

		We next go on to find lattice representations of these boundaries where the respective label subsets condense.

	\subsection{Lattice representations of the color code boundaries}
		\label{sec:CCLatticeRepBoundaries}

		We have identified six Lagrangian subgroups of the color code anyon model.
		However, to the best of our knowledge, only three of these subgroups have been explored previously.
		In what follows we review the boundaries that have already been studied and we find and discuss the new boundaries that correspond to the other three Lagrangian subgroups of the color code model.

\begin{figure}
			\centering
			\includegraphics[width=.9\linewidth]{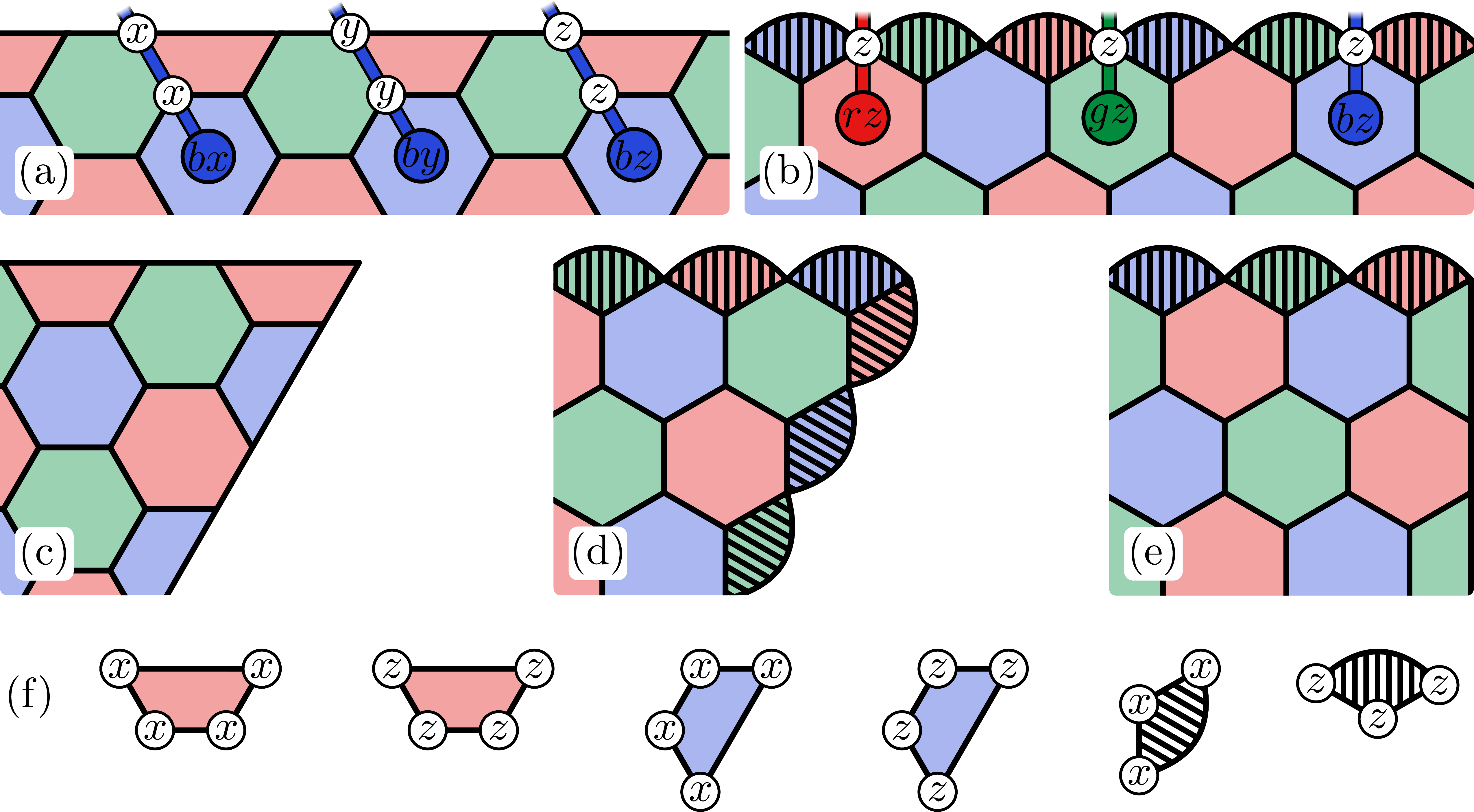}
			\caption{
				The color code boundaries and corners between them.
				\textbf{(a)} shows a blue boundary as an example of a color boundary, as well as the condensation of the three blue bosons.
				\textbf{(b)} An example of a Pauli boundary.
				Here the $z$ boundary is shown and the condensation of the three bosons with Pauli label $z$.
				\textbf{(c)} Corners between two color boundaries appear if a qubit is only part of a single plaquette of the third color.
				\textbf{(d)} A corner between two Pauli boundaries is characterized by a qubit which is only part of two striped boundary plaquettes.
				\textbf{(e)} Corners between Pauli and color boundaries are also possible, here shown a $z$ Pauli boundary meeting a blue color boundary.
				\textbf{(f)} shows the stabilizers.
				As usual, hexagons carry associated $X$- and $Z$-type weight-six stabilizer.
				The color boundaries shown have quadrilateral plaquettes which host weight-four $X$-type and $Z$-type stabilizers.
				The striped triangular plaquettes featured in Pauli boundaries are only stabilized by a single stabilizer.
				We indicate the basis in which the triangular plaquettes are measured by the direction of the stripes.
			}
			\label{fig:CCBoundaries}
		\end{figure}

		The Lagrangian subgroups that have been considered explicitly form the columns of Tab.~\ref{tab:CC3x3anyons}.
		Each of the columns share a common color label, as such, we find that we can obtain boundaries that absorb bosons with a common color label.
		We call these the red, green and blue color boundaries.
		These color boundaries were presented in the original work introducing the color code~\cite{Bombin06}.
		We show an example of a blue color boundary in Fig.~\ref{fig:CCBoundaries}~(a).
		We remark that the stabilizer generators at the boundary of the model respect the same rules presented in Sec.~\ref{sec:Review2DColorCode} where each plaquette $p$ supports two stabilizers $s^x_p = \prod_{j\in \partial p}X_j$ and $s^z_p = \prod_{j\in \partial p}Z_j$.
		The figure also shows string operators that create single blue excitations from the boundary.

		In addition to the three Lagrangian subgroups that correspond to the columns of Tab.~\ref{tab:CC3x3anyons}, we also find three Lagrangian subgroups that correspond to the rows of the table.
		We can therefore expect the existence of boundaries that absorb all of the bosonic excitations with a common Pauli label, independent of their color label.
		These boundaries will be referred to as the $x$, $y$ and $z$ Pauli boundaries.
		A similar conjecture has been made in Ref.~\cite{Kubica15} by consideration of the unfolded color code, i.e., two copies of the toric code.
		We discuss this in more detail in Sec.~\ref{sec:2xTC}.

		In Fig.~\ref{fig:CCBoundaries}~(b) we show a lattice realization of the $z$ Pauli boundary, which condenses bosons with the Pauli label $P = z$.
		At the boundary we show weight-three plaquettes.
		In principle we could choose any number of stabilizers acting at these plaquettes $p$, $s^x_p = \prod_{j \in \partial p} X_j$, $s^y_p = \prod_{j \in \partial p} Y_j$ or $s^z_p = \prod_{j \in \partial p} Z_j$.
		But since we require all stabilizers to commute, we see that we can only choose one stabilizer per plaquette.
		This is because $s^x_p$, $s^y_p$ and $s^z_p$ mutually anti commute as they each share three qubits.
		Similarly, two of these boundary stabilizers on adjacent plaquettes must be of a common Pauli type in order to commute with one another.
		Adjacent boundary stabilizers have common support only on one single qubit.
		In the figure we arbitrarily choose $s^z_p$ stabilizers, and we show examples of differently colored excitations with a $z$ Pauli label that are emitted from the boundary.
		To the best of our knowledge, the Pauli boundary has not been explicitly represented on the color code lattice before.

		The following table summarizes the discussion given above.
		It explicitly shows that the columns of the table represent the Lagrangian subgroups that give rise to color boundaries and the rows represent the Lagrangian subgroups that give rise to Pauli boundaries.
		\begin{table}
			\centering
			\begin{tikzpicture}
				\draw (-1,.22) -- (1,.22);
				\draw (-1,-.22) -- (1,-.22);
				\draw (-.35,-.6) -- (-.35,.6);
				\draw (.35,-.6) -- (.35,.6);
				\node (anyons) at (0,0) {
					$\begin{matrix}
						rx & gx & bx \\
						ry & gy & by \\
						rz & gz & bz
					\end{matrix}$
				};
				\draw [decorate,decoration={brace,amplitude=2pt},yshift=0pt]
				(-1.3,-.5) -- (-1.3,.5)node [black,midway,xshift=-30pt,align=center] {Pauli\\boundaries};
				\draw [decorate,decoration={brace,amplitude=2pt},yshift=0pt]
				(-.8,.85) -- (.8,.85)node [black,midway,yshift=15pt,align=center] {color\\boundaries};
			\end{tikzpicture}
			\caption{
				To visualize the color code boundaries, we can return to the arrangement of the bosonic anyons presented in Tab.~\ref{tab:CC3x3anyons}.
				The nine non-trivial bosonic anyons in the color code are arranged such that anyons in the same column or in the same row exchange trivially.
				Each row and column corresponds to a boundary.
				Anyons which can be condensed at a common color (Pauli) boundary share their color (Pauli) label.
			}
			\label{tab:Boundaries}
		\end{table}

		Finally, we present the interfaces between the different boundary types.
		General interfaces between boundaries have attracted some interest in the condensed matter community~\cite{Lindner12}.
		For quantum information applications, points where boundary type changes are referred to as corners~\cite{Brown17}.

		We show different corners in Fig.~\ref{fig:CCBoundaries}~(c), (d) and~(e).
		The first interface separates two color boundaries, see Fig.~\ref{fig:CCBoundaries}~(c).
		This interface has already been utilized in Ref.~\cite{Bombin06} to find a two-dimensional color code capable of transversally performing all Clifford operations.
		See also Refs.~\cite{Bombin15,Kubica15a} for a more detailed discussion on similar boundary interfaces.
		In Fig.~\ref{fig:CCBoundaries}~(d) we show an interface between two different Pauli boundaries.
		We show the corner between an $x$ Pauli boundary and a $z$ Pauli boundary which generalizes immediately to all Pauli-Pauli corners.
		At the top-right corner of the figure we show two boundary plaquettes that are commonly supported on two qubits where two Pauli boundaries can meet such that the boundary stabilizers commute.
		Such corners can be used to construct codes with Pauli boundaries, as discussed in Sec.~\ref{sec:FourQubitCC}.
		Finally, in Fig.~\ref{fig:CCBoundaries}~(e), we show a corner where a Pauli boundary meets a color boundary.
		Once again, it is readily checked that the stabilizers represented in the figure commute.
		This example can also be generalized easily to obtain all nine color/Pauli corners.

\section{Small color codes and fault-tolerant quantum computation}
	\label{sec:ApplicationsPauliBdry}

	In addition to giving us a better understanding of the fundamental physics that underlies the color code phase, a second benefit of systematically laying out all of the different features of the color code is that it allows us to discover new modes of fault-tolerant quantum computation with the model.
	In what follows we give two such examples of applications of Pauli boundaries.
	Specifically, we show that the new boundaries we have introduced give rise to better methods of performing code deformations, and we also find a new example of a four-qubit quantum error correcting code which can be interpreted as the smallest instance of a color code with Pauli boundaries.

	\subsection{Code deformation}
		\label{sec:PauliBoundariesAndCodeDeformation}

		We first consider code deformations, and argue that braiding punctures with Pauli-boundaries to perform fault-tolerant entangling gates will be advantageous over equivalent operations using punctures with colored boundaries as have been considered previously in the literature~\cite{Fowler11}.

		One of the leading paradigms for performing logical gates is through the braiding of punctures in some topological quantum error-correcting code.
		This, traditionally, put the color code at a disadvantage compared to the toric code.
		In the toric code, single qubit measurements are sufficient to create and move punctures.
		In contrast, the creation and movement of punctures featuring the color boundaries of the color code require two qubit Bell measurements~\cite{Fowler11}.
		We find that we can move punctures with Pauli boundaries using only single-qubit measurements.
		This is readily checked because we can measure a patch of individual qubits of the color code lattice using single-qubit Pauli measurements in a common Pauli basis to introduce or increase the size of a puncture.
		In doing so, we remain in a known eigenstate of the stabilizers at the boundary of the new puncture that has been produced by the measurements.
		We leave a detailed derivation of this argument as an exercise to the reader.
		This property of Pauli boundaries is advantageous for practical implementations for color code quantum computation as single-qubit measurements are invariably easier to perform than two-qubit parity measurements and are therefore less likely to introduce noise during deformations.
		This feature may also play an important role for measurement based fault-tolerant quantum computation~\cite{Raussendorf06} using color codes.

	\subsection{A four-qubit color code}
		\label{sec:FourQubitCC}

		Our extended toolbox has also enabled us to find a new four-qubit color code which we expect will be of interest for near-term experimental realization with small-scale quantum devices.
		They complement what has been seen in small-scale trapped ion quantum computers~\cite{Nigg14,Linke17} and relates to insights into the smallest interesting color code~\cite{CampbellBlog}.

\begin{figure}
			\centering
			\includegraphics[width=0.75\linewidth]{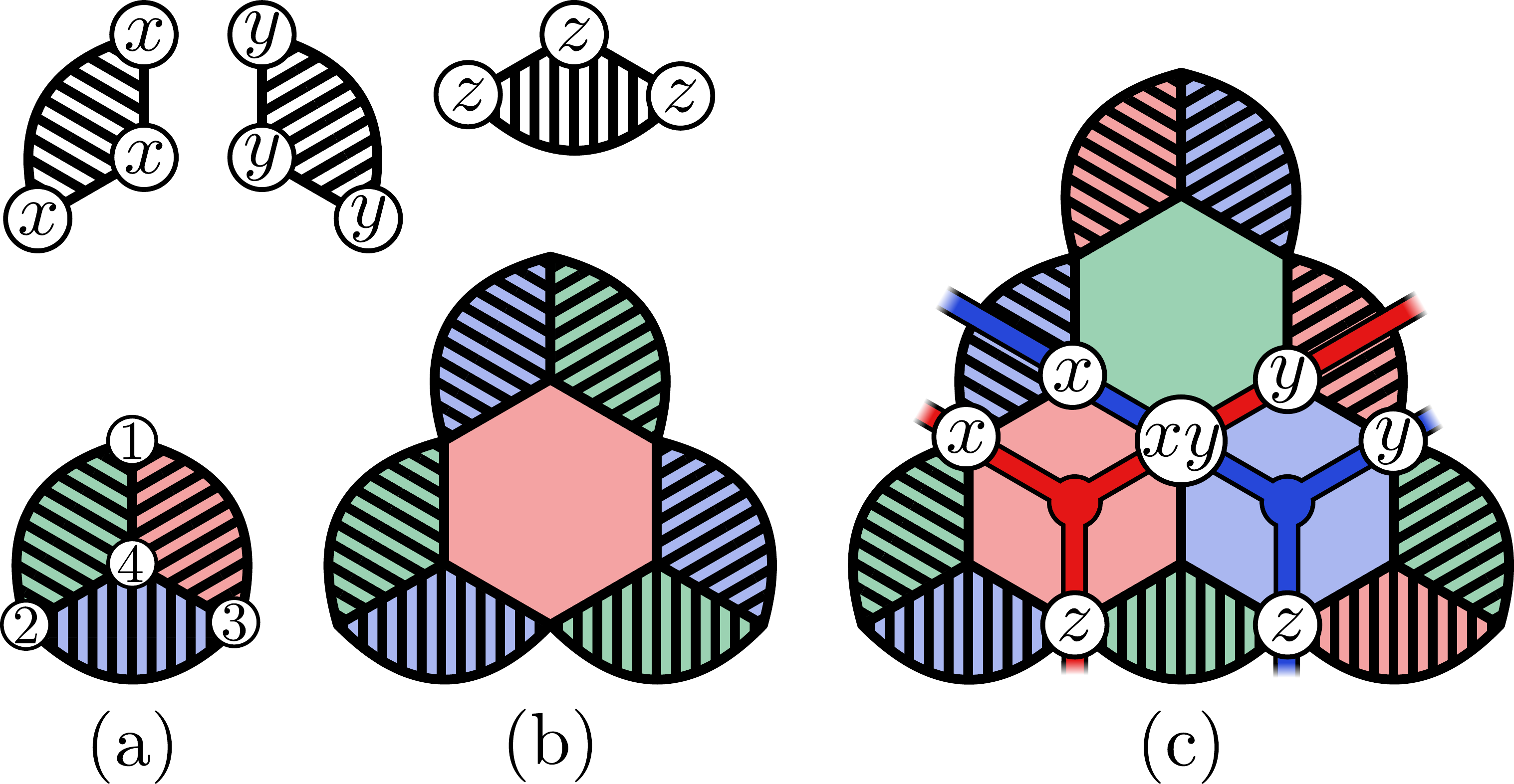}
			\caption{
				Examples of small triangular color codes with Pauli boundaries.
				The stabilizers acting on bulk plaquettes are the usual.
				On the boundary plaquettes, indicated by the stripes, only one stabilizer is measured.
				These boundary stabilizers are shown in the top left.
				\textbf{(a)} shows the smallest member of this family, a $\llbracket 4, 1, 2 \rrbracket$ code.
				It consists of four physical qubits, three weight-three stabilizers and hosts one logical qubit.
				Its logical operators are, for instance, $X_L = X_3 X_4$, $Y_L = Y_2 Y_4 $ and $Z_L = Z_1 Z_4 $.
				\textbf{(b)} and \textbf{(c)} show the next smallest codes with Pauli boundaries, consisting of nine and sixteen qubits respectively.
				In \textbf{(c)} examples of two logical operators, $X_L$ and $Z_L$, are shown.
			} 
			\label{fig:SmallCCPauliBoundaries}
		\end{figure}

		To describe the features of our new code, we first present the family of triangular color codes with Pauli boundaries.
		These codes feature an $x$ Pauli boundary, a $y$ Pauli boundary and a $z$ Pauli boundary.
		The three smallest members from this family of codes are shown in Fig.~\ref{fig:SmallCCPauliBoundaries}.
		Bulk plaquettes feature the two regular stabilizers given in Eq.~\eqref{eq:CCStabilizers}, the striped boundary stabilizers host a single weight-three stabilizer each, the basis in which it acts is indicated by the direction of its stripe pattern.
		These codes encode a single logical qubit each.
		Examples of their logical operators are shown in Fig.~\ref{fig:SmallCCPauliBoundaries}~(c).
		Remarkably, the smallest instance of this code, shown in Fig.~\ref{fig:SmallCCPauliBoundaries}~(a), 
		is a new example of a $\llbracket 4,1,2\rrbracket$ code with only weight-three stabilizers.

\section{Bulk domain walls}
	\label{sec:Domainwalls}

	In this section we introduce transparent domain walls, or invertible domain walls~\cite{Kitaev12}, to the model.
	These domain walls change the charge labels of anyons which cross them in a non-trivial way, such that fusion, braiding and self-exchange of the anyons is conserved.
	Specifically, we discuss closed domain walls here; open domain walls and in particular their terminal points are considered in Sec.~\ref{sec:CCTwists}.
	As we are interested in applying the theory to the color code, we restrict the discussion to abelian anyon models and refer the readers to Refs.~\cite{Barkeshli14,Tarantino16} for the general case.

	We first provide a comprehensive study of transparent domain walls for abelian anyon models in Subsec.~\ref{sec:GeneralDWSymmetries}.
	Then we apply this theory in our concrete example of the color code, whose symmetries we study in Subsec.~\ref{sec:CCSymmetries}.
	Lattice representations of domain walls are given in Subsec.~\ref{sec:DWLatticeRep}.

	\subsection{Symmetries in abelian anyon models}
		\label{sec:GeneralDWSymmetries}
		
\begin{figure}
			\centering
			\includegraphics[width=.75\linewidth]{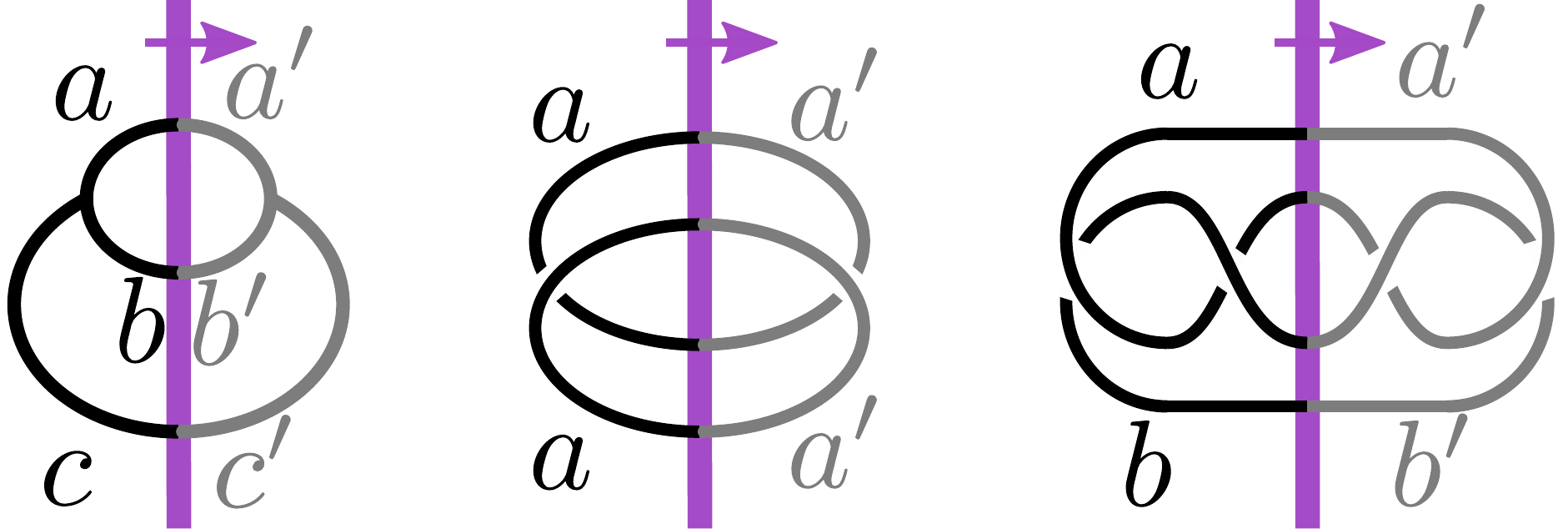}
			\caption{
				The consistency conditions on a domain wall can be derived by studying contractible loops crossing the domain wall.
				From left to right the figures correspond to 
				Eqs.~(\ref{eq:DWConditionFusion}-\ref{eq:DWConditionBraiding}).
			}
			\label{fig:ConditionDW}
		\end{figure}

		Transparent domain walls change the labels of anyons crossing them according to a map.
		This map is an automorphism, $\phi: \cC \rightarrow \cC$ which maps  the anyon model onto itself.
		Acting on an anyon $a \in \cC$ it returns again an anyon $\phi(a) = a' \in \cC$.
		We require these automorphisms to leave the basic data of the anyon model, i.e its fusion, self-exchange and braid statistics, invariant.
		See Ref.~\cite{Barkeshli14}.
		Explicitly, this means
		\begin{align}
			\phi \left( N_{a,b}^{c} \right) &= N_{a',b'}^{c'} \stackrel{!}{=} N_{a,b}^{c} ,
			\label{eq:DWConditionFusion}
			\\
			\phi \left( \theta_{a} \right) &= \theta_{a'} \stackrel{!}{=} \theta_{a} ,
			\label{eq:DWConditionExchange}
			\\
			\phi \left( M_{a,b} \right) &= M_{a',b'} \stackrel{!}{=} M_{a,b}.
			\label{eq:DWConditionBraiding}
		\end{align}
		These conditions can be obtained by requiring that anyons moved along contractible loops crossing the domain wall constitute stabilizers and hence leave the state invariant.
		Considering a branching loop, like on the left of Fig.~\ref{fig:ConditionDW}, we can follow Eq.~\eqref{eq:DWConditionFusion}.
		If an anyon $c$ on the left of a domain wall decays into $a$ and $b$ which are then moved over the domain wall, then on the right $a'$ and $b'$ must fuse to $c'$.
		And using that stabilizers commute with each other, we can deduce Eq.~\eqref{eq:DWConditionExchange} from the paths shown in the middle of Fig.~\ref{fig:ConditionDW}.
		We observe that any phase $\theta_a$ picked up by the self-exchange on the left must be canceled by the phase $\theta_{a'}^*$ obtained by the reversed exchange on the right.
		Lastly, to get Eq.~\eqref{eq:DWConditionBraiding}, consider the two unlinked loops on the right of Fig.~\ref{fig:ConditionDW}.
		Again, any phase $M_{a,b}$ picked up by the braid performed on the left must be canceled out by the phase $M_{a',b'}^*$ which is obtain by the reverse braid on the right.

		\begin{figure}[b!]
			\centering
			\includegraphics[width=.45\linewidth]{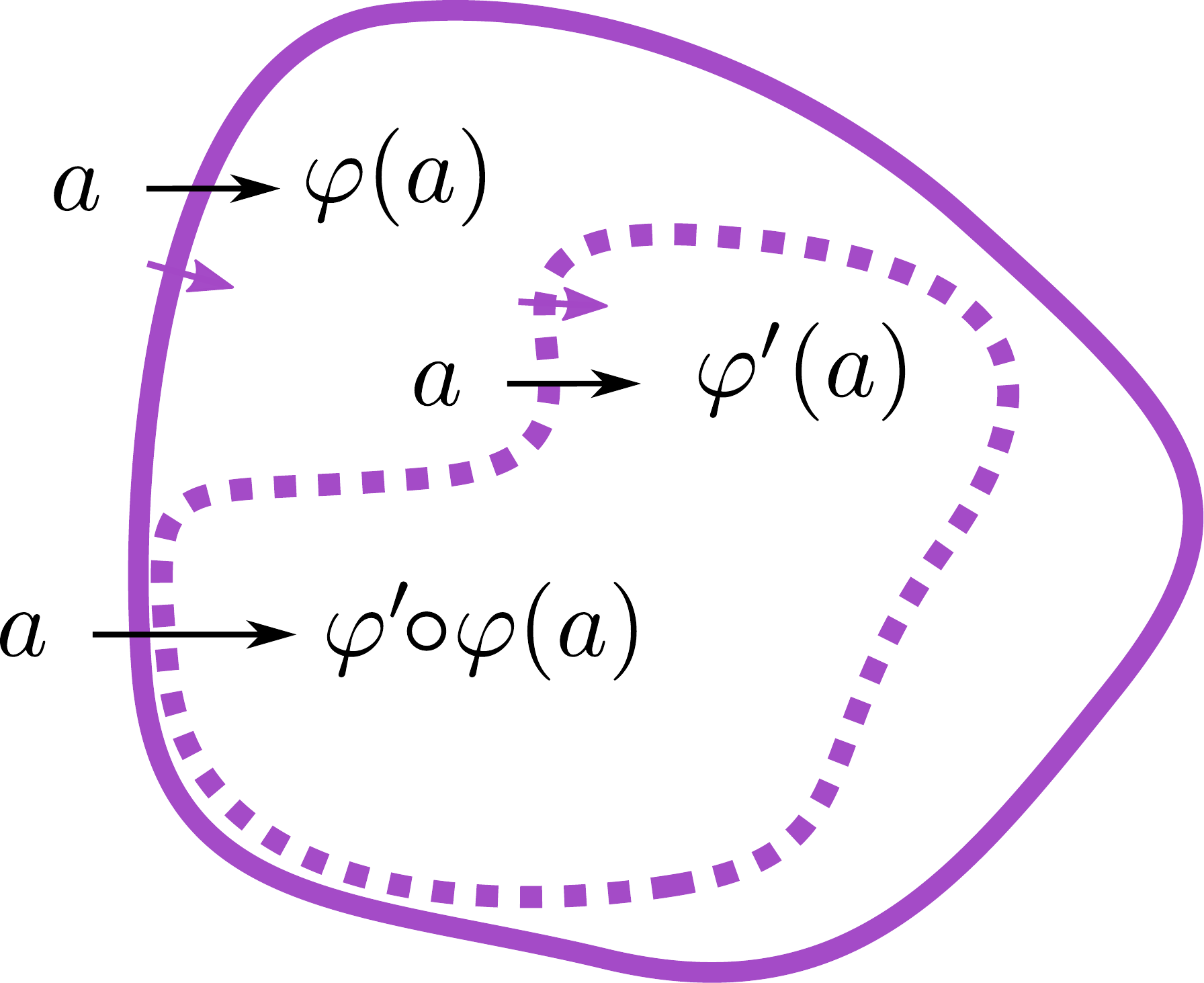}
			\caption{
				Two closed domain wall, drawn in purple.
				The arrows are to indicate the directionality of the domain wall.
				The domain wall drawn as a solid line is described by the automorphism $\phi$, the dashed one by $\phi'$.
				Where the domain walls are fused, a new domain wall is obtained which is described by $\phi' \circ \phi$.
				Note that domain walls have an orientation and changing it corresponds to taking the inverse $\overline{\phi}$ in the group of automorphisms.
			}
			\label{fig:closedDW}
		\end{figure}
		The automorphisms fulfilling these conditions for an anyon model $\cC$ are called the symmetries of $\cC$.
		And each transparent domain wall can be described and labeled by one such automorphism $\phi$.
		The set of symmetries of an anyon model forms a group~\cite{Barkeshli14}.
		The group identity is the trivial map, corresponding to the absence of a domain wall, and the group multiplication is the composition of maps.
		Thus, composition maps also fulfill the conditions and are therefore also symmetries of the anyon model.
		Physically, composing maps corresponds to multiple domain walls being combined or fused.
		See Fig.~\ref{fig:closedDW} for a schematic of how domain walls change anyons and how the fusion of domain walls corresponds to the composition map.

	\subsection{Symmetries in the color code}
		\label{sec:CCSymmetries}

		To classify the transparent domain walls in the color code, it proves once again useful to go back to  the $3 {\times} 3$ grid of bosonic color code anyons, shown in Tab.~\ref{tab:CC3x3anyons}.
		As mentioned earlier, fusion of two different anyons in a common row or column results in the third anyon in said row or column.
		As such we find that the fusion rules are preserved by all transformations on the grid that respect that all elements of a given row or column remain together in any other row or column.
		Likewise, we know that excitations braid trivially if and only if they share a column or row of the grid.
		Like fusion then, the mutual braid statistics of this generating set of excitations shown in the table are respected provided all excitations that share a common row or column remain in rows and columns with the same excitations.

		The above discussion implies that the symmetries of the generating set of bosonic excitations, and thus the excitations of the color code in general, are obtained by finding all the transformations that respect the rows and columns of the grid.
		Indeed, these are row permutations, column permutations, and the transposition of the grid.
		Given there are three rows and three columns, there are six different permutations of the rows, and of the columns, and finally one transpose.
		We thus arrive at $6 \cdot 6 \cdot 2 = 72$ symmetries of the color code.
		This agrees with the exhaustive numerical search given in Ref.~\cite{Yoshida15}.
	
		Further insight on the symmetry group can be gained using the following decomposition:
		$(S_3 \cross S_3) \ltimes \bZ_2$.
		One $S_3$ comes from permutations of the columns, i.e. colors, the second from the rows, i.e. the Pauli labels.
		Since permuting rows and columns commutes, the corresponding group is given by the direct product, $S_3 \cross S_3$.
		Finally, the $\bZ_2$ matrix transposition exchanging color and Pauli labels acts via the semi direct product ($\ltimes$).
		As a compact notation we can use the wreath product ($\wr$) to write the symmetry group as $S_3 \wr \bZ_2$.
		This decomposition follows directly from interpreting the color code anyon model as two copies of the three-fermion model, see App.~\ref{app:ThreeFermionModel} for more details.

		To label symmetries which exchange two colors, we use a capital letter of the third, unchanged color.
		For example, the symmetry which exchanges the green and the blue color label while leaving the red label, we denote as $R$.
		Equivalently, $G$ and $B$ respectively leave green and blue color labels invariant while exchanging the other two.
		Similarly, the exchanges of the Pauli labels we denote as $X$, $Y$ or $Z$, for symmetries which leave the $x$, $y$ or $z$ Pauli label invariant.
		Finally, we use $D$ for the duality symmetry between color and Pauli labels, corresponding to the transposition of the $3 {\times} 3$ anyon grid, shown in Tab.~\ref{tab:GeneratingTwists}.
		All of the symmetries mentioned so far are of order $2$, i.e. applying them twice leads to the trivial symmetry.

		Combining two of the above order $2$ color symmetries creates one of the two cyclic color permutations.
		Thus we can label them $RB$ and $BR$, for example.
		Conjugation of one color exchange with a second results in the third, e.g. $RGR = B$.
		The analogous statement holds for the permutations of Pauli labels;
		$XZ$ and $ZX$ are the cyclic Pauli label permutations and we can check that $XYX = Z$.
		Conjugating a color exchange with $D$ creates a symmetry which solely exchanges Pauli labels, for example $DBD = Z$, or conversely $DZD = B$.
		\begin{table}[!htbp]
			\centering
			\begin{tikzpicture}
				\draw (0.375,-.2) -- (0.375,1.2);
				\draw (1.125,-.2) -- (1.125,1.2);
				\draw (-.25,.25) -- (1.7,.25);
				\draw (-.25,.75) -- (1.7,.75);
				
				\node[anchor = base] (x) at (-.5,0.95) {$x$};
				\node[anchor = base] (y) at (-.5,0.45) {$y$};
				\node[anchor = base] (z) at (-.5,-.05) {$z$};
				
				\node[anchor = base] (r) at (0.0,1.4) {$r$};
				\node[anchor = base] (g) at (.75,1.4) {$g$};
				\node[anchor = base] (b) at (1.5,1.4) {$b$};
				
				\draw [<->] (0.0,-.25) to[in=-90,out=-90] (0.75,-.25);
				\node[anchor = base] at (0.375,-.75) {$B$};
				
				\draw [<->] (1.75,0.0) to[in=0,out=0] (1.75,1.0);
				\node[anchor = base] at (2.25,0.4) {$Y$};
				
				\draw [<->] (0,.5) to (.75,1);
				\draw [<->] (0,0) to (1.5,1);
				\draw [<->] (.75,0) to (1.5,.5);
				\node[anchor = base] at (0.575,0.45) {$D$};
			\end{tikzpicture}
			\caption{
				The effect of the three generating symmetries on the anyons.
				$B$ is the blue invariant color permutation which exchanges the red and green color label.
				$Y$ the $y$-invariant Pauli permutation which exchanges $x$ and $z$ Pauli labels.
				And $D$ the duality symmetry which exchanges color and Pauli labels as shown.
			}
			\label{tab:GeneratingTwists}
		\end{table}

		To generate all $72$ symmetries, three are already sufficient.
		A minimal choice might be, for instance, $X$, $Y$ and $D$, two different row permutations and the duality.
		However, here, we choose $B$, $Y$ and $D$, a row exchange, a column exchange and the transposition, as shown in Tab.~\ref{tab:GeneratingTwists}.
		This is to demonstrate all three different types of domain walls in what follows.

	\subsection{Lattice representation of domain walls}
		\label{sec:DWLatticeRep}

		Now, we give lattice representations of color code domain walls which correspond to the generating symmetries above.
		In particular, these are the $B$, $Y$ and $D$ symmetries.
		They can generate all $72$ color code domain walls, as discussed in Sec.~\ref{sec:CCSymmetries}.
		And they also constitute examples for three different classes of twists, one only acting on the color label, one on the Pauli label and duality transformation which exchanges the two.
		To obtain these closed domain walls, we transform the stabilizers along a closed path.
		This provides an apparent physicality to the location of the domain walls, which although artificial provides a way to later emphasize their terminal points whose locality is physical.
		These terminal points are called twist defects and are the focus of Sec.~\ref{sec:CCTwists}.
		We emphasize that neither of the transformations can be achieved by means of a local unitary with support only where the domain wall will come to lie.
		
\begin{figure}
			\centering
			\includegraphics[width=1\linewidth]{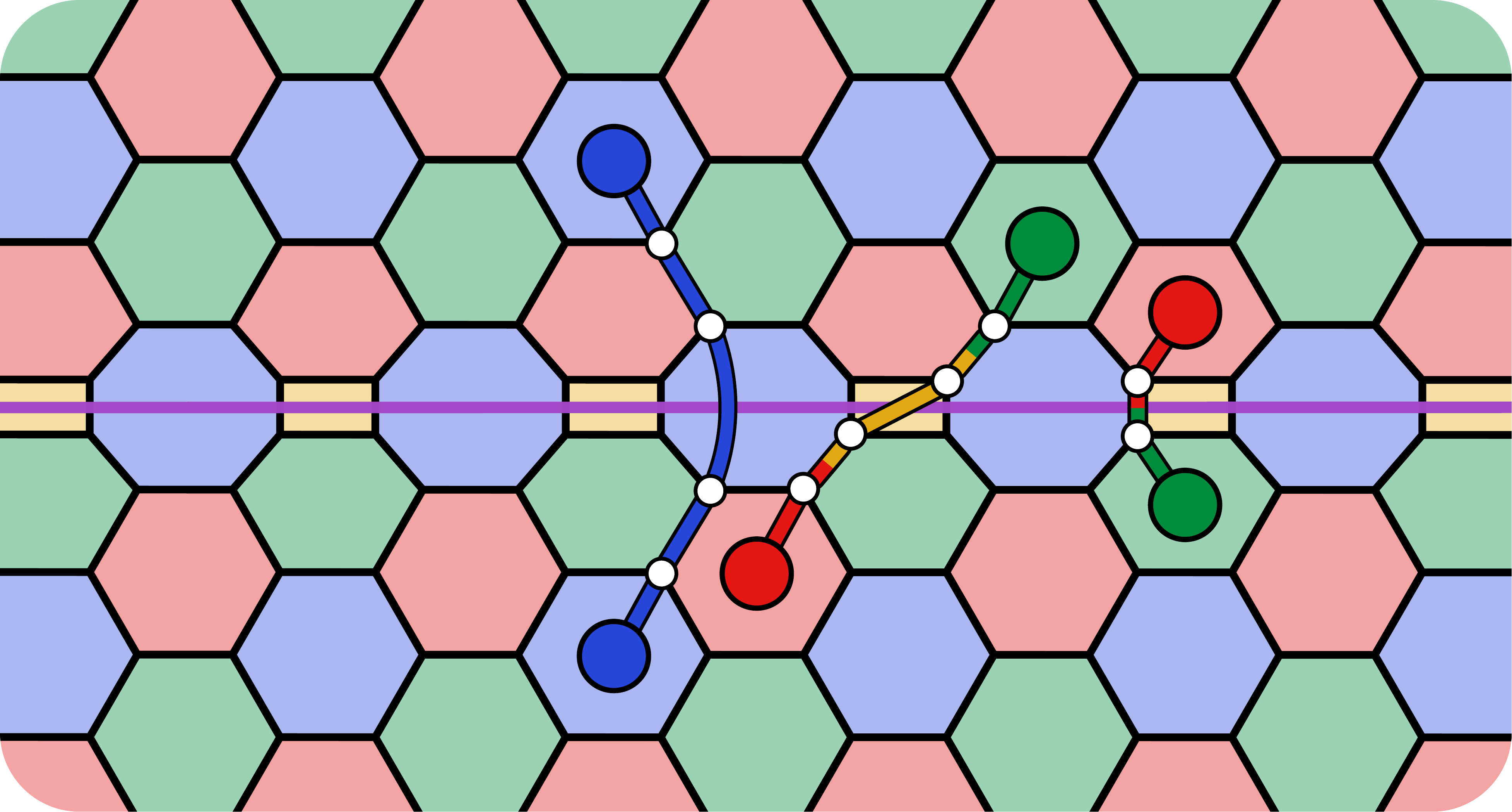}
			\caption{
				A 6.6.6 color code lattice with a blue-invariant domain wall $B$, marked by the purple line.
				Along the domain wall a fourth color, yellow, is introduced.
				Two commuting stabilizers are measured on each plaquette, one $x$-type and one $z$-type.
				Anyons crossing the domain wall keep their Pauli label, but red and green color labels are exchanged.
				Note how we omitted the arrow indicating the orientation of the domain walls, as it corresponds to an order two symmetry $B = \overline{B}$.
			}
			\label{fig:B_DW}
		\end{figure}

		An example of a domain wall changing the color label is shown in Fig.~\ref{fig:B_DW}.
		It exchanges solely the red and the green color label of anyons crossing it.
		It corresponds to the symmetry $B$, shown in Tab.~\ref{tab:GeneratingTwists}.
		Along the domain wall, each plaquette hosts two stabilizers, an $x$- and a $z$-type.
		Notice how in this example a fourth color is used to color the faces.
		The braking of three-colorability is in fact a general feature of color permuting domain walls.
		In App.~\ref{app:PachnerMoves} we show how color permuting domain walls can be introduced by applying a series of Pachner moves on the dual lattice.
		Using the Pachner move technique also allows us to generate domain walls which permute the three color labels in a cyclic fashion, see Fig.~\ref{fig:PachnerMoveTwists666}~(a) and Fig.~\ref{fig:PachnerMoveTwists488}~(a).
		The two other color exchanging domain walls, $R$ and $G$ can be drawn equivalently.
		In Refs.~\cite{Bombin11, Teo14, Teo16}, similar domain walls have been demonstrated for equivalent models.

\begin{figure}
			\centering
			\includegraphics[width=1\linewidth]{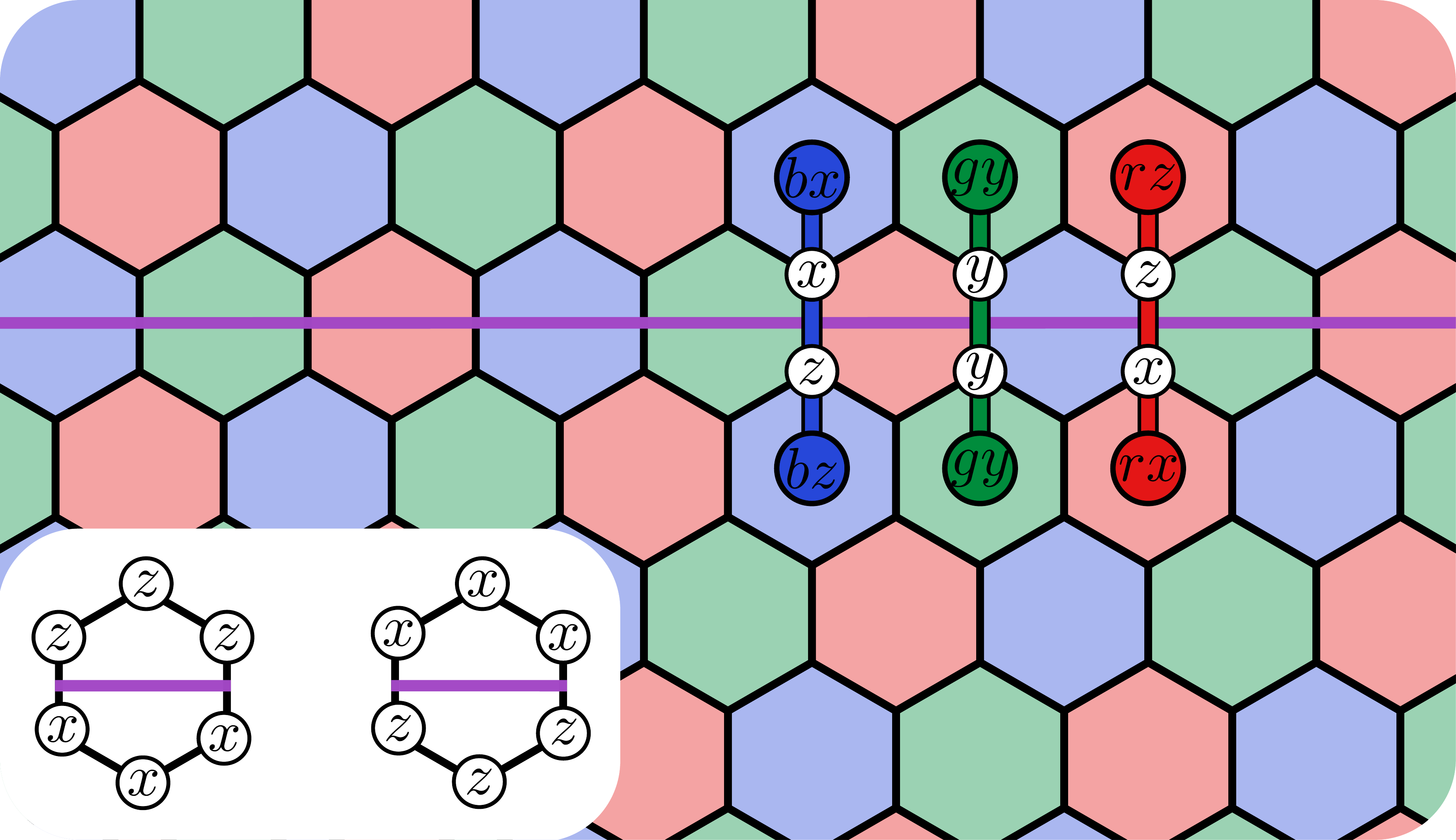}
			\caption{
				A 6.6.6 color code with a $y$-invariant Pauli domain wall $Y$ represented by a purple line.
				Along the line, every plaquette hosts two stabilizers, they are shown in the bottom left.
				They can be obtained by conjugating the basis in which the stabilizers act by a Hadamard on one side of the domain wall.
			}
			\label{fig:Y_DW}
		\end{figure}

		Next, we turn our attention to domain walls permuting Pauli labels.
		They can be constructed without changing the underlying lattice.
		Instead, just the basis in which stabilizers act along the domain wall are measured is altered.
		Such a domain wall is represented by a violet line in Fig.~\ref{fig:Y_DW}.
		To generate a $y$-invariant domain wall $Y$, the two stabilizers for each plaquette lying on top of the domain are changed.
		The first one acts in the $x$-basis on qubits on one side of the domain wall, and in the $z$-basis on qubits on the other side.
		The second one acts in the $z$-basis on one side and the $x$-basis on the other side.
		This is equivalent to conjugating the basis in which the domain wall stabilizers act with a Hadamard on one side of the domain wall, as discussed in Ref.~\cite{Yoshida15}.
		These also correspond to locality preserving gates in the topological quantum field theory, which were studied in Ref.~\cite{Beverland16}.
		Domain walls which permute the Pauli label of anyons in a cyclic manner can also be constructed in a similar fashion by conjugating them partially by a unitary which cyclically permutes $x$, $y$ and $z$.

		\begin{figure}[b!]
			\centering
			\includegraphics[width=1\linewidth]{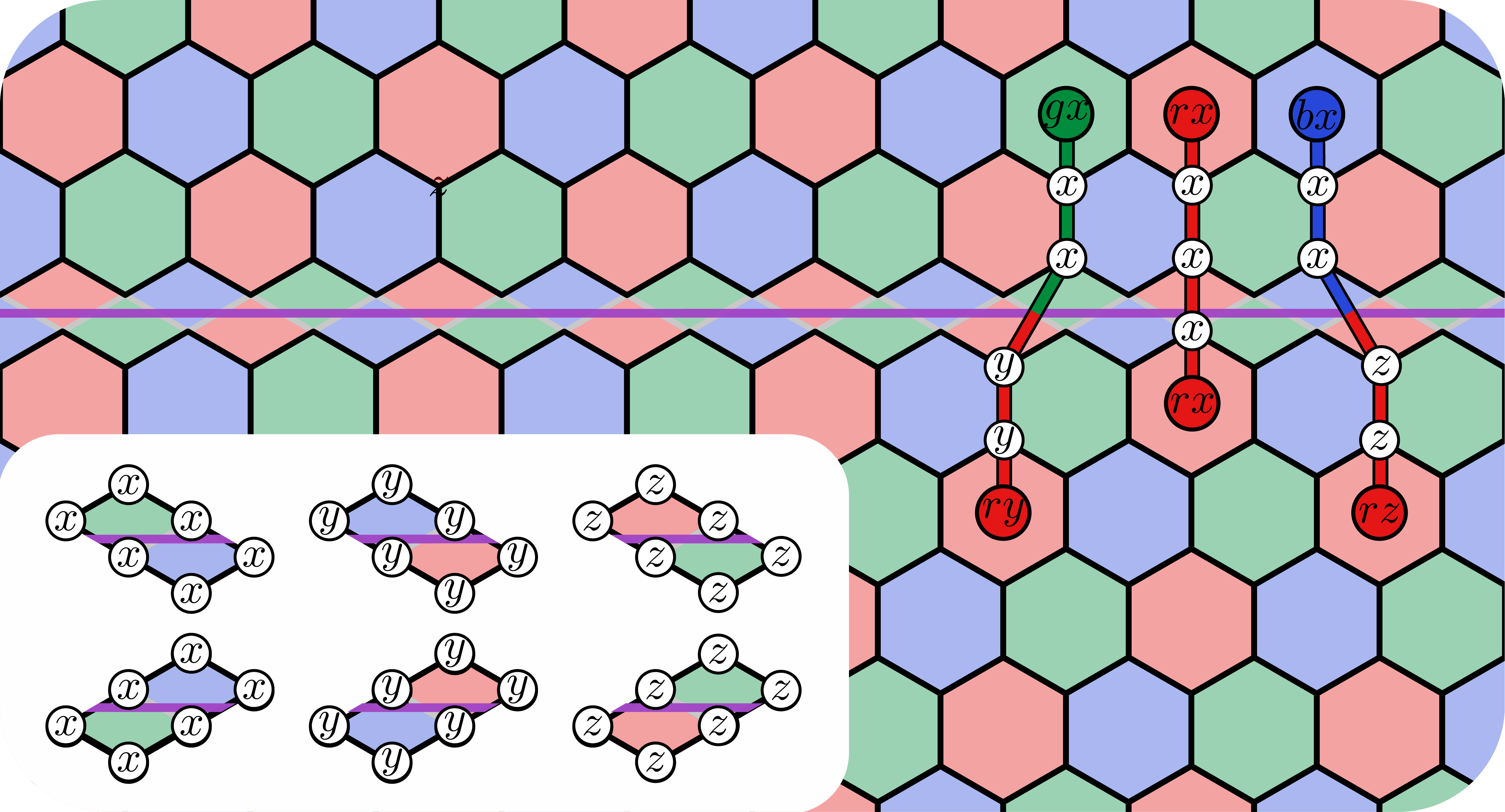}
			\caption{
				An example of an domino domain wall $D$.
				The dislocation line hosts rectangular stabilizers supporting six qubits.
				Like domino bricks, they are colored in a different color on either half.
				Their colors dictate the basis in which they act, shown in the bottom left.
			}
			\label{fig:D_DW}
		\end{figure}
		Lastly, we give a lattice representation of the domain wall which transposes the $3 {\times} 3$ arrangement of anyons.
		It corresponds to the symmetry $D$ and we refer to it as the domino domain wall, since it features stabilizers that resemble domino bricks, see Fig.~\ref{fig:D_DW}.
		Such a domain wall can be constructed by pulling a zigzag line of edges apart which duplicates all of the qubits lying along the line.
		In the created space we color squares of the lattice which define rectangular stabilizers that each act on six qubits which are supported on two adjacent squares that make a rectangle.
		Each of these new stabilizers has two colors, one on either side, resembling domino bricks.
		These colors are determined by the colors of the neighboring plaquettes.
		To each of these domino-brick shaped cells we assign one weight-six stabilizer, where the colors of the brick determine which operator acts on rectangle.
		Stabilizers colored green and blue are measured in the $x$-basis, red and blue in the $y$-basis, and red and green in the $z$-basis.
		This results in the symmetry $D$ which transposes the $3 {\times} 3$ grid.
		This domain wall it leaves the bosons $rx$, $gy$ and $bz$ that cross it invariant.
		See Fig.~\ref{fig:D_DW} for a lattice representation of the domain wall and the effect it has on anyons.
		Changing what pair of colors is associated with which Pauli basis, six different domain walls can be obtained.
		To the best of our knowledge, this is the first time a lattice representation of this domain wall has appeared in the literature.

\section{Twists}
	\label{sec:CCTwists}

	In the previous section, we considered continuous transparent domain walls that change the charge labels of an abelian anyon model while preserving the fusion rules and braid statistics between the charges.
	In what follows we build on our discussion by terminating the domain walls.
	Again, we restrict our discussion to abelian anyon models.
	At the terminal points of these domain walls, we realize twist defects or simply twists~\cite{Barkeshli10, Bombin10, Kitaev12, Barkeshli14}.
	These remarkable point-like objects have many properties in common with non-abelian anyons.
	This is especially intriguing considering that we started with an abelian theory.
	As such they are of particular interest for applications in quantum information processing.

	We begin by introducing twist defects in Subsec.~\ref{sec:TwistsDWEndpoints}.
	In Subsec.~\ref{sec:TwistsFusion} the fusion of twists is discussed.
	We then study the twist defects of the color code and discuss how they interact with its anyons in Subsec.~\ref{sec:TwistsAndAnyons}, with domain walls in Subsec.~\ref{sec:TwistsBraiding} and, lastly, with boundaries and corners in Subsec.~\ref{sec:CornerTwists} and Subsec.~\ref{sec:TwistCondensation}.

	\subsection{Twist defects}
		\label{sec:TwistsDWEndpoints}

		We start by labeling the twist defects.
		To this end, we consider an anyon $a$ that is moved once around a twist before returning to its initial position.
		The anyon will cross the domain wall $\phi$ which the twist terminates once under this monodromy, and as such is mapped onto the excitation $\phi(a)$.
		We label a twist $\phi$ if it acts on $\cC$ with the mapping $\phi$ under a single clockwise monodromy.

		\begin{figure}
			\centering
			\includegraphics[width=.55\linewidth]{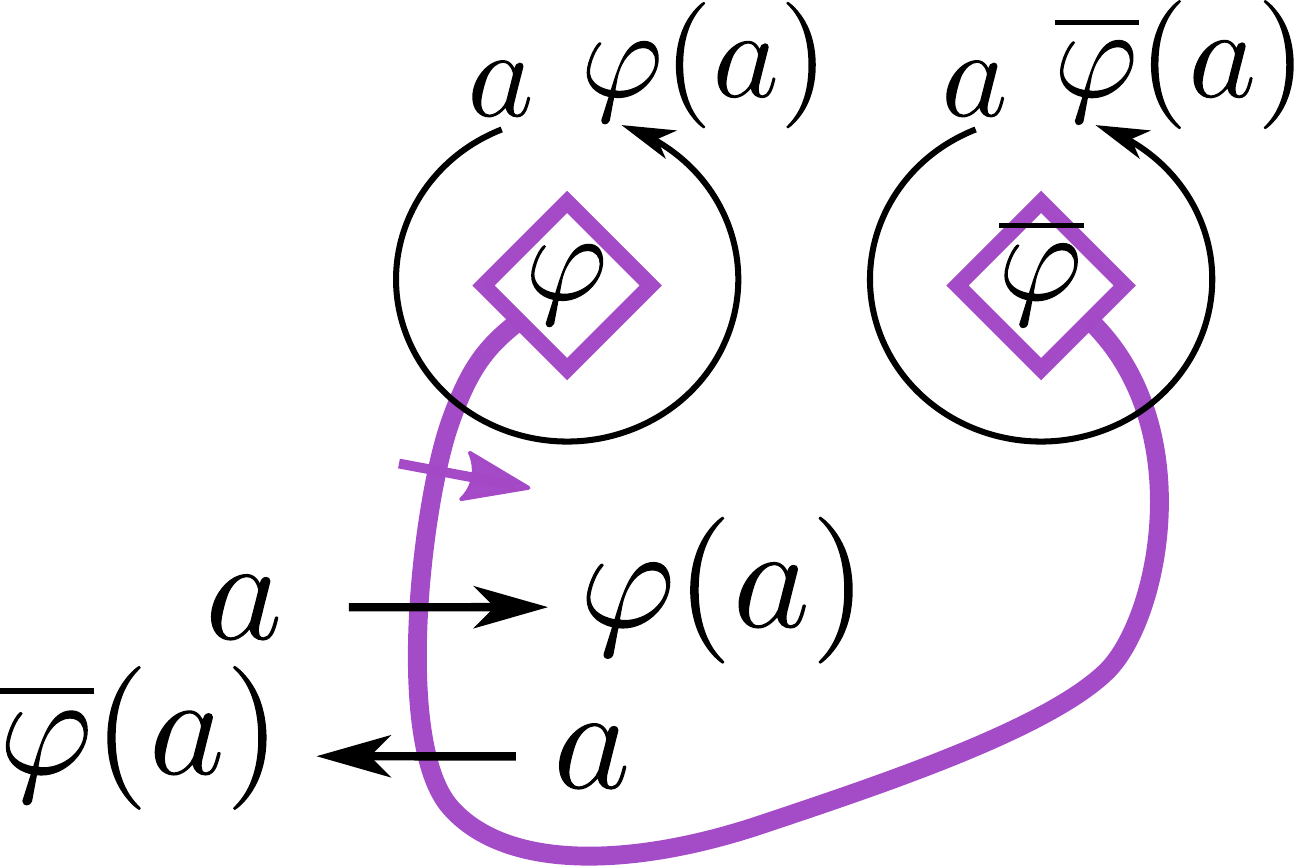}
			\caption{
				Twists terminating a domain wall correspond to the automorphism as the domain wall.
				Consider a domain wall which maps an anyon $a$ according to $\phi$ if the anyon gets transported in one direction across.
				If the anyon goes the other way, the inverse $\overline{\phi}$ is applied.
				The two twists lying on the terminal point of the domain wall are drawn as purple squares.
				Moving the anyon $a$ counter-clockwise around the twists applies the automorphism $\phi$ for one twist and $\overline{\phi}$ for the other twist terminating the domain wall.
			}
			\label{fig:TwistsTerminateDW}
		\end{figure}

		With this definition we observe particle anti-particle like creation of pairs of twists when adding an open domain wall to the bulk of a phase, see Fig.~\ref{fig:TwistsTerminateDW}.
		Indeed, we observe a twist $\phi$ at one terminal point of the domain wall and $\overline{\phi}$ at the other.

		We now go on to discuss the explicit example of twist defects in the color code.
		Specifically, we consider a generating set of twist defects, $B$, $Y$ and $D$, which are the terminal points of their corresponding domain walls.
		This generating set will be sufficient to generate all of the twists of the color code using twist composition rules we discuss later in Sec.~\ref{sec:TwistsFusion}.

\begin{figure}[b!]
			\centering
			\includegraphics[width=1\linewidth]{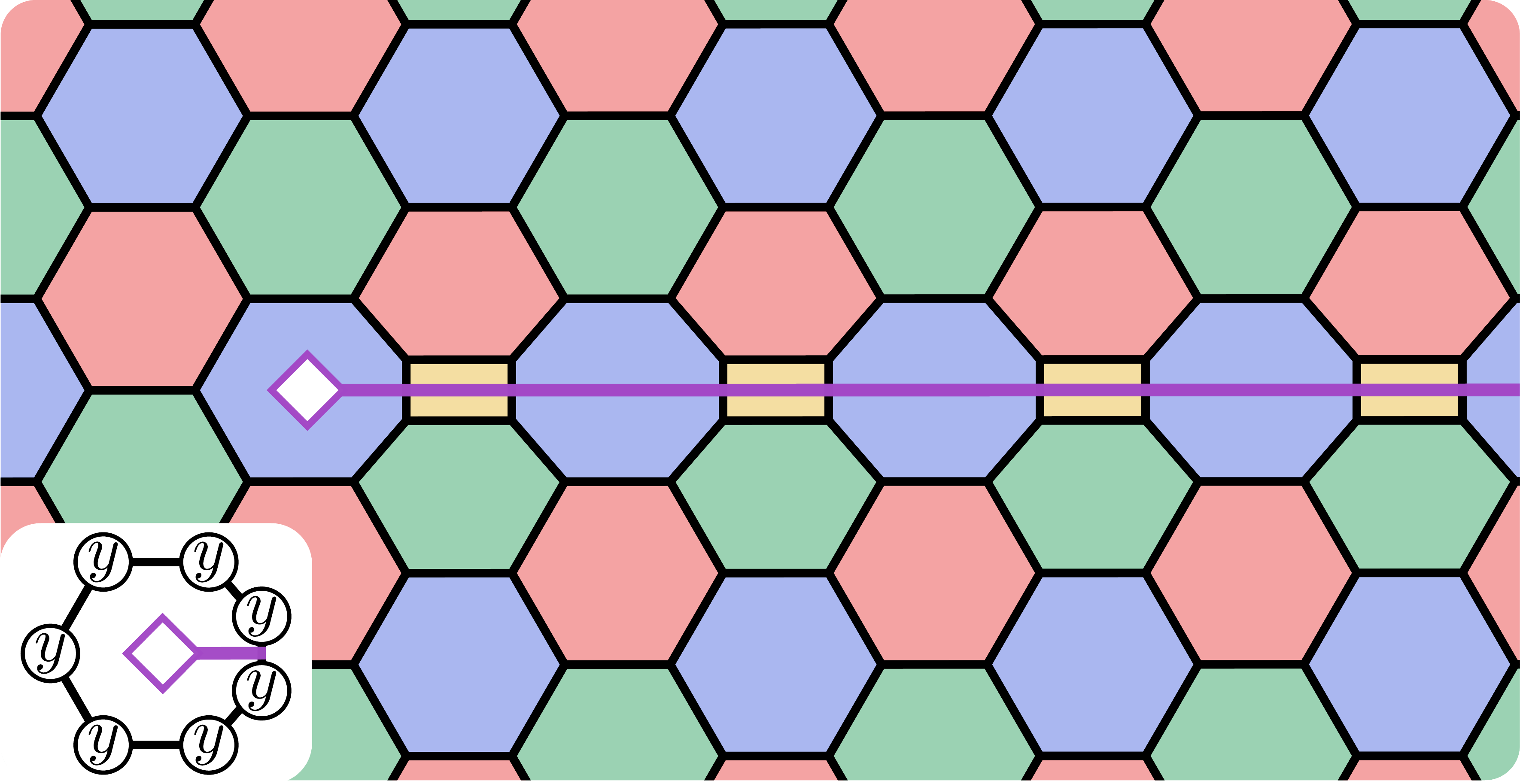}
			\caption{
				Shown is a lattice representation of the $B$ twist.
				It features a weight-seven stabilizer which is just measured in one basis, the $y$-basis for example.
				If anyonic charges are moved once around it, their color label changes according to the automorphism $B$.
			}
			\label{fig:BTwist}
		\end{figure}

		We first look at a lattice realization of the blue-invariant twist $B$ lying at the terminal point of the blue-invariant domain wall, see Fig.~\ref{fig:BTwist}.
		On the lattice, the terminal point of the domain wall lies at a plaquette with an odd number of vertices.
		This means it can no longer support both an $x$-type and a $z$-type stabilizer, as they would not commute.
		Instead, a single basis has to be chosen in which the parity is measured.

		\begin{figure}
			\centering
			\includegraphics[width=1\linewidth]{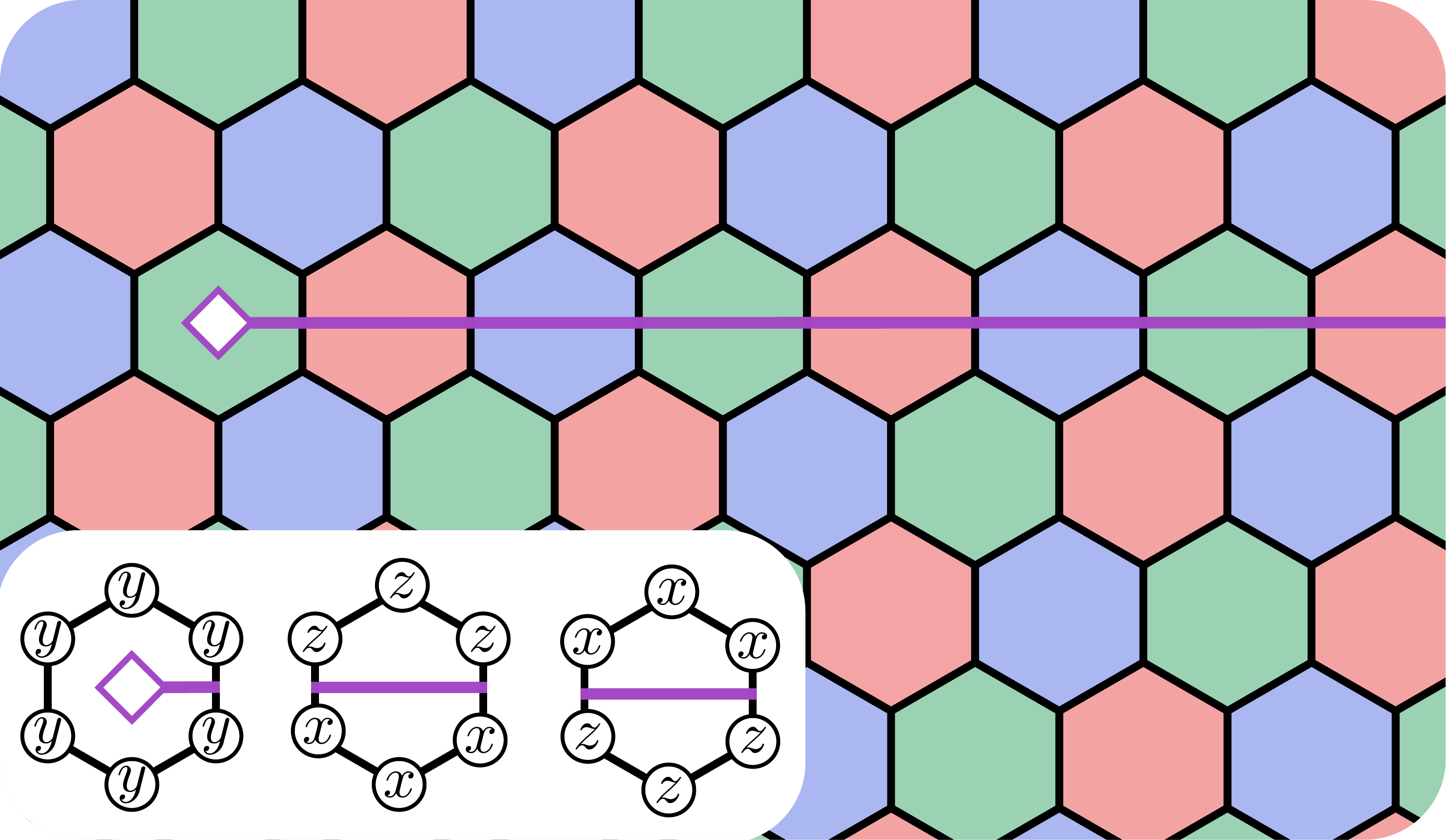}
			\caption{
				A $y$-invariant domain wall can be terminated by a plaquette which hosts one single stabilizer acting in the $y$-basis.
				This constitutes a $Y$ twist which exchanges $x$ and $z$ Pauli labels of color code anyons under monodromy.
			}
			\label{fig:YTwist}
		\end{figure}
		A $Y$ twist lies at the terminal point of the domain wall that permutes the Pauli-X and Pauli-Z label of the color code bosons.
		Unlike a $B$ twist we do not need to change the structure of the lattice to introduce a $Y$ twist.
		However, at the terminal plaquette of the domain wall we can impose only a single stabilizer of Pauli-Y terms, as is depicted in Fig.~\ref{fig:YTwist}.

\begin{figure}[b!]
			\centering
			\includegraphics[width=1\linewidth]{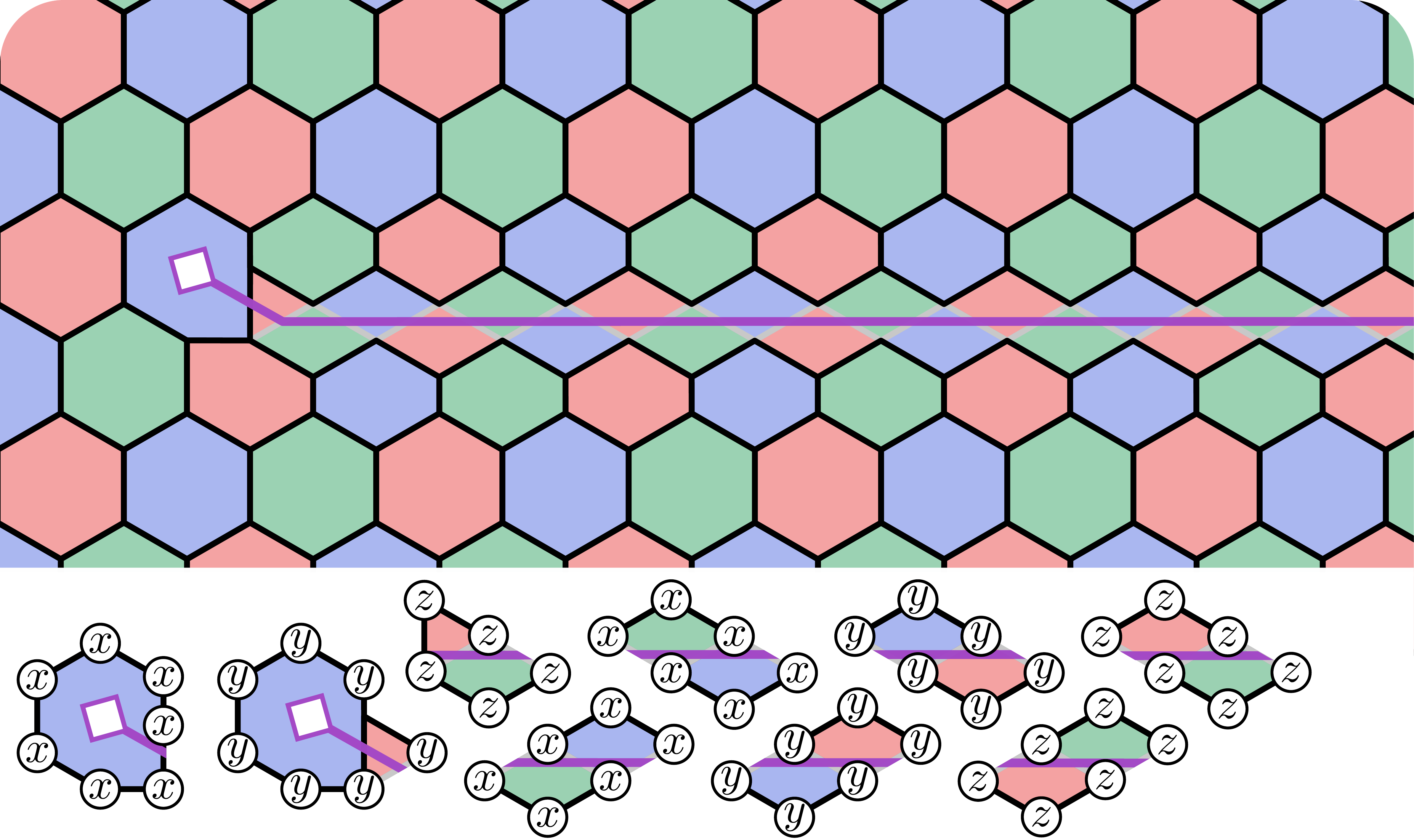}
			\caption{
				The domino twist $D$ has three changed stabilizers.
				One of the Domino brick stabilizers is cut so that it only supports five qubits.
				Additionally we have two weight-seven stabilizers, overlapping on six oft their qubits.
			}
			\label{fig:DTwist}
		\end{figure}

		We finally look at the twist $D$ that terminates the domino-brick domain wall.
		As shown in Fig.~\ref{fig:DTwist}, at the terminal point of the domain wall we have one weight-five and two weight-seven stabilizers.

		While in some cases it is possible to propose twist defect stabilizers via an educated guess, the stabilizers can be rather complex;
		this is particularly true in the case of $D$ twists.
		In fact, it is possible to essentially eliminate any guess work and derive the stabilizers for the twist defect by considering a sequence of Pauli operations which by construction preserves the ground space.
		First, we create a pair of anyons, $a$ and $ \bar{a}$.
		Excitation $a$ is then transported around the defect $\ell$ times, following its label change every time it crosses the domain wall $\phi$ such that  $\phi^\ell(a)=a$.
		Upon completing this repeated monodromy we can annihilate the two excitations to obtain a new stabilizer operator from the product of the Pauli operators used to realize this process.
		In order to find low-weight stabilizers, the resulting string operator can be contracted by multiplying it with conventional stabilizers that lie on the lattice.

	\subsection{Fusion between twists}
		\label{sec:TwistsFusion}

\begin{figure}
			\centering
			\includegraphics[width=.7\linewidth]{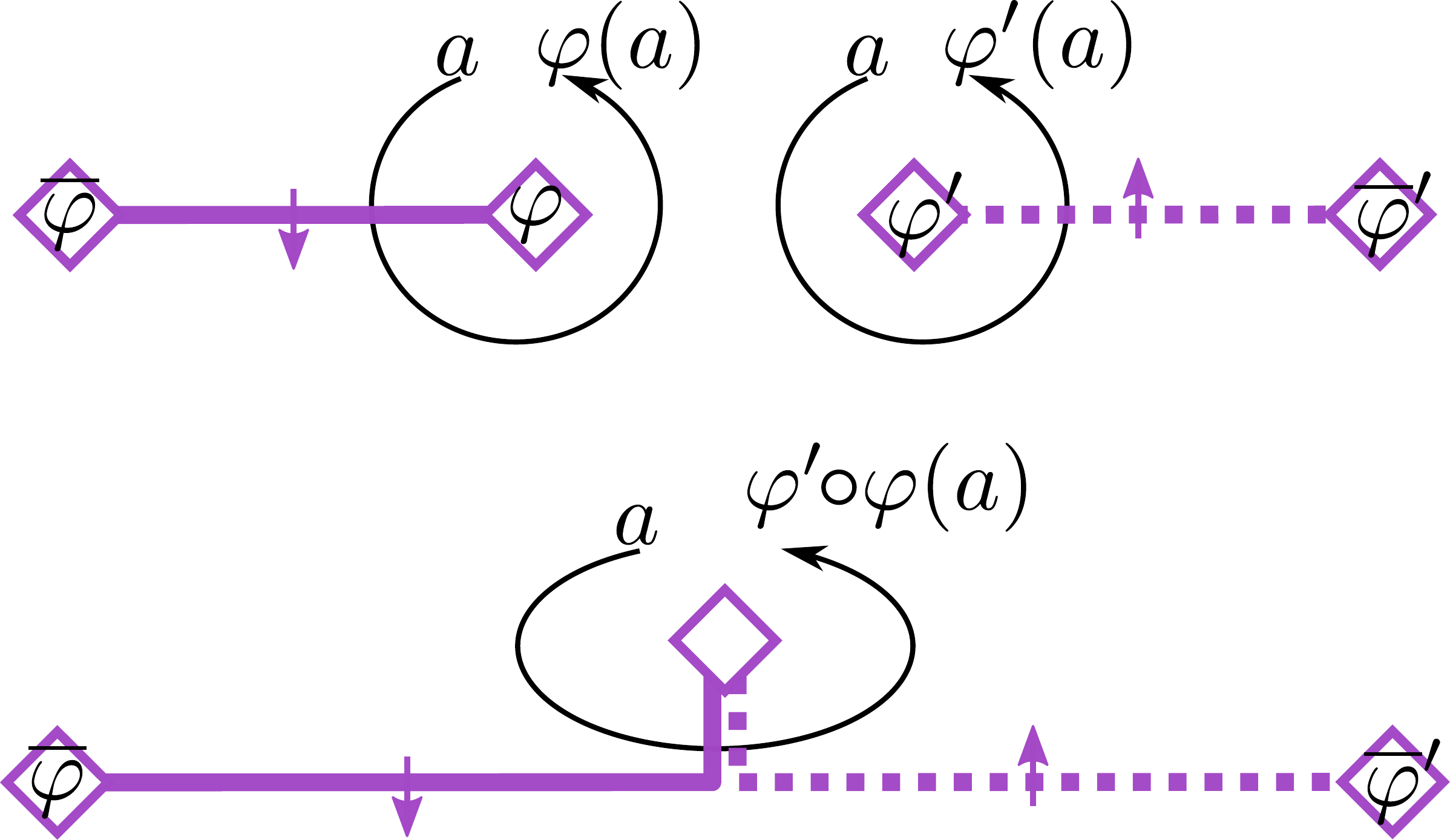}
			\caption{
				Similar to how the fusion of domain walls corresponds to composing maps, the same holds for twists.
				In this example, two twists described by the maps $\phi$ and $\phi'$ are combined to create a twists described by $\phi'\circ \phi$.
			}
			\label{fig:TwistFusion}
		\end{figure}

		The fusion of domain walls can be described by composition maps, see Sec.~\ref{sec:GeneralDWSymmetries}.
		Twists correspond to terminal points of domain walls, and hence the fusion of twists is also described by the composition of maps, see Fig.~\ref{fig:TwistFusion}.
		Given that the symmetry of an anyon model forms a group, the fusion of pairs of twists result in other twists.
		This means places where a domain wall changes type or where multiple domain walls meet are also twists.

\begin{figure}[b!]
			\centering
			\includegraphics[width=0.85\linewidth]{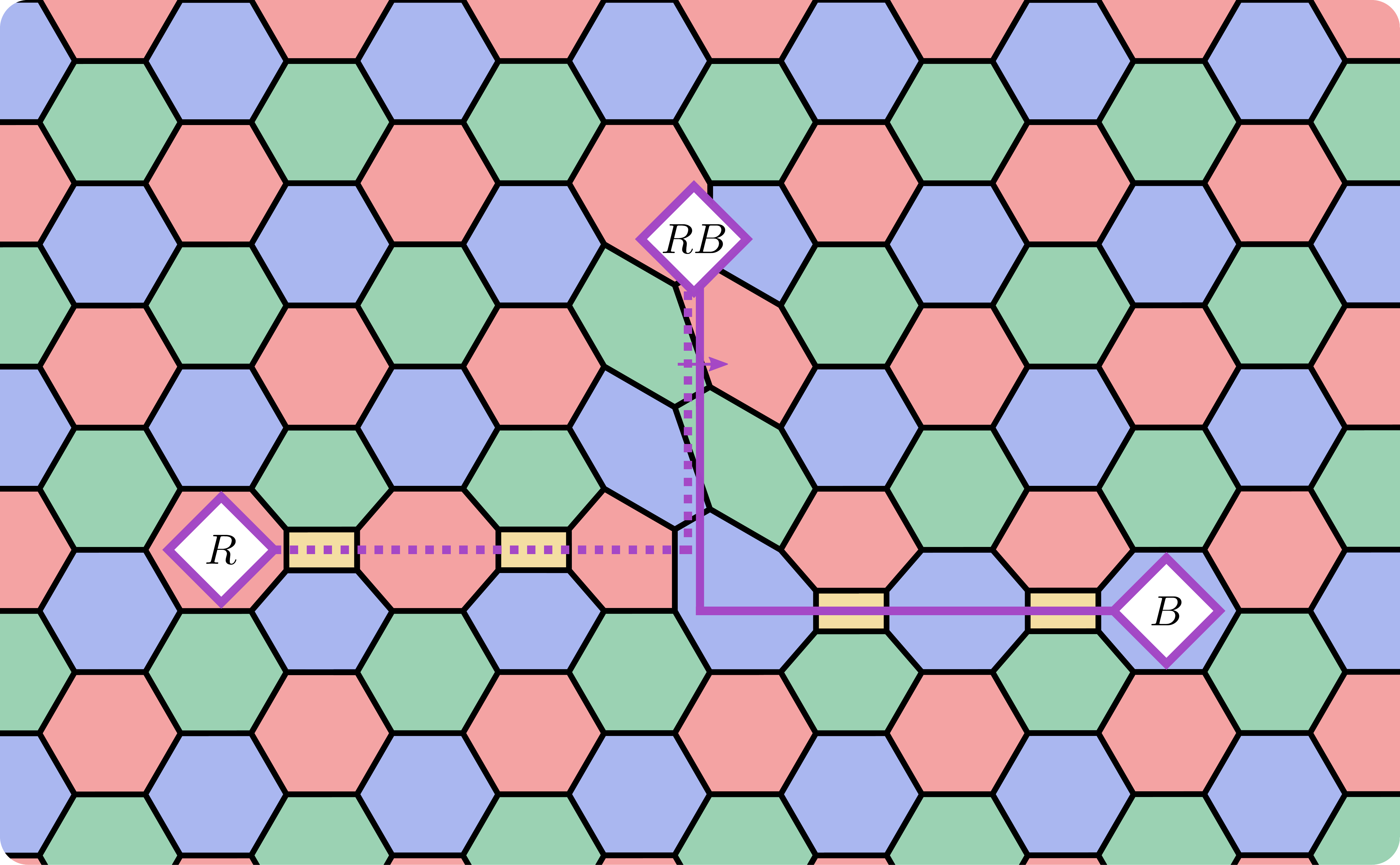}
			\caption{
				An $R$ twist on the left and a $B$ twist on the right are fused to form an $RB$ twist.
				The $RB$ twist features two plaquettes with odd weight, where only one stabilizer each is measured.
			}
			\label{fig:666TwistFusionPrimal}
		\end{figure}

		For an example, in the color code we can consider the fusion of two color exchanging twists, $R$ and $B$, to form a cyclic color permuting twist, $RB$, see Fig.~\ref{fig:666TwistFusionPrimal}.
		Note that the lattice representation of the $RB$ twist is not presented in the main text, we refer the reader to App.~\ref{app:PachnerMoves}, in particular to Fig.~\ref{fig:PachnerMoveTwists666}~(a).
		This fusion can be visualized nicely on the dual lattice using Pachner moves, see Fig.~\ref{fig:666TwistFusionDual} in App.~\ref{app:PachnerMoves}.

\begin{figure}
			\centering
			\includegraphics[width=0.9\linewidth]{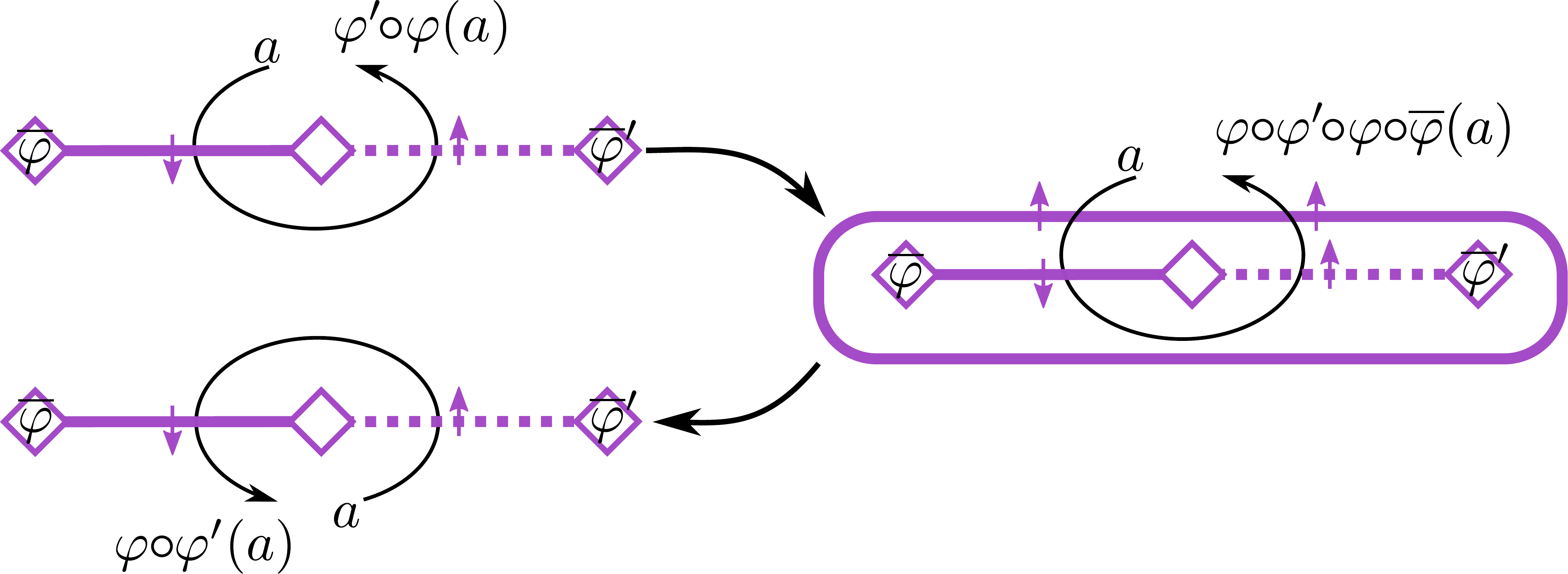}
			\caption{
				This figure depicts the fusion of two twists described by the symmetries $\phi$ and $\phi'$.
				On the left, we illustrate how the reference point used to label the fused twist matters.
				If the reference point is chosen to lie above the twist, as in the top left, we obtain a map $\phi' \circ \phi$.
				Whereas a reference point below the twist results in the description $\phi \circ \phi'$, as shown in the bottom left.
				However, one description can be changed to the other by wrapping a domain wall around the twists.
				This is seen on the right, where the lower description is obtained by enclosing the three twist with a closed $\phi$ domain wall.
			}
			\label{fig:ConjugacyTwists}
		\end{figure}

		In Fig.~\ref{fig:TwistFusion} we have deformed the domain walls upwards at the point where the twists meet.
		We thus obtain a twist $\phi' \circ \phi$.
		Instead we could have chosen to bend the domain walls downwards, such that the twist maps $a$ according to $\phi \circ \phi'(a)$.
		This ambiguity in the label is a general feature of twist fusion.
		To rectify this, we need to consider conjugacy classes of the symmetry group.
		The label used to describe a twists does in general depend on an arbitrarily chosen starting point.
		Consider Fig.~\ref{fig:ConjugacyTwists}, where a twist is labeled by two different maps on the left half of the figure.
		Shown on the right, we see that the two maps are equivalent up to conjugation.
		Here, conjugating the twist is equivalent to changing gauge or engulfing the entire twist compound inside a closed domain wall.
		If we do this in the example of Fig.~\ref{fig:ConjugacyTwists}, as shown on the right, we can convince our selves that the two maps are equivalent up to conjugation with $\phi$.

		In a general example, where multiple domain walls meet, even more than two different but valid descriptions might be found.
		And in general, without fixing a global gauge and precise conventions, twists are only identified up to conjugacy.

		\begin{table}[b!]
			\centering
			\begin{tabular}{|c||c|c|c|c|c|c|}
				\hline
				& $\one$ & $X$ & $Y$ & $Z$ & $XZ$ & $ZX$ \\
				\hhline{|=||=|=|=|=|=|=|}
				$\one$ & A & B & B & B & D & D \\
				\hline
				$R$ & B & C & C & C & E & E  \\
				\hline
				$G$ & B & C & C & C & E & E  \\
				\hline
				$B$ & B & C & C & C & E & E  \\
				\hline
				$RB$ & D & E & E & E & F & F  \\
				\hline
				$BR$ & D & E & E & E & F & F  \\
				\hline
			\end{tabular}
			\\
			\vspace{1mm}
			\centering
			\begin{tabular}{|c||c|c|c|c|c|c|}
				\hline
				$D \circ ()$ & $\one$ & $X$ & $Y$ & $Z$ & $XZ$ & $ZX$ \\
				\hhline{|=||=|=|=|=|=|=|}
				$\one$ & G & H & H & H & I & I \\
				\hline
				$R$ & H & G & I & I & H & H \\
				\hline
				$G$ & H & I & G & I & H & H \\
				\hline
				$B$ & H & I & I & G & H & H \\
				\hline
				$RB$ & I & H & H & H & I & G  \\
				\hline
				$BR$ & I & H & H & H & G & I \\
				\hline
			\end{tabular}
			\caption{
				The equivalence classes of the $72$ color code twists.
				Each entry corresponds to a twist obtained by fusion of the twist labeling the row and column.
				In the second table, every twist is additionally fused with the domino twist $D$.
				We find nine conjugacy classes, labeled A to I here.
			}
			\label{tab:CCConjClasses}
		\end{table}
		In the color code, twists belong to one of nine equivalence classes.
		Tab.~\ref{tab:CCConjClasses} lists the conjugacy classes A to I of all $72$ twists.
		Within an equivalence class twists share many properties, such as their order or their quantum dimension (see Tab.~\ref{tab:CCQuantumDims}).

	\subsection{Anyon localization and the quantum dimensions of twists}
		\label{sec:TwistsAndAnyons}

		Similar to how certain anyons can condense at boundaries, twists can also absorb anyons.
		This process is called localization of an anyon at a twist.
		We can express the localization of an anyon $a$ at twist $\phi$ with the fusion rule
		\begin{equation}
			\phi \times a = \phi.
			\label{eq:FusionTwistLocalizations}
		\end{equation}
		
\begin{figure}
			\centering
			\includegraphics[width=0.5\linewidth]{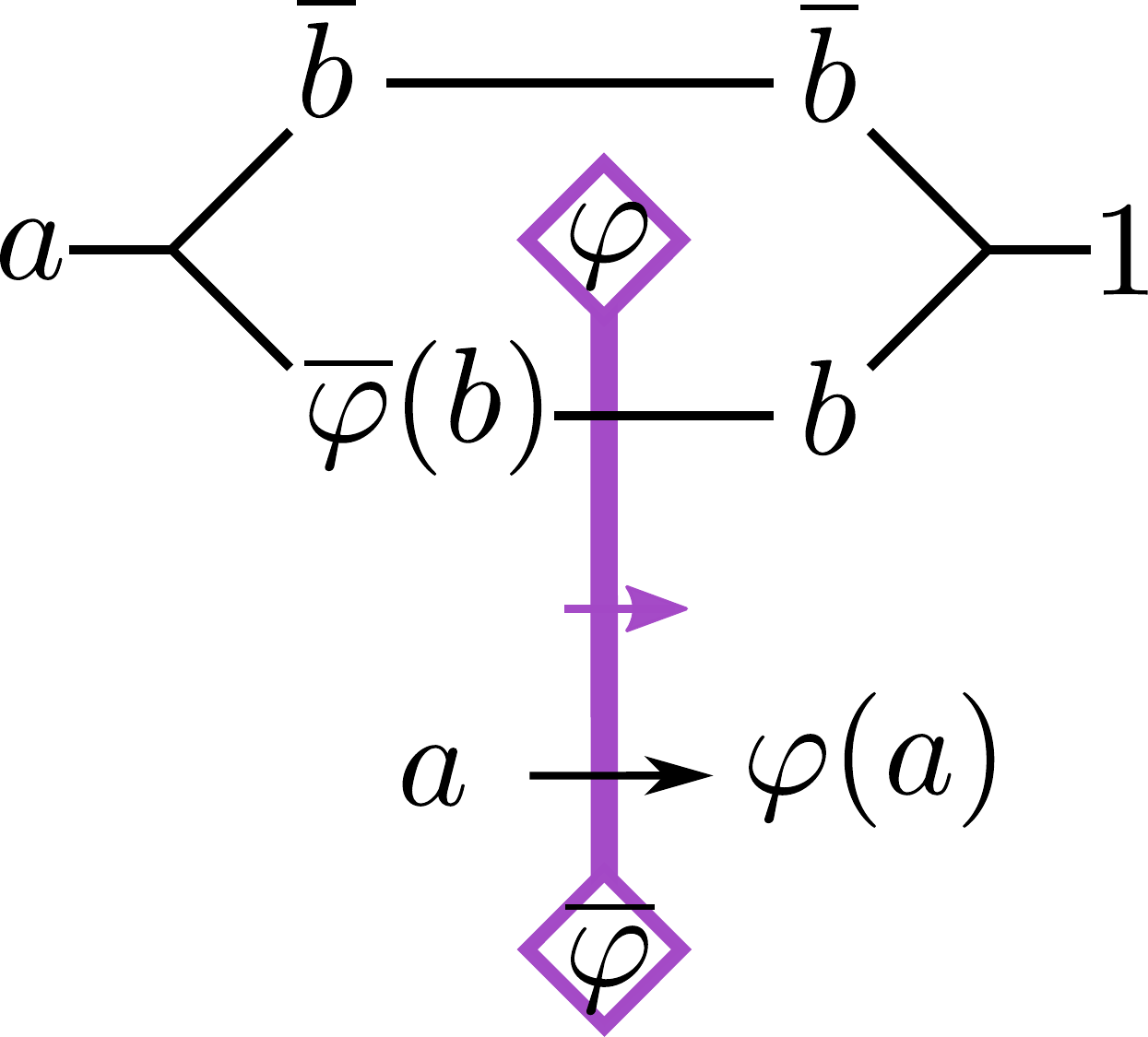}
			\caption{
				Shown is a twist described by the map $\phi$.
				We consider an anyon $a$ which can decay into $\overline{\phi}(b)$ and $\overline{b}$.
				If the $\overline{\phi}(b)$ anyon is now moved counter-clockwise around the twist, it crosses the domain wall described by $\phi$.
				Hence it turns into a $b$ anyon, which can fuse with $\overline{b}$ to form the trivial particle $1$.
				This process is called localization of anyon $a$ at twist $\phi$.
			}
			\label{fig:AnyonLocalizationGeneral}
		\end{figure}

		To determine the anyons that can localize at a twist, we study its automorphism $\phi$ together with the fusion multiplicities of the anyon model $N_{a,b}^c$.
		In particular, if there exists some excitation $b \in \cC$ such that
		\begin{equation}
			N_{\overline{\phi}(b),\overline{b}}^a = 1,
		\end{equation}
		then we have that $\phi \times a = \phi$.
		To show why this is true, we study Fig.~\ref{fig:AnyonLocalizationGeneral}.
		The figure shows a twist $\phi$ and an anyon $a$ which decays into a pair of particles $\overline{\phi}(b)$ and $\overline{b}$.
		$\overline{\phi}(b)$ is then moved counter-clockwise around the twist such that the resulting excitation $\phi \circ \overline{\phi}(b) = b$ annihilates with $\overline{b}$ and restores the system to the ground state.

\begin{figure}
			\centering
			\includegraphics[width=0.9\linewidth]{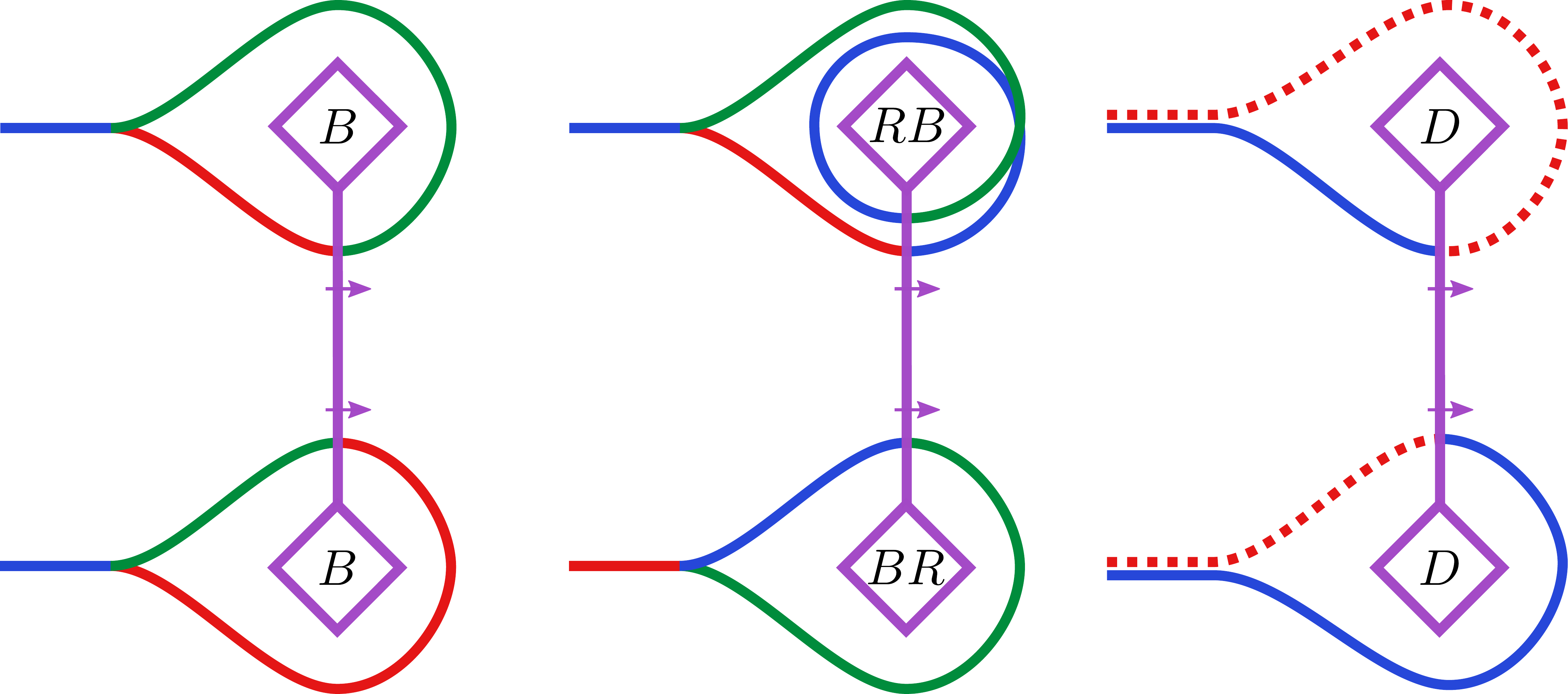}
			\caption{
				\textbf{Left:} A $B$ twist, which exchanges the red and the green color labels of anyons is shown.
				A blue anyon with any Pauli label can localize at the twist.
				We decay it into a red and a green anyon, which when moved around the twist match color and annihilate.
				\textbf{Middle:} A cyclic color twist can localize anyons of any color.
				This is because an anyon can obtain all three color labels if moved around the twist sufficiently many times.
				\textbf{Right:} The $D$ twist can condense a single fermionic anyon.
				In this case, the $f_1 = rx \times bz$ anyon gets localized.
				The dashed line indicates the different Pauli label of the red anyon.
			}
			\label{fig:AnyonLocalizationCC}
		\end{figure}

		To illustrate anyon localization in the color code, consider the three examples shown in Fig.~\ref{fig:AnyonLocalizationCC}.
		The $B$ twist can localize all blue bosonic anyons by decaying them into red and green anyons.
		If one is moved around the twist it changes color and can annihilate with the other anyon.
		Cyclic color permuting twists can annihilate all sixteen color code anyons.
		This is because all anyons are composed of the nine bosonic anyons, which in turn can decay into two anyons of different color.
		If one is moved around the twist one or two times, it can be brought to match color with the other one, so that they annihilate.
		Lastly, the $D$ twist can condense one of the fermionic particles.
		Fermions in the color code are composed of two bosonic particles with differing color and Pauli labels, see App.~\ref{app:CCFermions}.
		The domino twist $D$ as we defined it can localize the fermion $f_1 = rx \times bz$.
		To see this, we take one of its constituent bosons, say $rx$, move it around the twist once to transform it into $bz$ which annihilates with the other constituent part of $f_1$.

		The process of anyon localization at twists, as described by Eq.~\eqref{eq:FusionTwistLocalizations}, is reminiscent of the fusion in non-abelian anyon models.
		Further, as with non-abelian anyons, we can encode logical qubits in the fusion space of twists.
		To increase the number of logical qubits, more twists can be added to the system.
		The quantum dimension of a twist defect is the factor by which the fusion space dimension grows, as a twist of that type is added.
		We calculate the quantum dimension of twist $\phi$, denoted $d_\phi$, with a formula that is analogous to that of non-abelian anyons given in Eq.~\eqref{eq:GeneralQuantumDimensionAnyons}
		\begin{align}
			d_{\phi}^2 = {\sum_{a \in \cC} d_a N_{\phi,a}^{\phi}} ,
			\label{eq:QuantumDimTwist}
		\end{align}
		where $N_{\phi,a}^{\phi}$ tells us if the anyon $a$ can localize at the twist $\phi$.
		In addition to calculating the quantum dimension of a twist defect directly from its fusion rules, we also point out work where, like with non-abelian anyons~\cite{Kitaev06a, Dong08}, the quantum dimension of twist defects can be evaluated by studying the entanglement entropy of the ground state of topological phases that include twist defects~\cite{Brown13, Liu17, Bonderson17}.
 
		Since all anyons in the color code are abelian, their dimension is $1$ and hence the quantum dimension of a twist is just the square root of the number of anyons it can localize.
		As mentioned above, the $B$ twist for example can, apart from the trivial charge, localize all three blue bosonic anyons.
		Thus, its quantum dimension is $d_B = 2$.
		Cyclic color twists can localize all sixteen anyons and have quantum dimension $d_{BR} = 4$.
		The domino twist $D$ can localize only a single fermion and the vacuum and has dimension $d_D = \sqrt{2}$.
		The quantum dimension of all $72$ color code twists is given in Tab.~\ref{tab:CCQuantumDims}.
		\begin{table}
			\centering
			\begin{tabular}{|c||c|c|c|c|c|c|}
				\hline
				& $\one$ & $X$ & $Y$ & $Z$ & $XZ$ & $ZX$ \\
				\hhline{|=||=|=|=|=|=|=|}
				$\one$ & 1 & 2 & 2 & 2 & 4 & 4 \\
				\hline
				$R$ & 2 & 2 & 2 & 2 & 4 & 4  \\
				\hline
				$G$ & 2 & 2 & 2 & 2 & 4 & 4  \\
				\hline
				$B$ & 2 & 2 & 2 & 2 & 4 & 4  \\
				\hline
				$RB$ & 4 & 4 & 4 & 4 & 2 & 2  \\
				\hline
				$BR$ & 4 & 4 & 4 & 4 & 2 & 2 \\
				\hline
			\end{tabular}
			\\
			\vspace{1mm}
			\centering
			\begin{tabular}{|c||c|c|c|c|c|c|}
				\hline
				$D \circ ()$ & $\one$ & $X$ & $Y$ & $Z$ & $XZ$ & $ZX$ \\
				\hhline{|=||=|=|=|=|=|=|}
				$\one$ & $\sqrt{2}$ & $\sqrt{8}$ & $\sqrt{8}$ & $\sqrt{8}$ & $\sqrt{8}$ & $\sqrt{8}$ \\
				\hline
				$R$ & $\sqrt{8}$ & $\sqrt{2}$ & $\sqrt{8}$ & $\sqrt{8}$ & $\sqrt{8}$ & $\sqrt{8}$ \\
				\hline
				$G$ & $\sqrt{8}$ & $\sqrt{8}$ & $\sqrt{2}$ & $\sqrt{8}$ & $\sqrt{8}$ & $\sqrt{8}$ \\
				\hline
				$B$ & $\sqrt{8}$ & $\sqrt{8}$ & $\sqrt{8}$ & $\sqrt{2}$ & $\sqrt{8}$ & $\sqrt{8}$ \\
				\hline
				$RB$ & $\sqrt{8}$ & $\sqrt{8}$ & $\sqrt{8}$ & $\sqrt{8}$ & $\sqrt{8}$ & $\sqrt{2}$ \\
				\hline
				$BR$ & $\sqrt{8}$ & $\sqrt{8}$ & $\sqrt{8}$ & $\sqrt{8}$ & $\sqrt{2}$ & $\sqrt{8}$ \\
				\hline
			\end{tabular}
			\caption{
				The quantum dimension of the $72$ color code twists.
				Each entry corresponds the twist obtained by fusing the twists labeling the row and column.
				In the second table, every twist is additionally fused with the domino twist $D$.
			}
			\label{tab:CCQuantumDims}
		\end{table}

		Anyons which do not localize at a twist might get identified with other non-trivial anyons.
		Say anyon $a$ localizes at twist $\phi$ and hence $a \times \phi = \phi$, and an anyon $c$ can decay to $a$ and $b$, i.e. $c = b \times a$.
		Then, we cannot distinguish the two twist-anyon compounds, $c \times \phi = b \times a \times \phi = b \times \phi$.
		Microscopically, twist-anyon compounds are anyons located close to the twist defect.
		Mathematically, the twist-anyon compounds form a $G$-crossed braided tensor category, for more details see Ref.~\cite{Barkeshli14}.
		In this manuscript we are solely concerned with the ground-state space, or code-space when viewed as a quantum code.
		This means we will focus on twists which are locally equivalent to the twist with the trivial particle $1$ localized.

	\subsection{Interplay between twists and domain walls}
		\label{sec:TwistsBraiding}

		\begin{figure}
			\centering
			\includegraphics[width=0.65\linewidth]{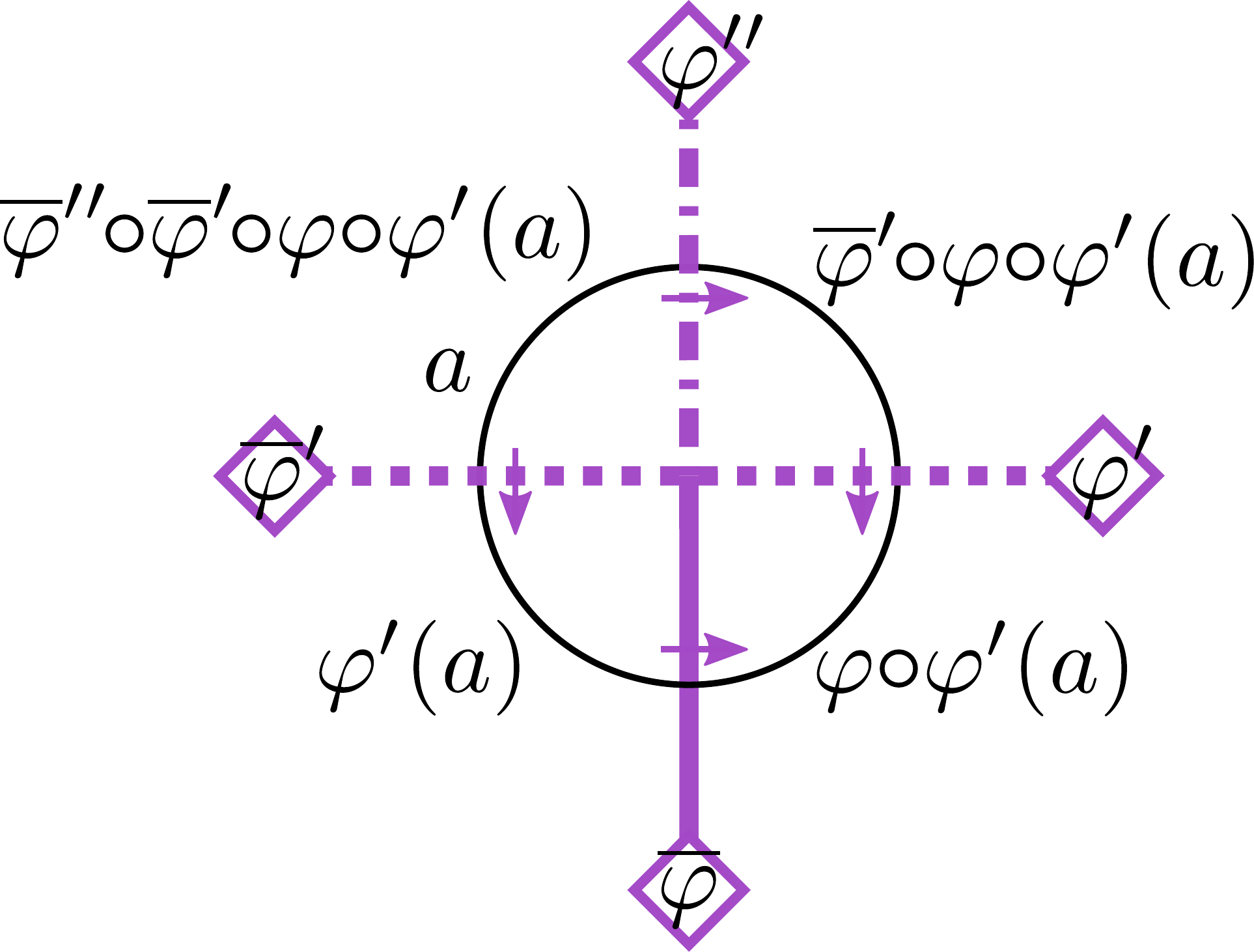}
			\caption{
				Consider a twist described by $\phi$, here drawn at the bottom, terminating a solid domain wall.
				It crosses a dotted domain wall $\phi'$.
				This transforms the twist to $\phi''$.
				An anyon $a$ starting in the top-left quadrant and traveling counter-clockwise around the crossing point gets finally mapped to $\overline{\phi}'' \circ \overline{\phi}' \circ \phi \circ \phi'(a)$.
				Since its path is contractible, the final label has to be equal to its original label $a$.
				Following this equivalence for a generating set of anyons identifies the twist $\phi'' = \overline{\phi}' \circ \phi \circ \phi'$ as the conjugation of $\phi$ with $\phi'$.
			}
			\label{fig:TwistCrossingDWGeneral}
		\end{figure}
		We now study crossing domain walls and the twists which terminate them.
		This is relevant when twists are moved over domain walls using, for instance, code deformations~\cite{Bombin10,Bombin11,Wootton15a,Teo14,You12}.
		To figure out how the twist changes after crossing a domain wall we consider closed contractible loops engulfing the crossing point of the domain walls.
		If an anyon was moved along such a contractible path until it reaches its original position, it should return to its initial charge label.
		Consider a twist which acts as the map $\phi$ on the anyon labels, as depicted in Fig.~\ref{fig:TwistCrossingDWGeneral}.
		After it crosses a domain wall $\phi'$ it transforms into a twist acting as $\phi''= \overline{\phi}' \circ \phi \circ \phi'$.
		An anyon $a$ which moves on the aforementioned path then gets mapped by $\overline{\phi}'' \circ \overline{\phi}' \circ \phi \circ \phi' = \one $.
		Remember, the equality stems from the fact that the loop followed by the anyon is contractible.
		This means the symmetry describing the transformed twist is obtained by conjugating its original symmetry with the one of the crossed domain wall.
		Hence, twists crossing domain walls will stay within the same equivalence class, see Sec.~\ref{sec:TwistsFusion}.

\begin{figure}[b!]
			\centering
			\includegraphics[width=0.5\linewidth]{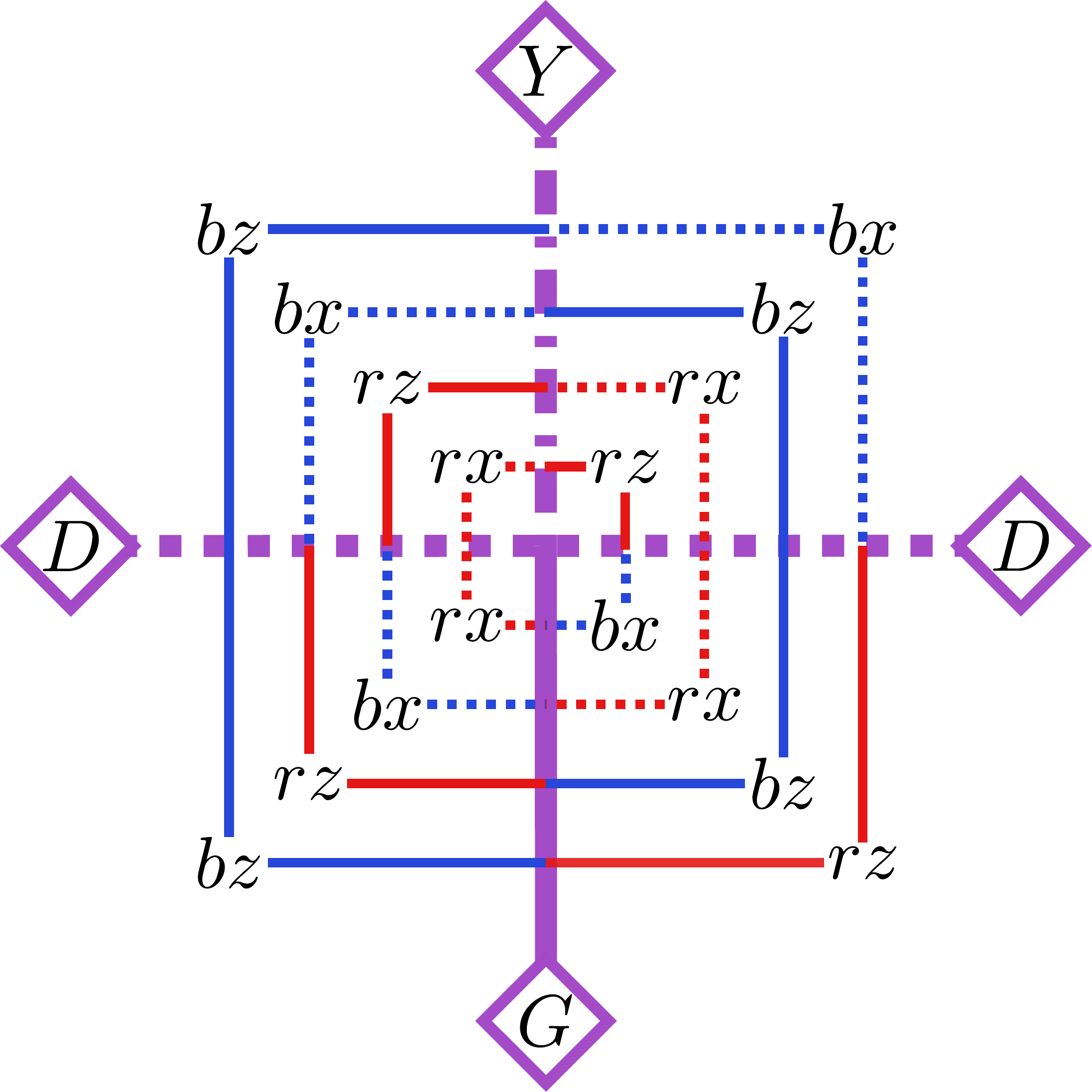}
			\caption{
				A $G$ twist (bottom) transforms to a $Y$ twist (top) if moved over a domino domain wall $D$ (middle).
				This can be seen by starting with a generating set of four anyons, $rx$, $rz$, $bx$ and $bz$, in the top-left quadrant.
				Moving them counter-clockwise around the crossing point to the top right, they transform to $rz$, $rx$, $bz$ and $bx$, respectively.
				Moving them back to their initial position over the top half of the vertical domain wall should return them to their initial charge.
				The only domain wall with this effect is the one terminated by a $Y$ twist.
			}
			\label{fig:CCTwistCrossing}
		\end{figure}

		Again, let us illustrate this with a color code example.
		Consider a color exchanging twist, say $G$ which crosses the domino domain wall described by the duality symmetry $D$.
		Using the procedure described above for a generating set anyons, $rx$, $rz$, $bx$ and $bz$ shows that the twist transforms to a Pauli exchanging twist $Y$, see Fig.~\ref{fig:CCTwistCrossing}.
		This is to show the duality transformation $D$ does affect twists in very much the same way as it does anyons.
		The Pauli label of twists and the color labels are exchanged.

		\begin{figure}
			\centering
			\includegraphics[width=.8\linewidth]{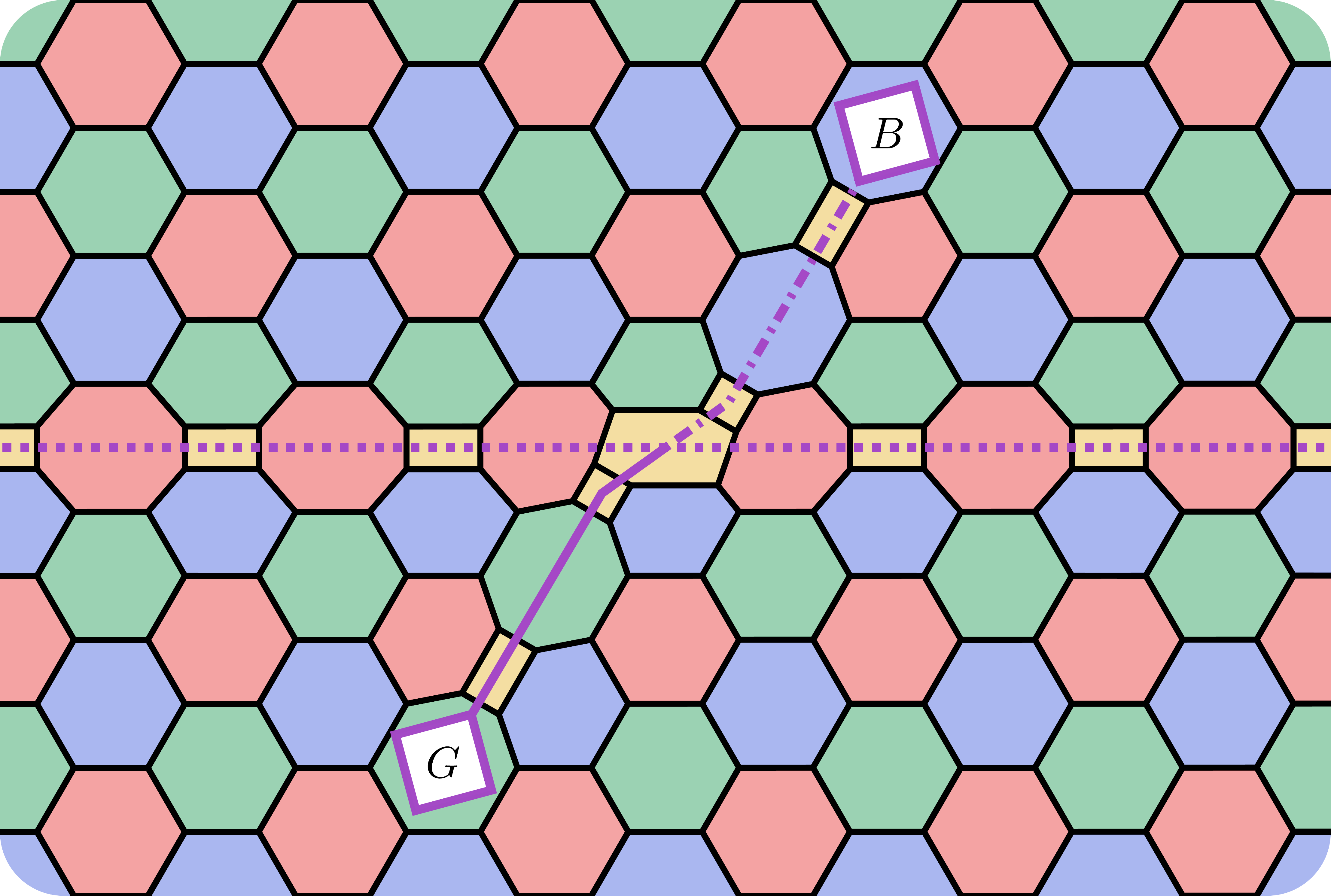}
			\caption{
				Lattice representation of a twist crossing a domain wall.
				A $G$ twist (bottom) is moved over an $R$ domain wall (middle) and transforms to a $B$ twist (top).
				Apart from the plaquettes hosting the twists marked by purple squares, all plaquettes host two independent stabilizers.
			}
			\label{fig:TwistCrossingDWPrimal}
		\end{figure}
		As a second example we consider a color exchanging twist $G$ crossing a color exchanging domain wall $R$ which turns it into a $B$ twist.
		This follows from the fact that $GRG = B$.
		Again, we see that the domain wall has an equivalent action on the twist label as it would on anyon labels, see Fig.~\ref{fig:TwistCrossingDWPrimal} for a lattice representation.
		This example can again be understood naturally using the Pachner move construction for twists and domain walls, see Fig.~\ref{fig:TwistCondensationDual} in App.~\ref{app:PachnerMoves}.

	\subsection{Confinement of twists at boundaries}
		\label{sec:CornerTwists}

		In this subsection, we explore how twists and domain walls interact with boundaries.
		In particular, a domain wall covering a boundary can change the boundary type.
		This implies that a twist defect confined at the boundary corresponds to a corner.
		A corner is the point where two different boundaries meet, as described in Sec.~\ref{sec:Boundaries}.
		If the boundary type is not changed by the automorphism applied by the domain wall then the corresponding twist can condense at said boundary.

\begin{figure}
			\centering
			\includegraphics[width=.55\linewidth]{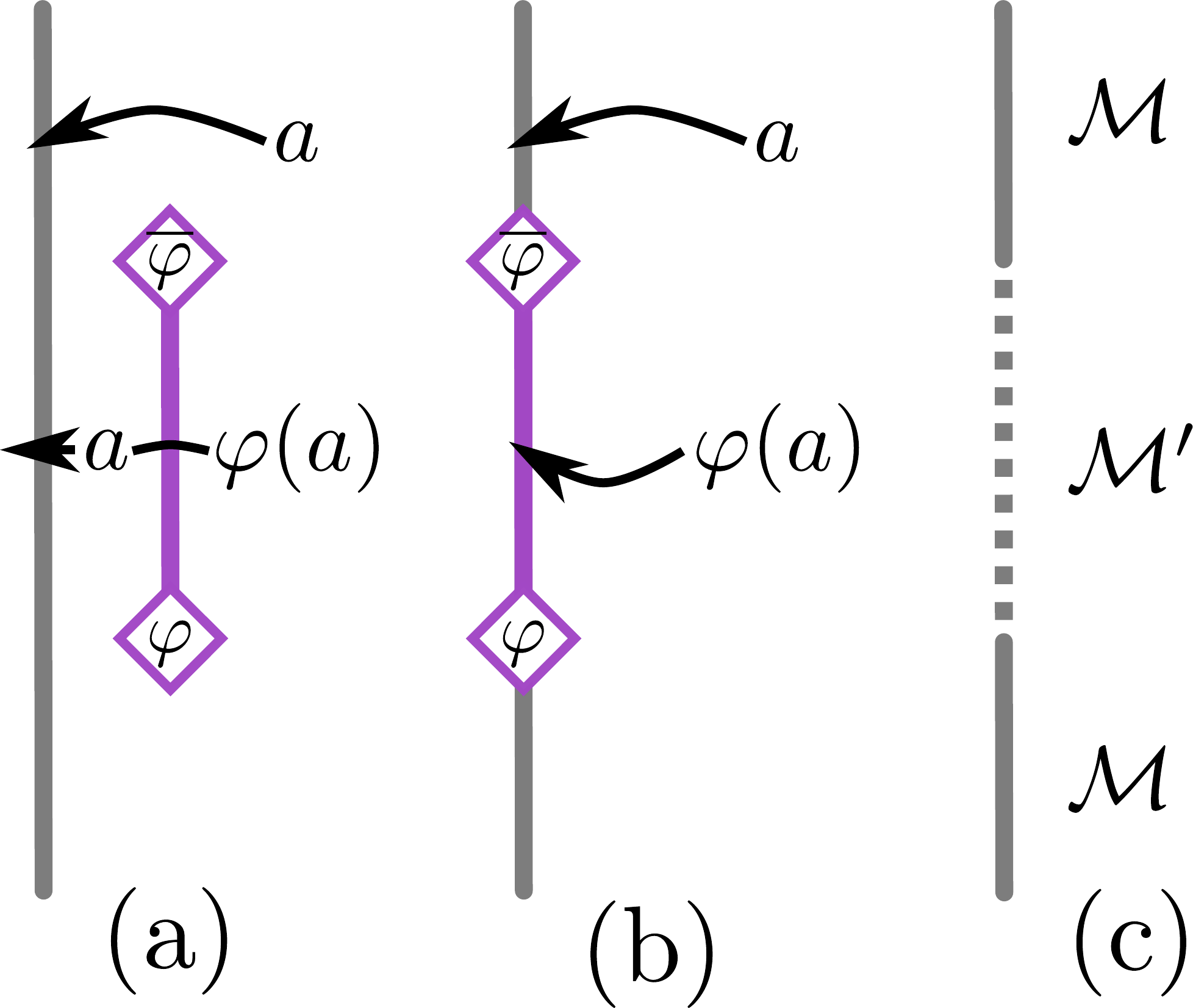}
			\caption{
				Corners, points where the boundary type changes, can be identified with confined twist defects.
				In \textbf{(a)}, a pair of twists $\phi$ and $\overline{\phi}$ sit close to a boundary $\cM$ where anyons $a \in \cM$ can condense.
				The boundary behind the domain wall is able to condense anyons $\phi(a)$, if moved over the domain wall first.
				\textbf{(b)} shows same situation as (a), but the twists and their connecting domain wall are now covering the boundary.
				This corresponds to \textbf{(c)}, where boundary covered by twists and the domain wall now correspond to corners and a changed boundary denoted by the dashed gray line.
				If the original boundary is described by the Lagrangian subgroup $\cM$, then the dashed domain wall is described by $\cM' \defeq \phi(\cM)$.
				Note, this encapsulates trivial corners, i.e. the case where $\cM = \cM'$.
				We call twists $\phi'$ for which $\phi'(\cM) = \cM$ condensible at the boundary $\cM$.
			}
			\label{fig:TwistsAndCorners}
		\end{figure}

		Let us begin by investigating how boundaries change when domain walls are present.
		Consider a domain wall described by the automorphism $\overline{\phi}$ close to a boundary which is characterized by its Lagrangian subgroup $\cM$, see in Fig.~\ref{fig:TwistsAndCorners}~(a).
		If an anyon $a$ is condensible at $\cM$, i.e. $a \in \cM$, then the anyon $\phi(a)$ can condense at the boundary by first moving over the domain wall.
		This transforms it to $\overline{\phi} \circ \phi(a) = a \in \cM$.
		Clearly, an anyon $\phi(a)$ can still condense when the domain wall covers the boundary, as depicted in Fig.~\ref{fig:TwistsAndCorners}~(b).

		In fact, the anyons $\phi(a)$ form a Lagrangian subgroup $\cM' \defeq \phi(\cM)$ which may condense at the resulting boundary.
		This follows from the properties of Lagrangian subgroups, discussed in Sec.~\ref{sec:GeneralLagrSubgroups}, and the conditions put on the symmetries of the anyon model, see Eqs.~(\ref{eq:DWConditionFusion}-\ref{eq:DWConditionBraiding}).
		To illustrate this, the covering domain wall was removed in Fig.~\ref{fig:TwistsAndCorners}~(c) and the new boundary $\cM'$ drawn as a dashed line.
		In this picture the twists correspond to the corners between the different boundaries.
		In an alternative reading of the above we have that the corners correspond to confined twists.
		In contrast, if an automorphism $\phi'$ leaves the boundary type unchanged, i.e. $\phi'(\cM) = \cM$, we say the twist $\phi'$ can condense at the boundary $\cM$.
		Twist condensation is discussed in more detail in Sec.~\ref{sec:TwistCondensation}.

		This leads to an ambiguity, corners can not, in general, be uniquely identified with a single twist, as they might fuse with condensible twists.
		If $\phi(\cM) = \cM'$ and $\phi'(\cM) = \cM$, then $\phi \circ \phi'(\cM) = \cM'$.
		This means a twist $\phi \circ \phi'$, the fusion outcome of a condensible twist $\phi'$ with a corner twist $\phi$ is again a valid interpretation of the corner.

\begin{figure}
			\centering
			\includegraphics[width=.85\linewidth]{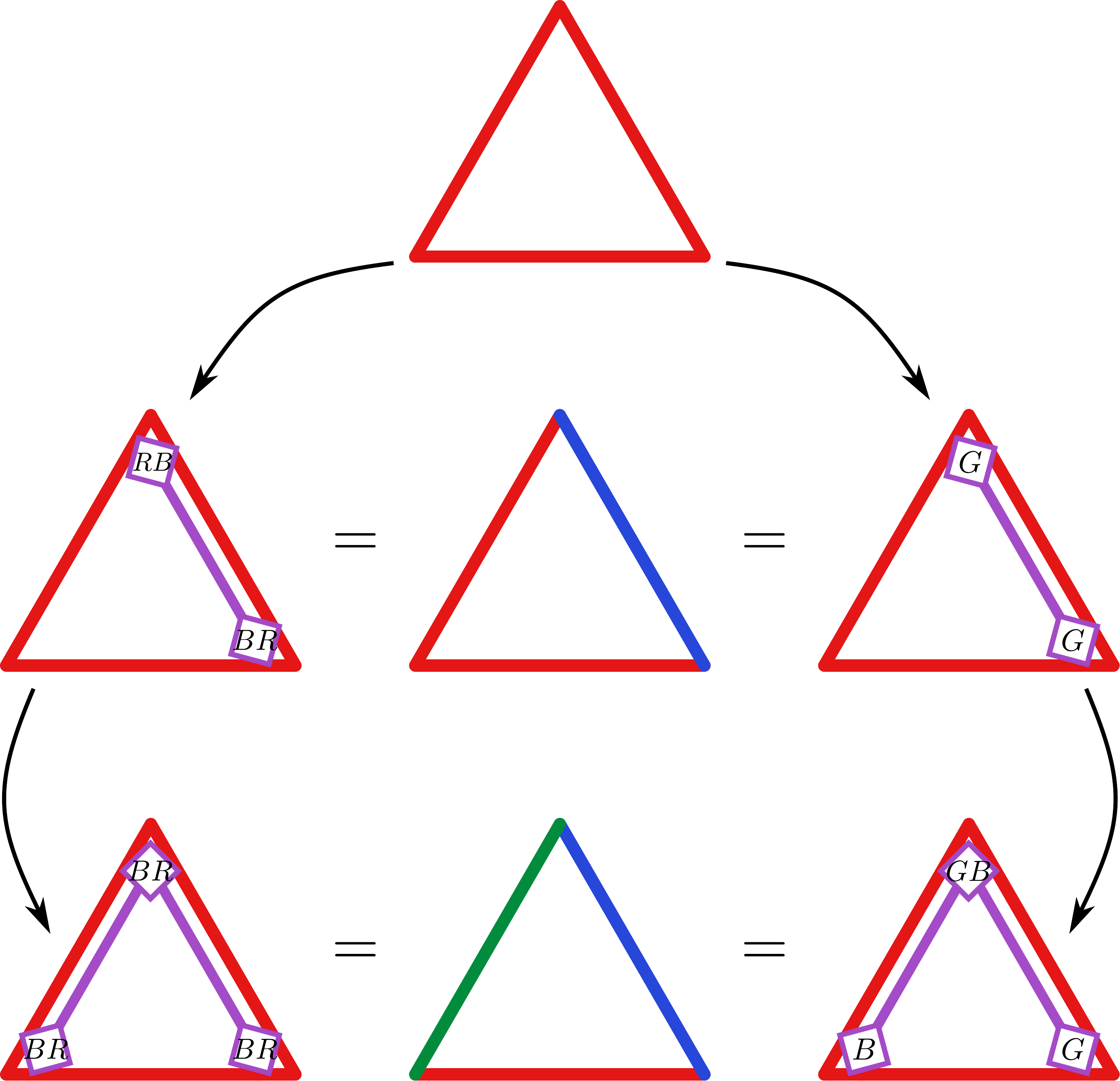}
			\caption{
				The identification of color code corners with twists is ambiguous.
				We start with a color code with just a single red boundary, as shown in the first row.
				Consecutively, we transform one boundary after the other by covering them with domain walls, until the common triangular color code in the bottom center is obtained.
				This construction shows the non-uniqueness of the interpretation of corners as twists.
				On the left, we end up with three $BR$ twist, one in each corner, where on the right only one corner is identified with a cyclic twist.
			}
			\label{fig:TwistsAndCornersCC}
		\end{figure}

		Let us now turn to the corners in the color code for concrete examples of the aforementioned ambiguity in the association of corners with twists.
		Consider the common triangular color code with the three color boundaries, red, green and blue.
		This code can be constructed in the several ways from a disk with a single red boundary and adding open domain walls to segments of the uniform red boundary.
		We show two such ways in Fig.~\ref{fig:TwistsAndCornersCC}.
		We first consider applying a domain wall that implements the $G$ symmetry to cover one red colored boundary segment which transforms it to the blue color boundary.
		Similarly, the $B$ domain wall creates a green boundary when covering the second part of the red boundary.
		Hence, the corner that lie between the red-blue(red-green) boundary can be interpreted as a $G$ ($B$) twist.
		The blue-green corner then is the fusion of the two twists which we call $GB = BR$.
		This is depicted on the right in Fig.~\ref{fig:TwistsAndCornersCC}.
		Alternatively we can use the cyclic domain walls $RB$ and $BR$ to transform a segment of the red boundary onto a blue or green boundary, respectively.
		Hence, as shown on the left in Fig.~\ref{fig:TwistsAndCornersCC}, we can label each of the three corners as a cyclic color permuting twist $BR$.
		The two approaches to realize the color code with three differently colored boundaries differ by an $R$ twist fused with the two corner twists that lie at the red boundary.
		However, given the red boundary is invariant if we cover it with a red domain wall, we see that the physics of the two models we have produced is invariant under the fusion of the $G$ and $B$ twist of Fig.~\ref{fig:TwistsAndCornersCC} with red twists.
		We thus see that the two pictures we have produced are consistent.

		Corners between Pauli boundaries can be identified equivalently with Pauli permuting twists.
		Corners between Pauli and color boundaries host twists shown in the second half of Tab.~\ref{tab:CCConjClasses}, who involve the $D$ twist when generated from the standard set of symmetries.
		This is shown in more detail in Fig.~\ref{fig:TwistCondensationLattices}, only that there, the corners are detached and pulled into the bulk as twists.
		This process is discussed in Sec.~\ref{sec:TwistCondensation}.

	\subsection{Detaching corners and twist condensation}
		\label{sec:TwistCondensation}

\begin{figure}
			\centering
			\includegraphics[width=.8\linewidth]{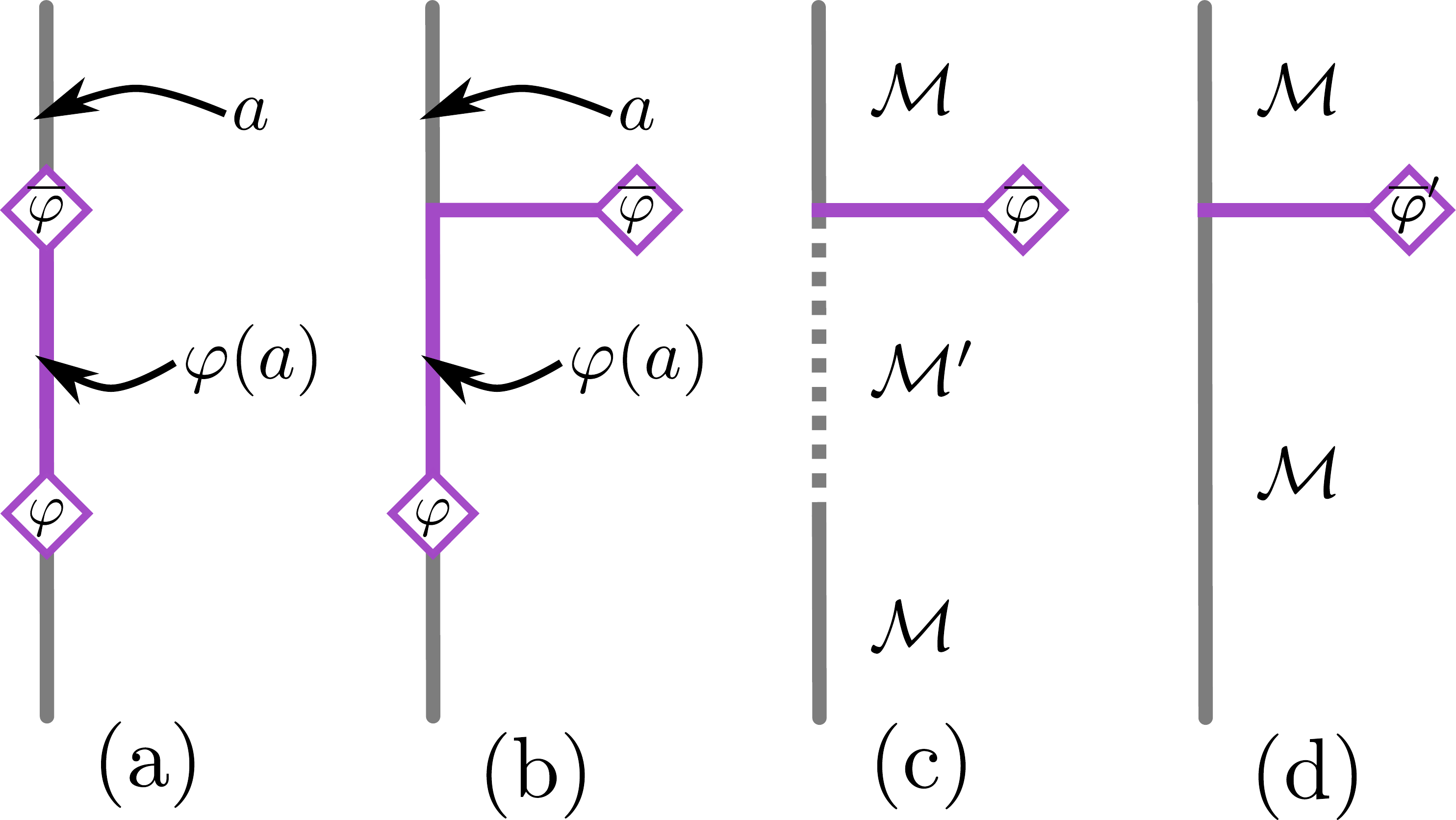}
			\caption{
				\textbf{(a)} Shows the same situation as depicted in Fig.~\ref{fig:TwistsAndCorners}~(b), where the $\cM'$ boundary is seen as a domain wall covering the $\cM$ boundary, and the corners as twists.
				In \textbf{(b)}, the domain wall is extended such that the $\overline{\phi}$ twist now lies in the bulk.
				This corresponds to \textbf{(c)}, where the domain wall covering the boundary is omitted to illustrate that it constitutes a different boundary.
				Now, the situation looks like a single twist whose domain wall terminates between the $\cM$ and $\cM'$ boundaries.
				This can be read as a detached corner which is pulled into the bulk as a $\overline{\phi}$ twist.
				\textbf{(d)} shows the case where $\cM = \phi'(\cM)$, the trivial corner.
				Twists which map the Lagrangian subgroup of a boundary back to itself are able to condense at and being emitted from said boundary.
			}
			\label{fig:TwistCondensationGeneral}
		\end{figure}

		Not only can we identify corners with certain twists, we can also detach corner twists and move them into the bulk.
		Consider Fig.~\ref{fig:TwistCondensationGeneral}~(a), where we interpret the $\cM'$ boundary as a domain wall $\phi$ covering the $\cM$ boundary, and the corners as twists.
		These twists can be moved, for example through code deformation, such that they come to lie in the bulk, see Fig.~\ref{fig:TwistCondensationGeneral}~(b).
		This leads to Fig.~\ref{fig:TwistCondensationGeneral}~(c), the twist is detached from the corner and pulled into the bulk.

		Notice how this framework also encapsulates the trivial corner, i.e. the case $\cM = \cM'$.
		This means a twist $\phi'$ which maps the Lagrangian subgroup describing a boundary back onto itself, i.e. $\phi'(\cM) = \cM$, can condense at or be emitted from said boundary, see Fig.~\ref{fig:TwistCondensationGeneral}~(d).

\begin{figure}[t!]
			\centering
			\includegraphics[width=0.95\linewidth]{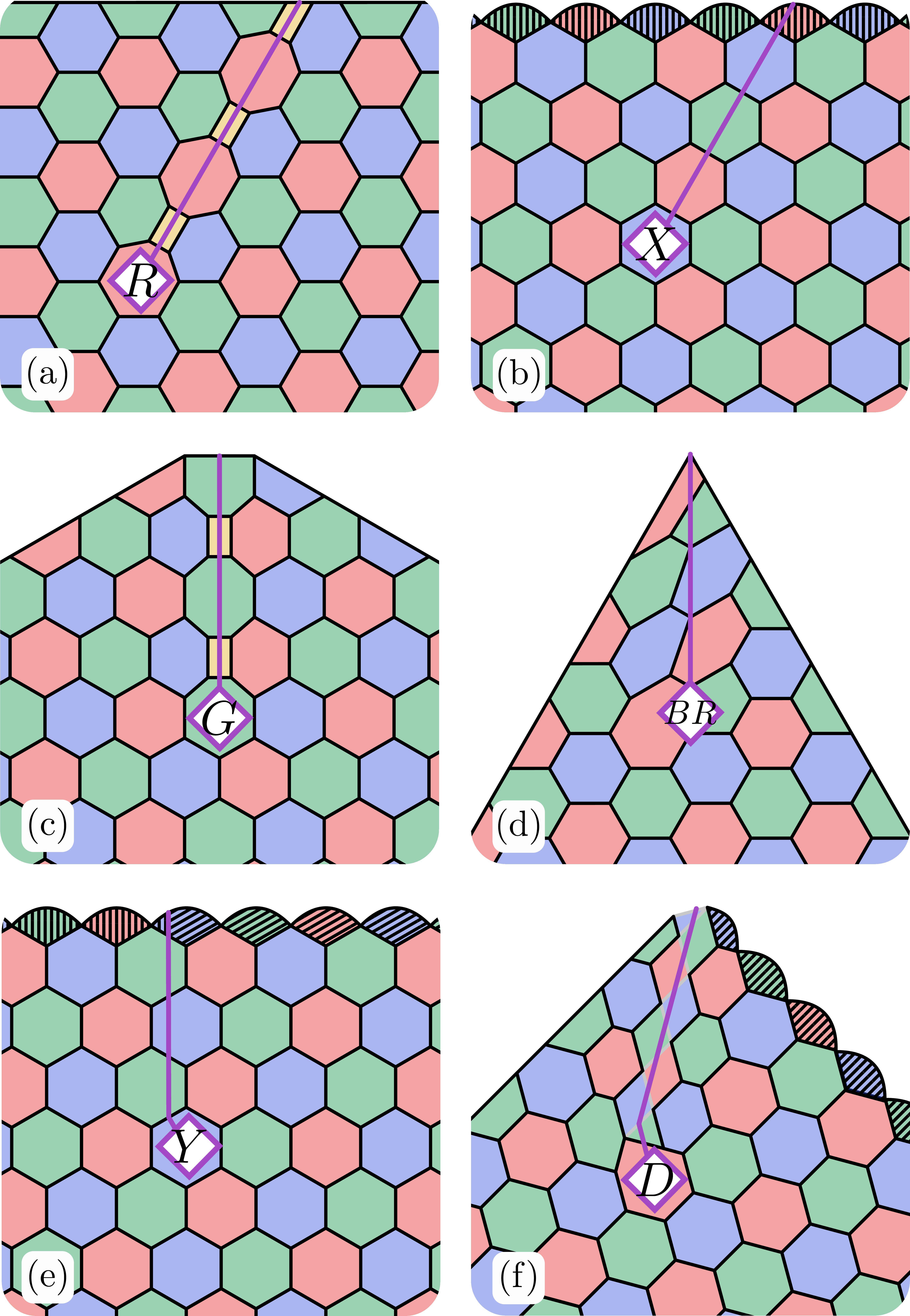}
			\caption{
				Lattice representations of twists condensing at boundaries and detached corner twists.
				\textbf{(a)} an $R$ twist emitted from a red color boundary.
				\textbf{(b)} an $X$ twist emitted from $x$ Pauli boundary.
				\textbf{(c)} a $G$ twist as a detached blue-red color/color corner.
				\textbf{(d)} a $BR$ twist as a detached blue-red color/color corner.
				\textbf{(e)} a $Y$ twist as a detached $x$-$z$ Pauli/Pauli corner.
				Notice, on the boundary plaquettes the changed orientation of the stripe pattern indicates a change in basis.
				\textbf{(f)} a $D$ twist as a detached red-$x$ color/Pauli corner.
			}
			\label{fig:TwistCondensationLattices}
		\end{figure}

		To illustrate the above with practical examples, let us turn our focus again to the color code model.
		Starting with twists condensation, we can observe that all Pauli twists can condense at color boundaries, since they map the Lagrangian subgroups of color boundaries back onto themselves.
		Additionally, color exchanging twists which leave one color invariant can condense at and be emitted from the boundary of said color.
		As an example, consider Fig.~\ref{fig:TwistCondensationLattices}~(a), where an $R$ twist is emitted from a red boundary.
		Pauli boundaries can emit Pauli exchanging twists which leave the relevant label invariant, as well as all purely color permuting twists and twists which are the fusion product of these examples.
		See Fig.~\ref{fig:TwistCondensationLattices}~(b) for an example of an $X$ twist emitted from an $x$ Pauli boundary.
		Let us now consider the different color code corners, discussed in Sec.~\ref{sec:CCLatticeRepBoundaries} and depicted in Fig.~\ref{fig:CCBoundaries}.
		Corners between two color boundaries can be interpreted as the twist which exchanges them.
		For instance, a corner between a blue and a red color boundary can be detached and pulled into the bulk as a $G$ twist, see Fig.~\ref{fig:TwistCondensationLattices}~(c).
		Alternatively, the cyclic color twist $RB$ can be pulled into the bulk as a detached blue-red corner, see Fig.~\ref{fig:TwistCondensationLattices}~(d).
		The same holds for Pauli/Pauli corners, consider Fig.~\ref{fig:TwistCondensationLattices}~(e) showing a $Y$ twist as a detached $x$-$z$ corner.
		The last type of corner where color and Pauli boundaries meet, corresponds to $D$ type twists.
		Fig.~\ref{fig:TwistCondensationLattices}~(f) shows a detached color/Pauli corner as a $D$ twist in the bulk.

		Twist condensation and detaching corner twists can be used in fault-tolerant quantum computation.
		Topological degrees of freedom associated to twist defects, in which twist encoded logical qubits can be stored, become local degrees of freedom when the twist is condensed at a boundary.
		Measuring these local degrees of freedom then corresponds to measurements on logical qubits.
		In the reverse process, where twists are emitted from corners or boundaries, the logical degrees of freedom increase.
		This is similar to considering non-abelian anyons and a boundary where such anyon excitations condense.
		This is similar to how merging and splitting codes using lattice surgery changes the number of logical qubits~\cite{Horsman12}.
		This is no coincidence, as discussed in Ref.~\cite{Brown17}, identifying corners with twists unifies encoding schemes relying on boundaries with twist encoding schemes.
		This offers a new understanding of lattice surgery used in recent proposals for efficient fault-tolerant quantum computation schemes~\cite{Nautrup17, Litinski18}.
		Lastly, identifying corners as twists can help in the search for hardware efficient quantum error-correcting codes.
		For instance, the code presented in Ref.~\cite{Yoder16} may be understood as a surface code where one corner twist is detached and pulled into the center of the lattice.
		This lowers the overhead of physical qubits needed to encode logical qubits at a given code distance.
		Even more efficient codes based the same insight are presented in App.~\ref{app:StellatedSurfaceCodes} and in the following Sec.~\ref{sec:TwistsApplications}, where $G$ twists in the center of the code can be interpreted as detached corners.

\section{Stellated color codes}
	\label{sec:TwistsApplications}

	We next present a new family of two-dimensional stabilizer codes that make use of the twist defects of the color code phase.
	The proposed codes reduce the number of physical qubits required per code qubit while preserving desirable properties such as code distance and geometric locality of stabilizer generators in a 2D euclidean lattice.

	For topological stabilizer codes on flat manifolds, the number of logical qubits $k$ one can encode using $n$ physical qubits with code distance $d$ is upper bounded by the holographic bound~\cite{Bravyi10} for codes defined by geometrically local commuting constraints embedded in 2D Euclidean space
	\begin{align}
		k \le c~\frac{n}{d^2}.
		\label{eq:BravyiTerhalPoulin}
	\end{align}
	Here, $c$ is a constant which depends only on the locality of the constraints defining the error-correcting code.
	One can compare these $c$-values of different codes, where a higher $c$ roughly corresponds to a more efficient encoding scheme.
	For the surface code with three-body boundary terms, as proposed by Kitaev~\cite{Kitaev03}, one obtains $c = \frac{1}{2}$.
	This value can be doubled to $c = 1$ by rotating the code and using two-body boundary terms, resulting in the rotated surface code~\cite{Wen03,Bombin06a,Hastings15}.
	The same $c$-value can be achieved by appropriately arranging punctures~\cite{Delfosse16b}.
	In surface codes, a further increase of the $c$ value is possible using twist defects.
	The triangular surface code~\cite{Yoder16} and certain configurations of twists and holes in hybrid encoding schemes~\cite{Brown17} achieve $c = \frac{4}{3}$.
	The same value, $c = \frac{4}{3}$, is found for the triangular 6.6.6 color code.
	Using a 4.8.8 lattice, the triangular color code with color boundaries reaches $c = 2$~\cite{Landahl11}.
	As far as we are aware of, this is the highest $c$-value of any known flat two-dimensional topological error correction code with bounded stabilizer weight $w \le 8$.
	Hyperbolic codes~\cite{Freedman02, Delfosse13, Breuckmann17} exceed the bound given in Eq.~\eqref{eq:BravyiTerhalPoulin}, but are expected to come along with significant experimental challenges, as they can not be embedded in two spatial dimensions while maintaining the locality of stabilizers.

\begin{figure}
		\centering
		\includegraphics[width=1\linewidth]{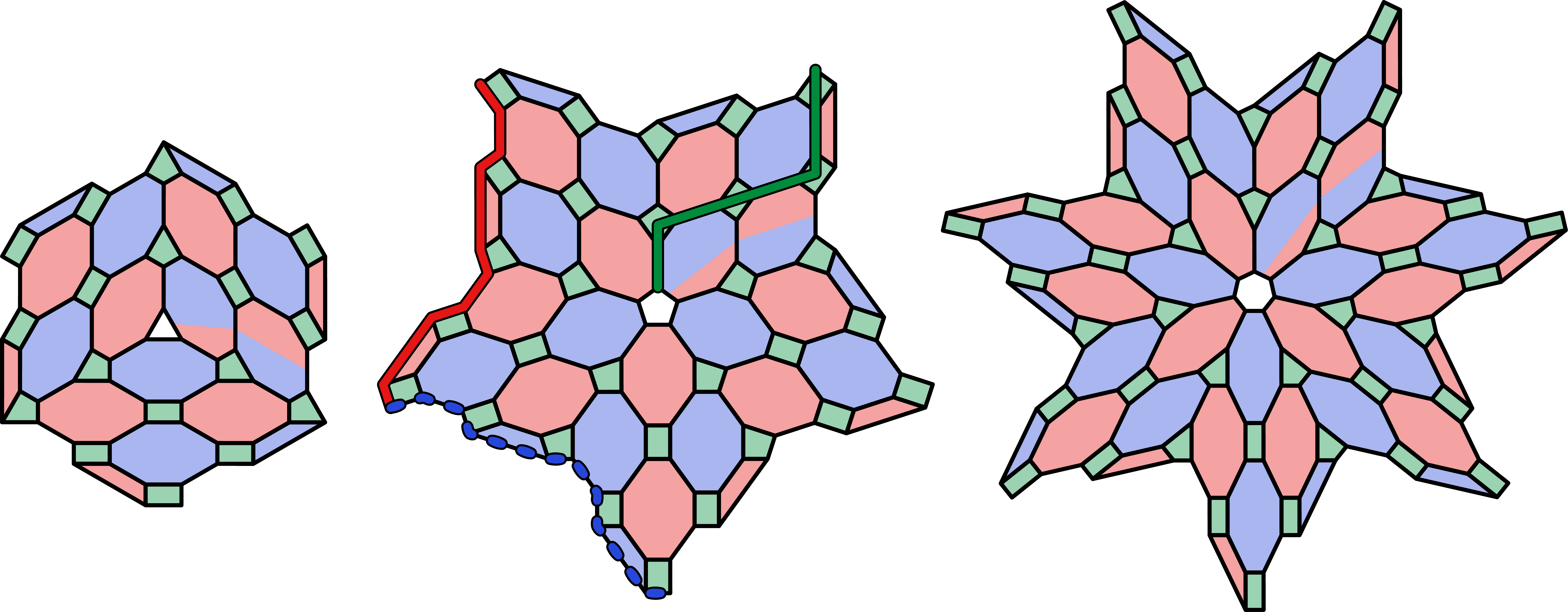}
		\caption{
			Three members of the family of stellated 4.8.8 color codes.
			The three shown examples have a code distance $d = 9$.
			\textbf{Left:} the $s = 3$ case is shown.
			This code encodes two logical qubits.
			The \textbf{middle} and the \textbf{right} show the cases $s = 5$ and $s = 7$, encoding four and six logical qubits, respectively.
			The logical operators apply $X$, $Y$ and $Z$ on qubits supported by strings connecting different boundaries.
			Some logical operators are shown in the middle code as examples.
		}
		\label{fig:StellatedColorCode}
	\end{figure}

	Inspired by the triangular surface code~\cite{Yoder16}, we start by presenting an analogous color code model that achieves $c = \frac{8}{3}$.
	It can be obtained by fattening~\cite{Bombin07b} the vertices of the triangular surface code to quadratic faces.
	Alternatively, this process can be understood in terms of code concatenation with the $\llbracket 4,2,2 \rrbracket$-code, as discussed in Ref.~\cite{Criger16}.
	It leads to the left most code shown in Fig.~\ref{fig:StellatedColorCode}.
	In the figure, a single qubit lies on each vertex of the lattice and each plaquette $p$ supports two stabilizers, $s_p^x = \prod_{j \in \partial p} X_j$ and $s_p^z = \prod_{j \in \partial p} Z_j$, where $\partial p$ denotes the set of qubits lying on the vertices adjacent to $p$.
	This is the same as the standard color code stabilizers, see Eq.~\eqref{eq:CCStabilizers}.
	We leave a puncture with a green boundary at the center of the model where no stabilizer is enforced.

	The properties of the model we have described are understood naturally from the perspective of non-abelian twist defects.
	We first point out that the color symmetry is broken when we undergo a full monodromy of the central puncture.
	As such, we find a single $G$-twist is condensed in the central puncture, connected to the boundary with a domain wall.
	Further, there are exactly three corner twists of type $G$ lying at the interfaces where red and blue boundaries meet.
	The ground-state degeneracy then is equivalent to the fusion space of four $G$-twists that are constrained to fuse to the vacuum.
	The central puncture encodes a qubit, which we treat as a gauge qubit.

	The described code is a representative of a whole code family we call stellated color codes.
	Codes in this family are parameterized by $s$, the order of rotational symmetry of the code.
	The representatives with $s = 3$, $5$ and $7$ of the stellated 4.8.8 color code family are shown in Fig.~\ref{fig:StellatedColorCode}.
	Again, each code supports $s+1$ color permuting twists for odd $s$.
	Note that in principle $s$ can be chosen to be even, in which case the trivial twist, i.e. no twist, is featured in the center.
	The case where $s = 4$, for example, is a rectangular 4.8.8 color code featuring alternating red and blue boundaries.
	For odd $s$, a stellated code encodes $k = s-1$ logical qubits and reaches a $c$-value of $c = 4 - \frac{4}{s}$.
	Stellated codes with even $s$ encode $k = s-2$ logical qubits and we find $c = 4 - \frac{8}{s}$.
	In both cases, $c$ approaches $4$ as $s$ grows.
	Allowing $s$ to diverge doubles the highest known values of $c$ for flat topological error-correcting codes such that $c \rightarrow 4$.
	The advantage is achieved by locally changing the curvature around the central plaquette which is known to enhance the encoding rate in hyperbolic codes \cite{Delfosse13, Breuckmann17}.

\begin{figure}
		\centering
		\includegraphics[width=1\linewidth]{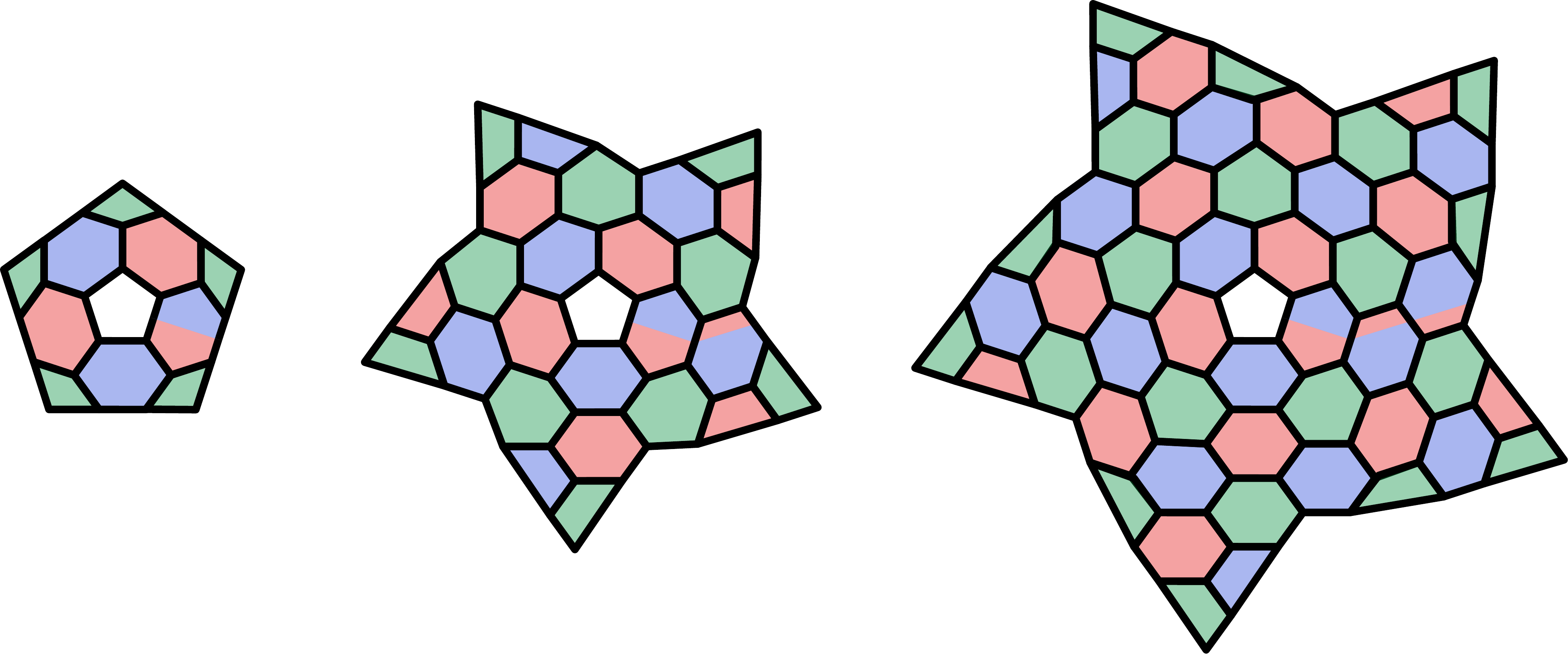}
		\caption{
			Examples of stellated color codes on a 6.6.6 lattice.
			These are the three smallest codes with $s = 5$, depicted with code distances $d = 3, \, 5, \, 7$ from left to right.
			The shown codes each encode two logical qubits.
			The encoding rate for the 6.6.6 stellated color codes approaches $c = \frac{8}{3} \approx 2.7$ for large $s$, which, to the best of our knowledge, is the highest encoding rate of any flat topological quantum code with $w \le 6$.
		}
		\label{fig:StellatedColorCode666}
	\end{figure}

	Given that the measurement of high-weight stabilizers can be challenging experimentally we also present the stellated color codes based on the 6.6.6 honeycomb lattice.
	Their encoding rate is $c = \frac{8}{3} - \frac{8}{3 s}$ ($c = \frac{8}{3} - \frac{16}{3 s}$) for odd (even) $s$.
	See Fig.~\ref{fig:StellatedColorCode666} for three $s = 5$ representatives.
	These codes are, to the best of our knowledge, the most efficient topological error correction codes with stabilizer weights $w \le 6$ which are embeddable in two spacial dimensions.

	Interestingly, the discussed increase of $s$ can also be applied to the twisted surface code model~\cite{Yoder16}.
	The encoding rates for these models halve compared to the stellated 4.8.8 color codes.
	However, the stellated surface codes we present may be of interest as typically surface code based architectures have higher fault-tolerance thresholds than those based on the color code model~\cite{Campbell17}.
	To substantiate such a claim, one would likely need to conduct numerical analyses which goes beyond the scope of the present work.
	We give a brief description of stellated surface codes in App.~\ref{app:StellatedSurfaceCodes}.
	One should also investigate different ways of performing logical operations on qubits encoded in stellated codes.
	As a first approach one could take the description of these models in terms of their twists and generalize the schemes presented in, for instance, Refs.~\cite{Landahl14, Brown16, Yoder16, Litinski18,Lavasani18}, where fault-tolerant approaches to braid and fuse twist defects are considered.
	We leave the problem of optimizing fault-tolerant protocols for quantum computation with these models to future work.

\section{Unfolding boundaries and twists}
	\label{sec:2xTC}

	It is well established~\cite{Bombin12, Kubica15, Bhagoji15, Criger16} that we can locally map the color code onto two copies of the toric code model~\cite{Kitaev03}.
	This is clear because the excitations of the color code have the same braid statistics and fusion rules of their analogous excitations in the mapped picture, and hence these two models are members of a common quantum phase~\cite{Bravyi06, Chen10}.
	So too then must two copies of the toric code share the Lagrangian subgroups and excitation symmetries of the color code, and thus give rise to analogous boundaries and twists.
	In this section we concisely describe these analogous objects.
	We study the different boundaries of the model, as well as a generating set of twist defects.

	To briefly review, all anyons in the toric code are abelian and take labels $1$, $e$, $m$ and $\epsilon = e \times m$, where $e$ and $m$ are bosons and $\epsilon$ is a fermion.
	Braiding two different non-trivial toric code anyons results in a phase, $M_{e,m} = M_{e,\epsilon} = M_{m,\epsilon} = -1$.
	The Lagrangian subgroups of the toric code are thus $\cM^e = \{ 1, e \}$ and $\cM^m = \{ 1, m \}$~\cite{Bravyi98,Dennis02,Kitaev12}.
	The boundaries that condense the excitations $\cM^e$ ($\cM^m$) are more colloquially known as rough (smooth) boundaries.
	Moreover, the model has a single non-trivial domain wall which exchanges $e$ and $m$ excitation~\cite{Kitaev06, Bombin10, Kitaev12}, and thus gives rise to a single twist defect that resembles a Majorana mode~\cite{Kitaev06, Bombin10, Wootton15a}.

	We specify the excitations of the system of two copies of the toric code as $a^\pm$ for $a = e,\, m,\, \epsilon$ where the $\pm$ symbol indicates which of the two copies the excitation lies on.
	We can consistently map the color code excitations onto the dual-layer toric code model under the following correspondence
	\begin{align}
	\begin{split}
		rx \leftrightarrow e^-, & \quad
		rz \leftrightarrow m^+, \\
		bx \leftrightarrow e^+, & \quad
		bz \leftrightarrow m^-,
	\end{split}
	\end{align}
	which succinctly describe a generating set for $\cC$ under fusion~\cite{Bombin12}.
	We can also populate Tab.~\ref{tab:CC3x3anyons} as shown in Tab.~\ref{tab:CCvsTCanyons}.
	\begin{table}[!htbp]
		\centering
		\begin{tikzpicture}
			\draw (0.625,-.1) -- (0.625,1.3);
			\draw (1.875,-.1) -- (1.875,1.3);
			\draw (-.55,.35) -- (3,.35);
			\draw (-.55,.85) -- (3,.85);

			\node[anchor = base] (x) at (-1,1.0) {$x$};
			\node[anchor = base] (y) at (-1,0.5) {$y$};
			\node[anchor = base] (z) at (-1,0.0) {$z$};

			\node[anchor = base] (r) at (0.0,1.5) {$r$};
			\node[anchor = base] (g) at (1.25,1.5) {$g$};
			\node[anchor = base] (b) at (2.5,1.5) {$b$};

			\node[anchor = base] (rx) at (0.00,1.0) {$e^-$};
			\node[anchor = base] (ry) at (0.00,0.5) {$e^-m^+$};
			\node[anchor = base] (rz) at (0.00,0.0) {$m^+$};
			\node[anchor = base] (gx) at (1.25,1.0) {$e^-e^+$};
			\node[anchor = base] (gy) at (1.25,0.5) {$\epsilon^-\epsilon^+$};
			\node[anchor = base] (gz) at (1.25,0.0) {$m^-m^+$};
			\node[anchor = base] (bx) at (2.50,1.0) {$e^+$};
			\node[anchor = base] (by) at (2.50,0.5) {$m^-e^+$};
			\node[anchor = base] (bz) at (2.50,0.0) {$m^-$};
		\end{tikzpicture}
		\caption{
			Color code anyons after unfolding the color code into two toric code layers.
			$e$ and $m$ refer to electric and magnetic charges, the superscript indicates on which toric code layer the anyon lives.
			Certain color code anyons correspond to combinations of two toric code anyons.
			The shown correspondence is found if the lattice is contracted on the blue and red lattices to form the $-$ and $+$ layer toric codes respectively, following the unfolding procedure laid out in Ref.~\cite{Kubica15}.
			Note that this relabeling may be defined up to an automorphism $\phi$ of the anyon model.
		}
		\label{tab:CCvsTCanyons}
	\end{table}

\begin{figure}
		\centering
		\includegraphics[width=1\linewidth]{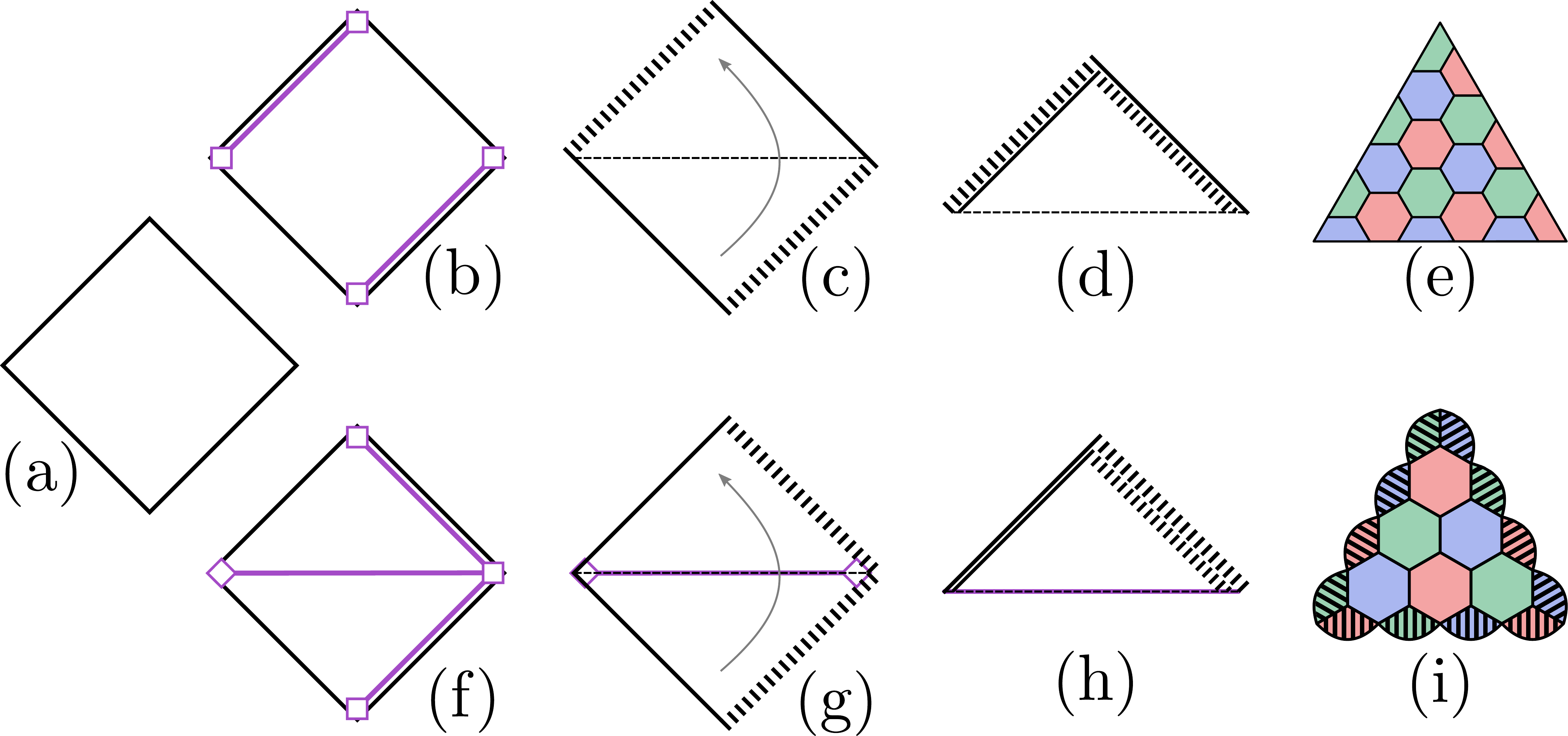}
		\caption{
			Illustrated is the folding of a toric code to obtain a triangular color code.
			The top, \textbf{(b)} - \textbf{(e)} row which starts with a toric code with a regular boundary configuration results in a triangular color code with color boundaries.
			The bottom row \textbf{(f)} - \textbf{(i)} starts with a toric code with a more unconventional domain wall arrangement and ends with a triangular color code with Pauli boundaries.
			The square \textbf{(a)} represents a patch of toric code with smooth boundaries.
			In \textbf{(b)} and \textbf{(f)} we introduce twists at the corners, connected by domain walls which change the boundary type from smooth to rough, see \textbf{(c)} and \textbf{(g)}.
			Now, the code is folded in half on top of itself to get \textbf{(d)} and \textbf{(h)}.
			Depending on the domain wall configuration, we get a code equivalent to a triangular color code with color boundaries \textbf{(e)} or Pauli boundaries \textbf{(i)}.
		}
		\label{fig:UnfoldingBoundaries}
	\end{figure}

	We next investigate the boundaries of the color code.
	The boundaries described by Lagrangian subgroups that correspond to the columns of Tab.~\ref{tab:CCvsTCanyons} have been studied in Ref.~\cite{Kubica15}.
	In particular, the system boundary where the top copy of the toric code is rough and the bottom copy is smooth corresponds to a red boundary in the color code picture.
	Likewise, the boundary where the top (bottom) copy of the toric code is smooth (rough) corresponds to a blue boundary.
	The green boundary corresponds to a fold where $e^+ \times e^- $ and $m^+ \times m^-$ are absorbed.
	In Fig.~\ref{fig:UnfoldingBoundaries}~(c) and (d), we show the planar code with two rough boundaries and two smooth boundaries being folded.
	In Ref.~\cite{Kubica15} it was argued that this model is analogous to the standard color code with a red, blue and green boundary~\cite{Bombin06}.

	Before describing the Pauli boundaries in this model, it will be helpful to understand the toric code model with rough and smooth boundaries as a model with only smooth boundaries but where twist defects lie at its corners.
	Indeed, in Ref.~\cite{Brown17} it was suggested that we should regard a rough boundary as a smooth boundary with a non-trivial domain wall stretched across its surface.
	Using this picture we see that twist defects lie at the corners of the planar code.
	In Fig.~\ref{fig:UnfoldingBoundaries}~(b) we show a configuration of domain walls that gives rise to the conventional planar code with two rough boundaries and two smooth boundaries.

	The color code with three different Pauli boundaries can also be understood from a folded system.
	In contrast, we can regard the color code with three distinct Pauli boundaries as the folded code model with a domain wall configuration shown in Fig.~\ref{fig:UnfoldingBoundaries}~(f).
	With this domain wall configuration we obtain through folding boundaries where both the top and bottom layers are both rough or both smooth, see Fig.~\ref{fig:UnfoldingBoundaries}~(h).
	These boundaries correspond to $x$ and $z$ Pauli boundaries, respectively.
	Finally, the $y$ Pauli boundary correspond to a fold with a non-trivial domain wall stretched across it.
	With this we complete the boundaries of the unfolded color code, as we have now explicitly realized boundaries that condense the Lagrangian subgroups of the rows of Tab.~\ref{tab:CCvsTCanyons}.

	We finally look at the domain walls and twist defects of the unfolded color code.
	Again, we consider a generating set of three domain walls.
	As discussed in Sec.~\ref{sec:Domainwalls} it is sufficient to find a domain wall that transposes Tab.~\ref{tab:CCvsTCanyons} together with two more domain walls that permute the rows of the table.

\begin{figure}
		\centering
		\includegraphics[width=0.75\linewidth]{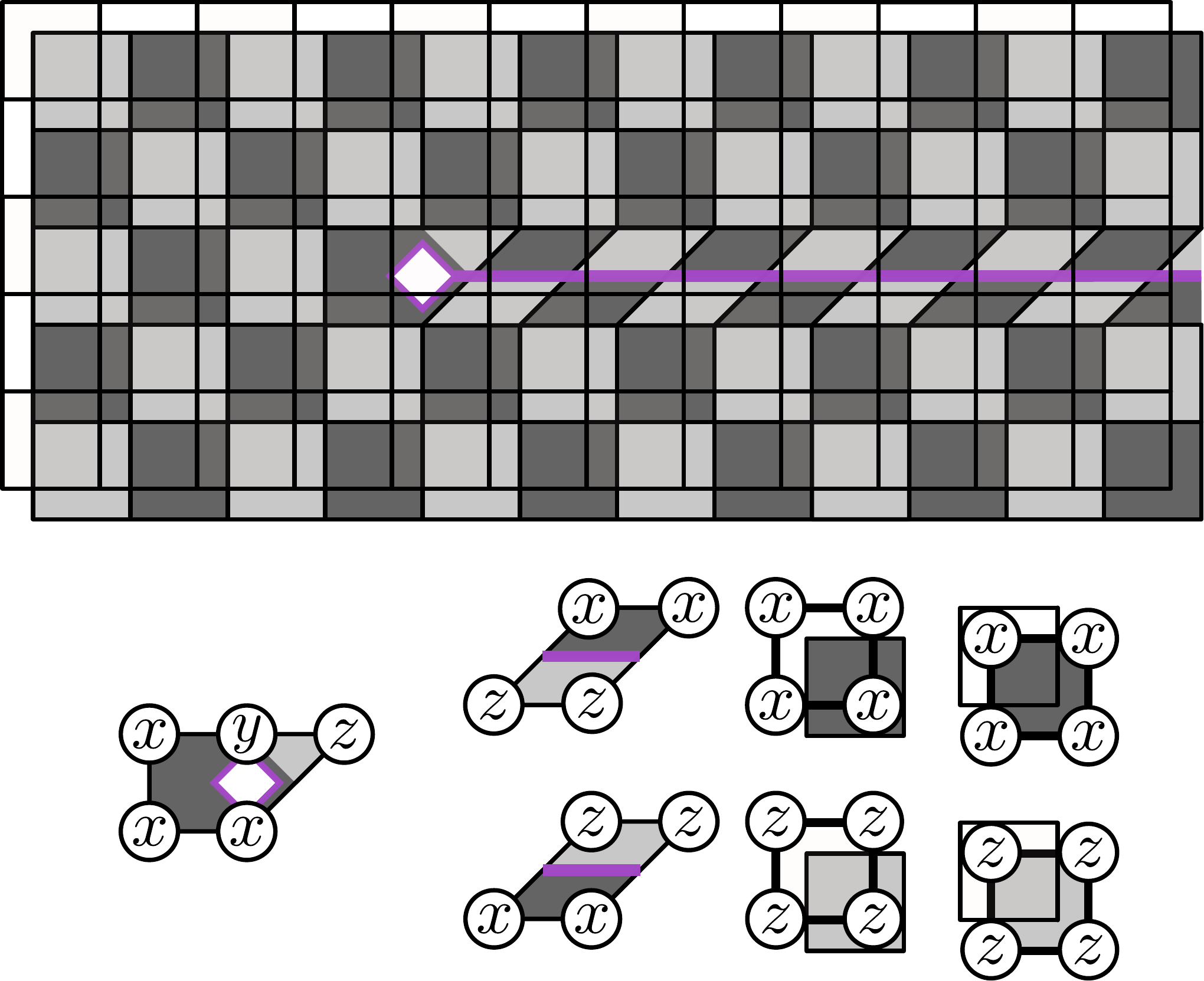}
		\caption{
			A domain wall and twist in two layers of toric code which corresponding to a $D$ symmetry in the color code.
			It consists of a toric code domain wall in the second layer.
			All stabilizers are shown in the legend below.
		}
		\label{fig:2TC_DW_TC}
	\end{figure}

	The domain wall that exchanges the excitations according to a transposition of Tab.~\ref{tab:CCvsTCanyons} is the mapping $e^- \leftrightarrow m^- $.
	Physically, this domain wall is the well studied domain wall of the toric code~\cite{Bombin10, Kitaev12} on the bottom layer of the two-copy system.
	We show this domain wall together with its termination twist defect in Fig.~\ref{fig:2TC_DW_TC}.

	\begin{figure}[b!]
		\centering
		\includegraphics[width=0.75\linewidth]{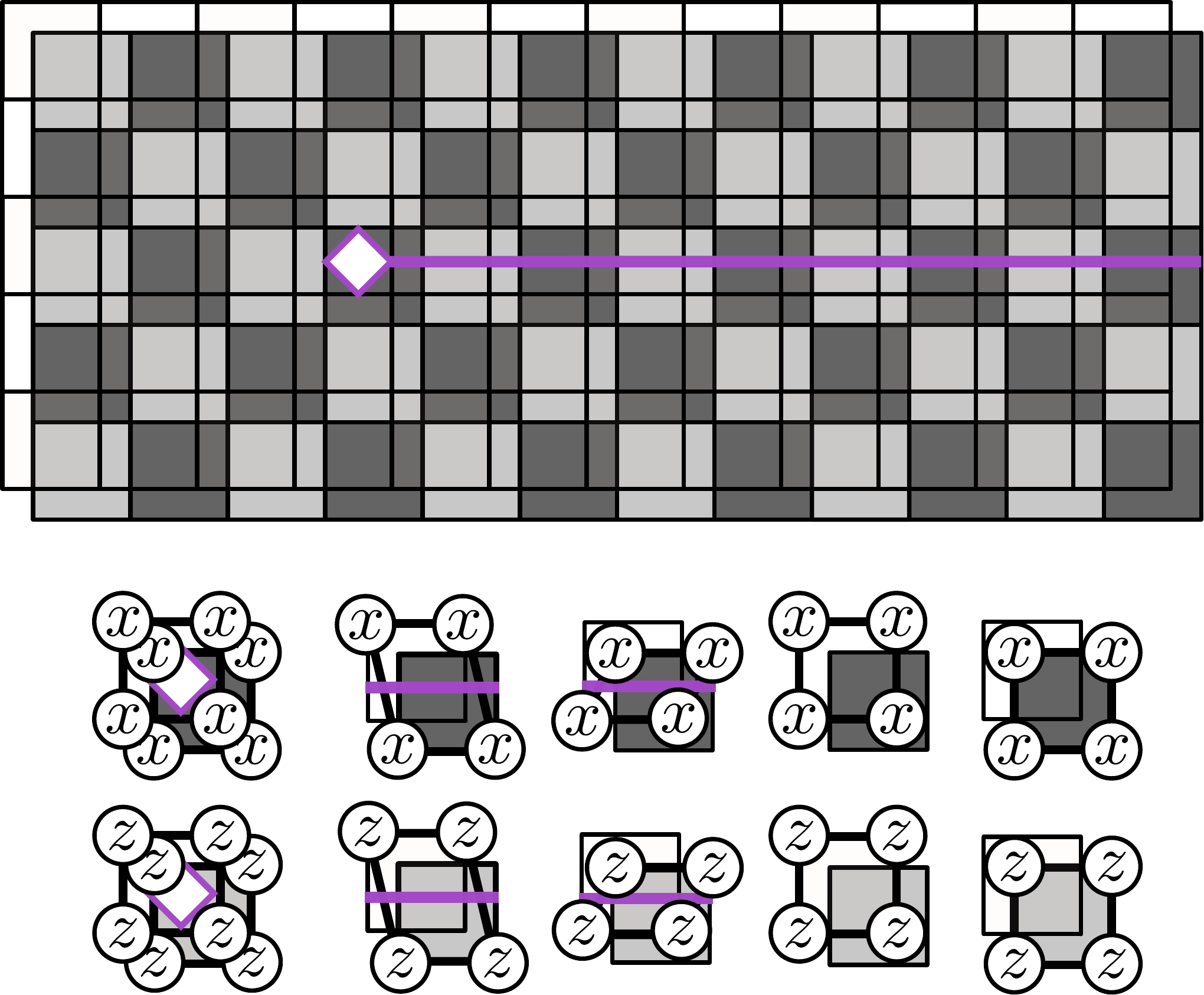}
		\caption{
			A layer swapping domain wall and twist in two layers of toric code.
			This corresponds to the $G$ symmetry in the color code.
			The stabilizers along the domain wall have support on both layers, see the legend below.
		}
		\label{fig:2TC_DW_LayerSwap}
	\end{figure}
	Next, we discuss the domain wall corresponding to the $G$ symmetry.
	In the unfolded color code it corresponds to a layer swap.
	This means toric code anyons moved over this domain wall get mapped according to $a^{\pm} \leftrightarrow a^{\mp}$, where $a = e,\, m,\, \epsilon$.
	See Fig.~\ref{fig:2TC_DW_LayerSwap} for a lattice representation of the layer swapping domain wall and its terminating twist.

\begin{figure}
		\centering
		\includegraphics[width=0.75\linewidth]{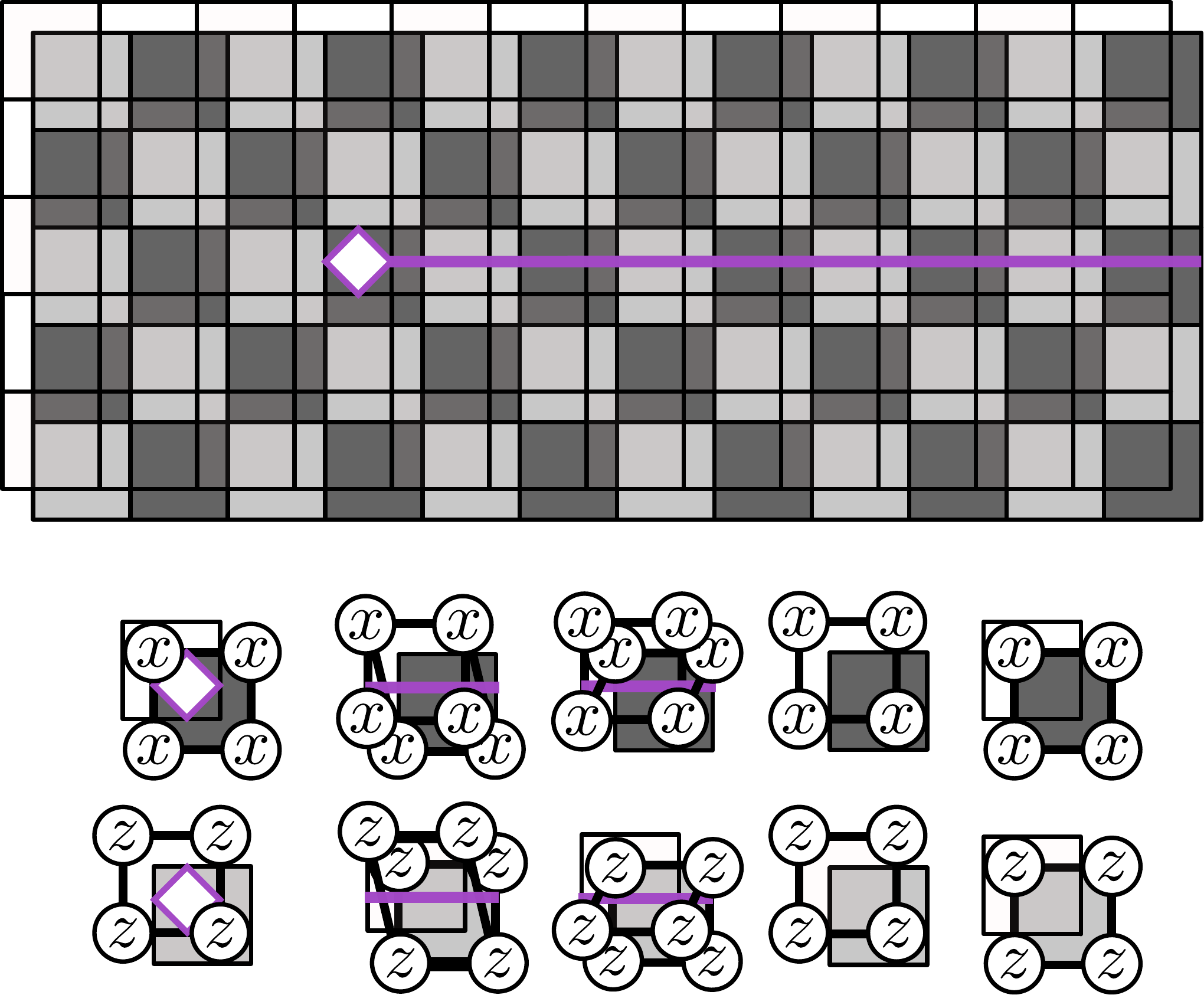}
		\caption{
			This domain wall and twist in two toric code layers correspond to the $B$symmetry in the color code.
			Certain anyons get cloned to the other layer when passing it.
		}
		\label{fig:2TC_DW_CNOT}
	\end{figure}

	The $B$ domain wall in the picture of two toric codes maps as follows, $e^- \leftrightarrow e^-e^+$ and $m^+ \leftrightarrow m^-m^+$.
	This means certain anyons get copied over to the other layer when passing the domain wall.
	On the lattice, it can be realized as shown in Fig.~\ref{fig:2TC_DW_CNOT}.
	It is finally worth remarking that the two domain walls analogous to the $G$ and $B$ maps that we have just presented can be realized in the two-copy system by, respectively, applying the swap gate and CNOT gate transversally between the two copies of the model on a region of the lattice.
	We thus realize a closed domain wall at the boundary of this region in the fashion of Ref.~\cite{Yoshida15}.

	Finally, we discuss an equivalence between the color code model and two copies of the three-fermion model~\cite{Wang17}.
	This connection is explored in App.~\ref{app:ThreeFermionModel}.
	The mapped system gives particularly natural way to view the symmetries, and thus the twist defects of the color code model.

\section{Conclusions and future work}
	\label{sec:SummaryAndOutlook}

	In this work, we have identified all of the boundaries, domain walls, and twist defects of the color code model, giving rise to a complete and comprehensive classification.
	On one hand, this endeavor has been motivated from the mindset of condensed matter physics.
	Specifically, using an exactly solvable toy model, we have been able to introduce a concrete example to give more substance to the abstract mathematical theory of phases.
	Not only this, but we have also described the model of focus in terms of other simpler phases to achieve a better grasp on the physics of the system.
	It is the hope that our study will encourage readers to explore other explicit phases to help build our understanding of the theory of phases.

	In addition to the exploration of fundamental condensed-matter physics, a second and equally important motivation of our work has been to aid the discovery of new applications of topological phases in quantum information science.
	Indeed, we have applied our analysis to the development of topological quantum codes.
	Topological quantum computing remains the most promising road towards realizing a scalable quantum computer.
	In order to contribute to the theory of quantum codes, we have introduced a new small code, a family of stellated codes with competitive encoding rates and improved methods of performing code deformations.
	These new applications clearly exhibit the potential to use our catalog to help find new codes and protocols for processing quantum information.
	However, the examples we have presented may only be the tip of the iceberg of quantum computational schemes that are yet to be discovered using the framework we have built for the color code model.
	In particular, the present study invites more research on how to devise fully fault-tolerant schemes for quantum computing within such a framework as well as on implementations.

	To extend this research path further, we encourage the reader to experiment with the different topological features we have laid out to find other new schemes for encoding and manipulating topologically protected quantum information.
	In particular we believe taking a more systematic approach~\cite{Delfosse16b, Delfosse16a} to combine the different objects we have described may be a fruitful avenue towards new discoveries.
	We hope that the present study stimulates such further research.

	\begin{acknowledgments}
		The authors thank A.\ Bauer, H.\ Bomb{\'\i}n, C.\ Brell, J.\ Bridgeman, T.\ Jochym-O'Connor, A.\ Kubica, D.\ Litinski, H.\ Poulsen Nautrup, A.\ Nietner, R.\ Raussendorf, N.\ Tarantino, and T.\ J.\ Yoder for helpful and encouraging conversations.
		BJB is also grateful to T.\ Adair, S.\ Bartlett, K.\ Korzekwa, and S.\ Roberts for discussions on code deformations.
		Further, we thank Zhenghan Wang for pointing out the connection between the color code and two copies of the three-fermion model.
		MSK is supported by the DFG (CRC183, project B02).
		FP was supported by the Alexander von Humboldt Foundation.
		JE is supported by DFG (CRC 183, project B02, EI 519/14-1, and EI 519/7-1), the ERC (TAQ), the Templeton Foundation,
		and the BMBF (Q.com).
		BJB is supported by the University of Sydney Fellowship Programme, the Australian Research Council via the Centre of Excellence in Engineered Quantum Systems(EQuS) project number CE170100009, and the Villum Foundation.

	\end{acknowledgments}

\appendix

\section{Fermions in the color code}
	\label{app:CCFermions}

	In the main text, we have discussed the existence of six fermionic anyons in the color code.
	As mentioned, fermions are the result of fusion of bosonic anyons which do not share either the color nor the Pauli label.
	For example, we obtain a fermion which we call $f_1$ by fusing $rz$ and $bx$, i.e. $f_1 = rz \times bx$.
	Since we can let $rz$ decay using the rule $rz = gz \times bz$, we can also identify a triple of bosonic anyons with the same fermion, $f_1 = gz \times bz \times bx$.
	And using $bz \times bx = by$, we obtain an alternative way of writing $f_1$ in terms of two bosons, $f_1 = gz \times by$.
	In fact, one can also write $f_1$ in a third way as the combination of two bosonic fermions, namely $f_1 = ry \times gx$.
	Once again, we can use the $3 {\times} 3$ grid of color code bosons to represent fermions, see Tab.~\ref{tab:CC3x3anyons}.
	Specifically, in Tab.~\ref{tab:f_1} we place a spot in each of the cells of the grid that represent the two bosons that constitute the fermion.

	\begin{table}[!htbp]
		\centering
		\begin{tikzpicture}
			\draw (-.8,3.2+.2) -- (.8,3.2+.2);
			\draw (-.8,3.2-.2) -- (.8,3.2-.2);
			\draw (-.25,3.2-.5) -- (-.25,3.2+.5);
			\draw (.25,3.2-.5) -- (.25,3.2+.5);
			\node (f_1) at (-1.3,3.2) {$f_1 = $};
			\node (anyons1) at (0,3.2) {
				$\begin{matrix}
					 &  & \bullet \\
					 $\quad$ & $\quad$ & $\quad$   \\
					 \bullet &  & 
				\end{matrix}$
			};
			
			\draw (2.5-.8,3.2+.2) -- (2.5+.8,3.2+.2);
			\draw (2.5-.8,3.2-.2) -- (2.5+.8,3.2-.2);
			\draw (2.5-.25,3.2-.5) -- (2.5-.25,3.2+.5);
			\draw (2.5+.25,3.2-.5) -- (2.5+.25,3.2+.5);
			\node (eq1) at (2.5-1.3,3.2) {$=$};
			\node (anyons2) at (2.5,3.2) {
				$\begin{matrix}
					$\quad$ & $\quad$ & $\quad$  \\
					&  & \bullet \\
					 & \bullet & 
				\end{matrix}$
			};
			
			\draw (5-.8,3.2+.2) -- (5+.8,3.2+.2);
			\draw (5-.8,3.2-.2) -- (5+.8,3.2-.2);
			\draw (5-.25,3.2-.5) -- (5-.25,3.2+.5);
			\draw (5+.25,3.2-.5) -- (5+.25,3.2+.5);
			\node (eq2) at (5-1.3,3.2) {$=$};
			\node (anyons3) at (5,3.2) {
				$\begin{matrix}
					 & \bullet &  \\
					 \bullet &  &  \\
					 $\quad$ & $\quad$ & $\quad$ 
				\end{matrix}$
			};
		\end{tikzpicture}
		\caption{The fermion $f_1$ written in three different ways as the fusion of two bosonic anyons.}
	\label{tab:f_1}
	\end{table}

	We can write all six color code fermions in three ways as the fusion product of two bosons,
	\begin{align}
	\begin{split}
		f_1 &= rz \times bx = gz \times by = ry \times gx, \\
		f_2 &= gy \times bz = rz \times gx = ry \times bx,\\
		f_3 &= gy \times bx = rx \times gz = gy \times bz, \\
		f_4 &= rx \times gy = gz \times bx = rz \times by, \\
		f_5 &= rz \times gy = gx \times bz = rx \times by, \\
		f_6 &= rx \times bz = ry \times gz = gx \times by.
		\label{eq:CCFermions}
	\end{split}
	\end{align}

	And again draw them in the $3 {\times} 3$ grid, see Tab.~\ref{tab:3x3CCfermions}.
	Here, different symbols where used for the different ways of composing the fermions.
	\begin{table}[!htbp]
		\centering
		\begin{tikzpicture}
			\draw (-.8,3.2+.2) -- (.8,3.2+.2);
			\draw (-.8,3.2-.2) -- (.8,3.2-.2);
			\draw (-.25,3.2-.5) -- (-.25,3.2+.5);
			\draw (.25,3.2-.5) -- (.25,3.2+.5);
			\node (f_1) at (-1.3,3.2) {$f_1 = $};
			\node (anyons1) at (0,3.2) {
				$\begin{matrix}
					& \bigstar & \bullet \\
					\bigstar &  & \blacktriangle \\
					\bullet & \blacktriangle &
				\end{matrix}$
			};
			\draw (3-.8,.2) -- (3.8,.2);
			\draw (3-.8,-.2) -- (3.8,-.2);
			\draw (3-.25,-.5) -- (3-.25,.5);
			\draw (3.25,-.5) -- (3.25,.5);
			\node (f_6) at (3-1.3,0) {$f_6= $};
			\node (anyons2) at (3,3.2) {
				$\begin{matrix}
					& \blacktriangle & \bigstar \\
					\bigstar & \bullet & \\
					\blacktriangle &  & \bullet 
				\end{matrix}$
			};
			\draw (-.8,1.6+.2) -- (.8,1.6+.2);
			\draw (-.8,1.6-.2) -- (.8,1.6-.2);
			\draw (-.25,1.6-.5) -- (-.25,1.6+.5);
			\draw (.25,1.6-.5) -- (.25,1.6+.5);
			\node (f_3) at (-1.3,1.6) {$f_3 = $};
			\node (anyons3) at (0,1.6) {
				$\begin{matrix}
					\blacktriangle &  & \bullet \\
					\bigstar & \bullet & \\
					& \blacktriangle & \bigstar 
				\end{matrix}$
			};
			\draw (3-.8,1.6+.2) -- (3.8,1.6+.2);
			\draw (3-.8,1.6-.2) -- (3.8,1.6-.2);
			\draw (3-.25,1.6-.5) -- (3-.25,1.6+.5);
			\draw (3.25,1.6-.5) -- (3.25,1.6+.5);
			\node (f_4) at (3-1.3,1.6) {$f_4 = $};
			\node (anyons3) at (3,1.6) {
				$\begin{matrix}
					\bullet &  & \blacktriangle  \\
					& \bullet & \bigstar \\
					\bigstar & \blacktriangle & 
				\end{matrix}$
			};
			\draw (-.8,.2) -- (.8,.2);
			\draw (-.8,-.2) -- (.8,-.2);
			\draw (-.25,-.5) -- (-.25,.5);
			\draw (.25,-.5) -- (.25,.5);
			\node (f_5) at (-1.3,0) {$f_5 = $};
			\node (anyons5) at (0,0) {
				$\begin{matrix}
					\bigstar & \blacktriangle & \\
					& \bullet & \bigstar \\
					\bullet &  & \blacktriangle 
				\end{matrix}$
			};
			\draw (3-.8,3.2+.2) -- (3.8,3.2+.2);
			\draw (3-.8,3.2-.2) -- (3.8,3.2-.2);
			\draw (3-.25,3.2-.5) -- (3-.25,3.2+.5);
			\draw (3.25,3.2-.5) -- (3.25,3.2+.5);
			\node (f_2) at (3-1.3,3.2) {$f_2= $};
			\node (anyons6) at (3,0) {
				$\begin{matrix}
					\bullet & \bigstar &  \\
					\blacktriangle &  & \bigstar \\
					& \blacktriangle & \bullet
				\end{matrix}$
			};
		\end{tikzpicture}
		\caption{
			The six fermionic anyons in the color code.
			Each one can be created by fusion of two bosonic anyons.
			Each table is the $3 {\times} 3$ grid introduced in Tab.~\ref{tab:CC3x3anyons}.
			The different shapes distinguish the different ways in which a fermion can be created by fusing two bosons, as per Eq.~\eqref{eq:CCFermions}.
		}
		\label{tab:3x3CCfermions}
	\end{table}

	Note, for fermions with even index, the bullets lie on the diagonal, for fermions with odd index, they lie on the anti-diagonal.
	Indeed, the distinction between fermions with even and odd index goes deeper.
	Braiding odd (even) fermions with even (odd) fermions is trivial, whereas braiding them with distinct odd (even) fermions results in a phase $-1$,
	\begin{align}
		M_{f_i,f_j} = (-1)^{\delta(i \neq j)} (-1)^{i-j}.
		\label{eq:CCFermionBraiding}
	\end{align}
	Furthermore, fusion of two different odd (even) fermions results in the third odd (even) fermion.
	Whereas fusing an odd fermion with an even fermion results in a bosonic anyon.

	Fermions can not condense at boundaries, but there are twists which localize fermionic anyons.
	For example, as mentioned in Sec.~\ref{sec:TwistsAndAnyons} and shown in Fig.~\ref{fig:AnyonLocalizationCC}, the $D$ twist can localize the $f_1$ anyon, and apart from the trivial charge $1$ only the $f_1$ anyon.
	The other five fermions also have a corresponding twist which only localizes them.
	These are the twists in the conjugacy class G in Tab.~\ref{tab:CCConjClasses}.

	As discussed in Sec.~\ref{sec:2xTC}, the color code is equivalent to two copies of the toric code model.
	In the toric code, there exists a single fermionic anyon which is the product of fusing an electric flux $e$ with a magnetic flux $m$, i.e. $\epsilon = e \times m$.
	Two copies of the toric code feature six fermions, as the fusion of a fermion on one layer and a boson on the other layer produces a fermion, see Eq.~\eqref{eq:CompositeSelfExchange}.
	Written as two toric codes, the fermions of the color code are hence $\epsilon^-,\,\epsilon^-e^+,\,\epsilon^-m^+,\,\epsilon^+,\,\epsilon^+e^-$, and $\epsilon^+m^-$.

	In App.~\ref{app:ThreeFermionModel}, we discuss the equivalence of two copies of the three-fermion model and the color code.
	This equivalence also makes the difference between what we here called even and odd fermions very apparent, as they correspond to the fermions on different layers of the three-fermion model.

\section{The three-fermion model and the color code}
	\label{app:ThreeFermionModel}

	In the main text in Sec.~\ref{sec:2xTC}, we have explored the equivalence between the color code ($\mathrm{CC}$) features and features in two copies of the toric code ($\mathrm{TC}$).
	In this appendix, we address a similar equivalence.
	Namely the equivalence between two copies of the three-fermion model ($\mathrm{3F}$) and the color code~\cite{Wang17}.

	The three-fermion model is discussed in Ref.~\cite{Rowell09} where it is called the $(D_4,1)$ modular tensor category.
	It can physically be obtained as low energy excitations of a topological subsystem code, as discussed in Ref~\cite{Bombin10b}.
	In the three-fermion model, there are four anyonic charges.
	Apart from the trivial charge $1$, there are three fermionic particles, $f_1$ $f_2$ and $f_3$.
	Fermionic here means $\theta_{f_i} = -1$.
	The fusion rules are the same as in the toric code, all particles are their own antiparticles and fusion of two distinct non-trivial anyons results in the third non-trivial anyon, $f_i \times f_j = f_k$ for $i \neq j \neq k \neq i$.
	Braiding two different(identical) fermions results in a phase of $-1$($+1$) such that $M_{f_i,f_j} = (-1)^{\delta(i \neq j)}$.
	Note that braiding and fusion in the three-fermion model is exactly the same as for the toric code, only the spin of two particles is changed.

	Let us now turn towards the features of $\mathrm{3F}$, its boundaries and symmetries.
	The three-fermion model does not have any boundaries.
	Due to the absence of any non-trivial bosons, the only Lagrangian subgroup is the trivial group.
	On the other hand, it has a richer symmetry compared to the toric code.
	All permutations between the fermions are valid symmetries, thus its symmetry group is $S_3$, the permutation of a set of three elements.
	Hence it has $|S_3| = 6$ domain walls and twists, including the trivial one.
	It has three order-two symmetries which exchange two fermions while leaving the third invariant, and two order-three symmetries which cyclically permute the three-fermions, one or the other way.

	Two copies of the three-fermion model is equivalent to the color code model, $\mathrm{3F} \otimes \mathrm{3F} = \mathrm{CC}$.
	To convince ourselves this is true, we show that the anyons in $\mathrm{3F} \otimes \mathrm{3F}$ are the same as the color code anyons.
	Braiding and fusion of anyons between layers is trivial;
	on a single layer, the braiding and fusion statistic is the same in $\mathrm{3F}$ as in $\mathrm{TC}$.
	Since we know $\mathrm{TC} \otimes \mathrm{TC} = \mathrm{CC}$, we only need to show how the self-exchange statistics in $\mathrm{3F} \otimes \mathrm{3F}$ is the same as in $\mathrm{CC}$.
	The spin of a joint particle $c = a^- \times b^+$ is given by the product of the self-exchanges on each layer, $\theta_{c} = \theta{a^-} \theta_{b^+}$, as per Eq.~\eqref{eq:CompositeSelfExchange}.
	In particular, the only way to obtain a fermionic anyon is to fuse a fermion on one layer with the trivial particle on the other layer.
	Hence, we obtain six fermions, the 10 other non-trivial anyons are bosonic, just as in the color code.
	One possible way of writing the color code anyons in terms of $\mathrm{3F}$ anyons is shown in Tab.~\ref{tab:CCvs3Fanyons}.
	Confirming the known $\mathrm{CC}$ fusion rules and braid statistics is left as an exercise to the reader.
	\begin{table}[!htbp]
		\centering
		\begin{tikzpicture}
			\draw (0.65,-.1) -- (0.65,1.3);
			\draw (1.95,-.1) -- (1.95,1.3);
			\draw (-.45,.35) -- (3.1,.35);
			\draw (-.45,.85) -- (3.1,.85);
			
			\node[anchor = base] (x) at (-.85,1.0) {$x$};
			\node[anchor = base] (y) at (-.85,0.5) {$y$};
			\node[anchor = base] (z) at (-.85,0.0) {$z$};
			
			\node[anchor = base] (r) at (0.0,1.5) {$r$};
			\node[anchor = base] (g) at (1.3,1.5) {$g$};
			\node[anchor = base] (b) at (2.6,1.5) {$b$};
		
			\node[anchor = base] (rx) at (0,1) {$f_1^- f_1^+$};
			\node[anchor = base] (ry) at (0,.5) {$f_2^- f_3^+$};
			\node[anchor = base] (rz) at (0,0) {$f_3^- f_2^+$};
			\node[anchor = base] (gx) at (1.3,1) {$f_2^- f_2^+$};
			\node[anchor = base] (gy) at (1.3,.5) {$f_3^- f_1^+$};
			\node[anchor = base] (gz) at (1.3,0) {$f_1^- f_3^+$};
			\node[anchor = base] (bx) at (2.6,1) {$f_3^- f_3^+$};
			\node[anchor = base] (by) at (2.6,.5) {$f_1^- f_2^+$};
			\node[anchor = base] (bz) at (2.6,0) {$f_2^- f_1^+$};		
		\end{tikzpicture}
		\caption{
			An example of writing the $3 {\times} 3$ grid of color code anyons (Tab.~\ref{tab:CC3x3anyons}) in terms of particles from two copies of the three-fermion model $\mathrm{3F}$.
			The superscript indicates the layer, the subscript the label of the fermion within a $\mathrm{3F}$ layer.
		}
		\label{tab:CCvs3Fanyons}
	\end{table}

	With the convention used in Tab.~\ref{tab:CCvs3Fanyons} is such that the fermion labels are related to those in App.~\ref{app:CCFermions} such that $f^+_j = f_{2j - 1}$ and $f^-_j = f_{2j}$.
	It is also interesting that the color label of a boson can be determined uniquely as the sum modulo 3 of the indices of the two particles from the three fermion layers.
	Equivalently, the Pauli label corresponds to the difference modulo three between said indices.

	Now, let us study the boundaries of the color code when read as two copies of the three-fermion model.
	As mentioned, the $\mathrm{3F}$ model does not have a non-trivial boundary, as it has not a single non-trivial bosonic particle.
	Instead, all six color code boundaries correspond to foldings between the two layers of the $\mathrm{3F}$ model, potentially with additional domain walls on one layer.
	The $x$ Pauli boundary is just the direct fold, where $f^-_i$ turns into $f^+_i$ and vice versa.
	This means particles of the form $f^-_i f^+_i$ can condense at said boundary.
	The $y$ Pauli boundary consists of a fold with a domain wall on the $-$ layer which permutes the fermions cyclically, $f^-_i \mapsto f^-_{i \oplus 1}$, where $\oplus$ is addition modulo 3.
	The $z$ Pauli boundary features the inverse domain wall, $f^-_i \mapsto f^-_{i \ominus 1}$.
	The color boundaries consists of folds together with domain walls of order two.
	For the red boundary it is the domain wall which exchanges fermions 2 and 3 in the $-$ layer, $f^-_2 \leftrightarrow f^-_3$.
	For the green and blue boundary the domain walls have the effect $f^-_1 \leftrightarrow f^-_3$ and $f^-_2 \leftrightarrow f^-_3$, respectively.

	The symmetries of the color code can be constructed using the $\mathrm{3F}$ symmetries.
	Each $\mathrm{3F}$ layer has the symmetry group $S_3$, this combines to $S_3 \times S_3$ for two layers.
	Additionally, we can introduce a layer swap symmetry, which acts as $f^-_i \leftrightarrow f^+_i$ on the anyons.
	This $\bZ_2$ action on the group leads to $(S_3 \times S_3) \ltimes \bZ_2 = S_3 \wr \bZ_2$, which as we know from the main text is the color code symmetry group.
	Again, this group can be generated by three elements, the layer swap with the generators on one of the $\mathrm{3F}$ layers for example.

	Through the folding of the three-fermion model, we can interpret the corners between color code boundaries as color code twist defects.
	This is a practical example of the theory given in Ref.~\cite{Barkeshli13a}.
	Introducing a domain wall in the folded $\mathrm{3F}$ model along the crease line changes the boundary type.
	Let us start with the color code $x$ boundary which corresponds to the trivial fold between the two $\mathrm{3F}$ layers.
	Adding a domain wall which exchanges the second and third fermion, i.e. $f_2 \leftrightarrow f_3$, changes the $x$ Pauli boundary to a red color boundary.
	This corresponds to a $D$ domain wall in the color code, which as discussed in the main text, changes an $x$ Pauli boundary to a red color boundary.

\section{Pachner moves and color code twists}
	\label{app:PachnerMoves}

	The goal of this appendix is to show a systematic way of deforming the color code lattice to create and move color exchanging twist defects and the domain walls they terminate.
	We will find that this can be achieved using a series of simple Pachner moves on the dual of the color code lattice.
	This construction also leads to a graphical understanding of some of the features of color twists we have seen in the main text.
	In particular, the fusion between color twists, the effect color domain walls have on color twists and condensation of color twists at color boundaries can be visualized nicely in this framework.

	Pachner moves~\cite{Pachner91, Nakahara} are a set of transformations that generate the mapping of a triangulation of a manifold onto any other triangulation of the same manifold.
	As the color code lattice is trivalent, its dual lattice is a triangulation~\cite{Bombin15, Kubica15a}.
	We find that in general, applying Pachner moves to said triangulation often introduces colored twists to the model as they typically break the three-colorability of the lattice.

\begin{figure}
		\centering
		\includegraphics[width=0.7\linewidth]{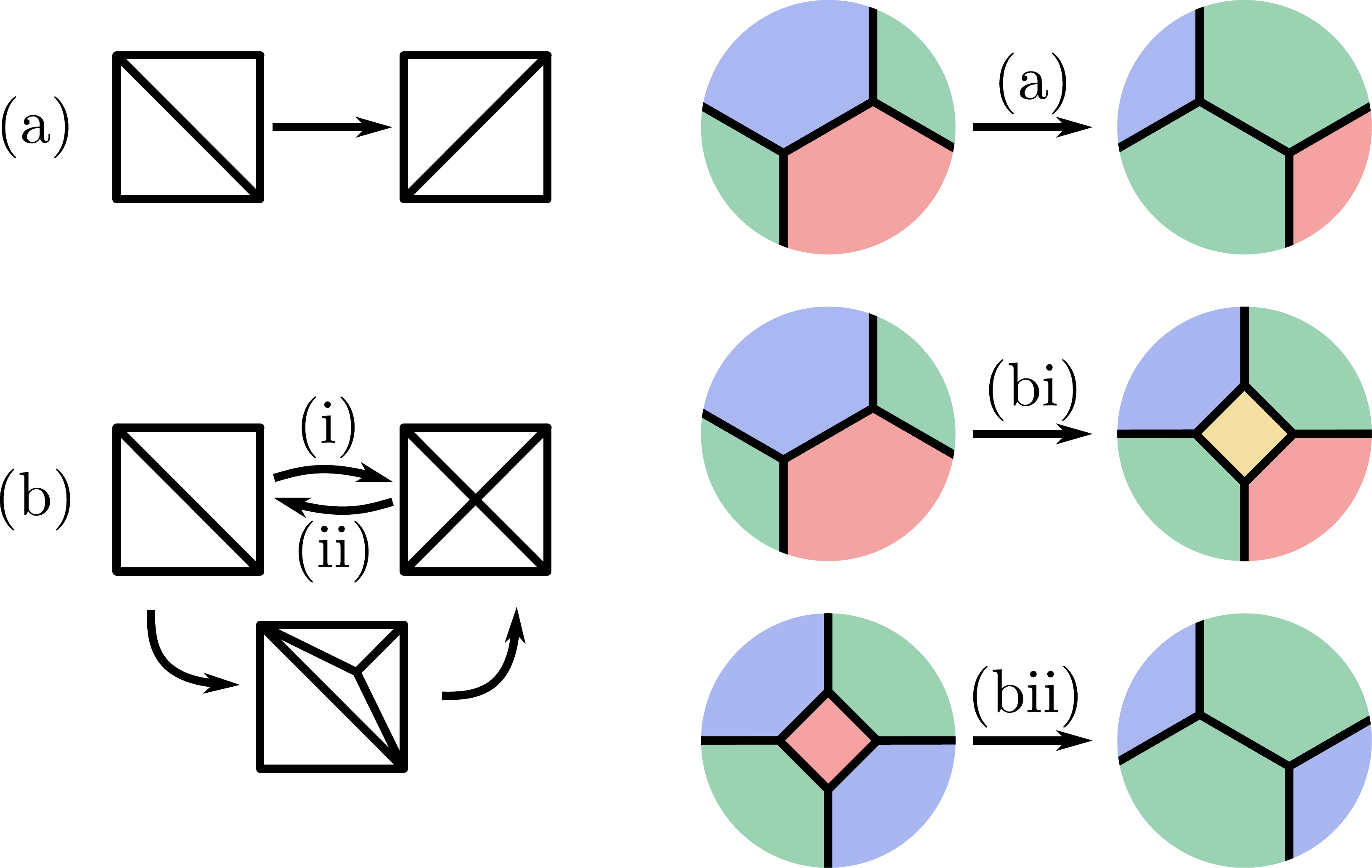}
		\caption{
			The Pachner moves needed to create and move color twists are shown on the dual lattice (left) and the primal color code lattice (right).
			(a) is a 2-2 Pachner move.
			Such a move changes the three-colorability of the lattice as shown on the right.
			After the move is applied, two plaquettes with the same color touch.
			(b) is a combination of a 1-3 and a 2-2 Pachner move.
			On the primal lattice, this move introduces (bi) or removes (bii) a square, depending on its direction.
			In either case, the three-colorability of the lattice is violated, either because a fourth color is introduced, or because two plaquettes of the same color share an edge.
		}
		\label{fig:PachnerMoves}
	\end{figure}

	To create twist defects, we distinguish between three Pachner moves, shown in Fig.~\ref{fig:PachnerMoves}.
	Shown in Fig.~\ref{fig:PachnerMoves}(a) is a 2-2 Pachner move, transforming two triangles joint at one edge into two other triangles by redrawing the diagonal.
	Fig.~\ref{fig:PachnerMoves}(bi) is a 1-3 followed by a 2-2 Pachner move, transforming two triangles into four by `adding a second diagonal'.
	While this is not typically among the generators of the Pachner moves, we find this move to be useful to discuss the introduction of twists to the dual lattice.
	Fig.~\ref{fig:PachnerMoves}(bii) also shows the inverse of (bi).
	The effect of these moves on the dual lattice are shown on the left in Fig.~\ref{fig:PachnerMoves}, the effect on the color code lattice is shown on the right.

\begin{figure}
		\centering
		\includegraphics[width=1\linewidth]{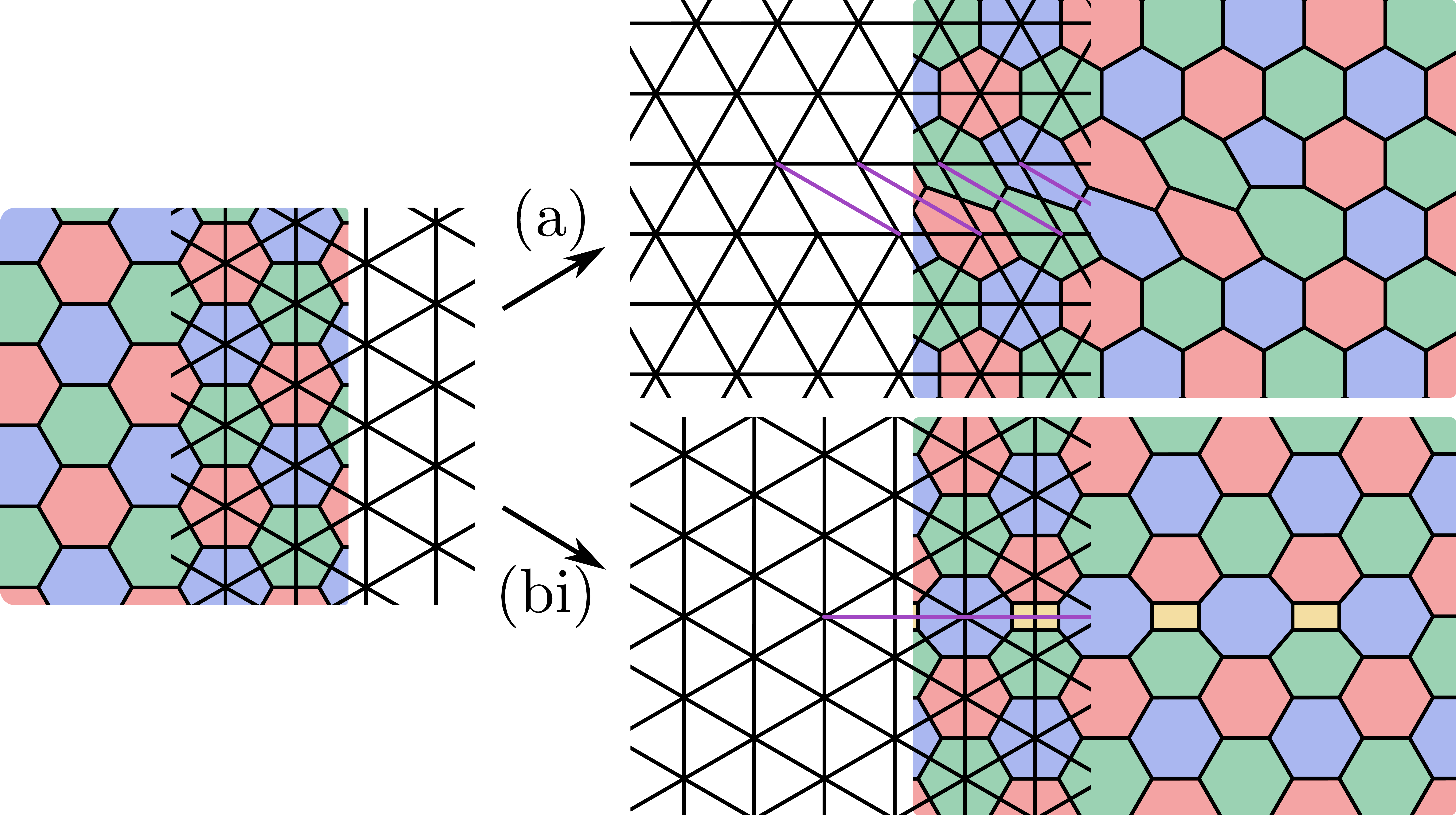}
		\caption{
			Only two of the Pachner moves shown in Fig.~\ref{fig:PachnerMoves} are possible in the 6.6.6 color code.
			Changed or added edges are colored in purple on the dual lattice.
			The resulting dislocation lines are shown here in the dual and primal lattice.
			Note that the top lattice on the right side is rotated $30^{\circ}$ compared to the rest.
		}
		\label{fig:PachnerMoveTwists666}
	\end{figure}

\begin{figure}[b!]
		\centering
		\includegraphics[width=1\linewidth]{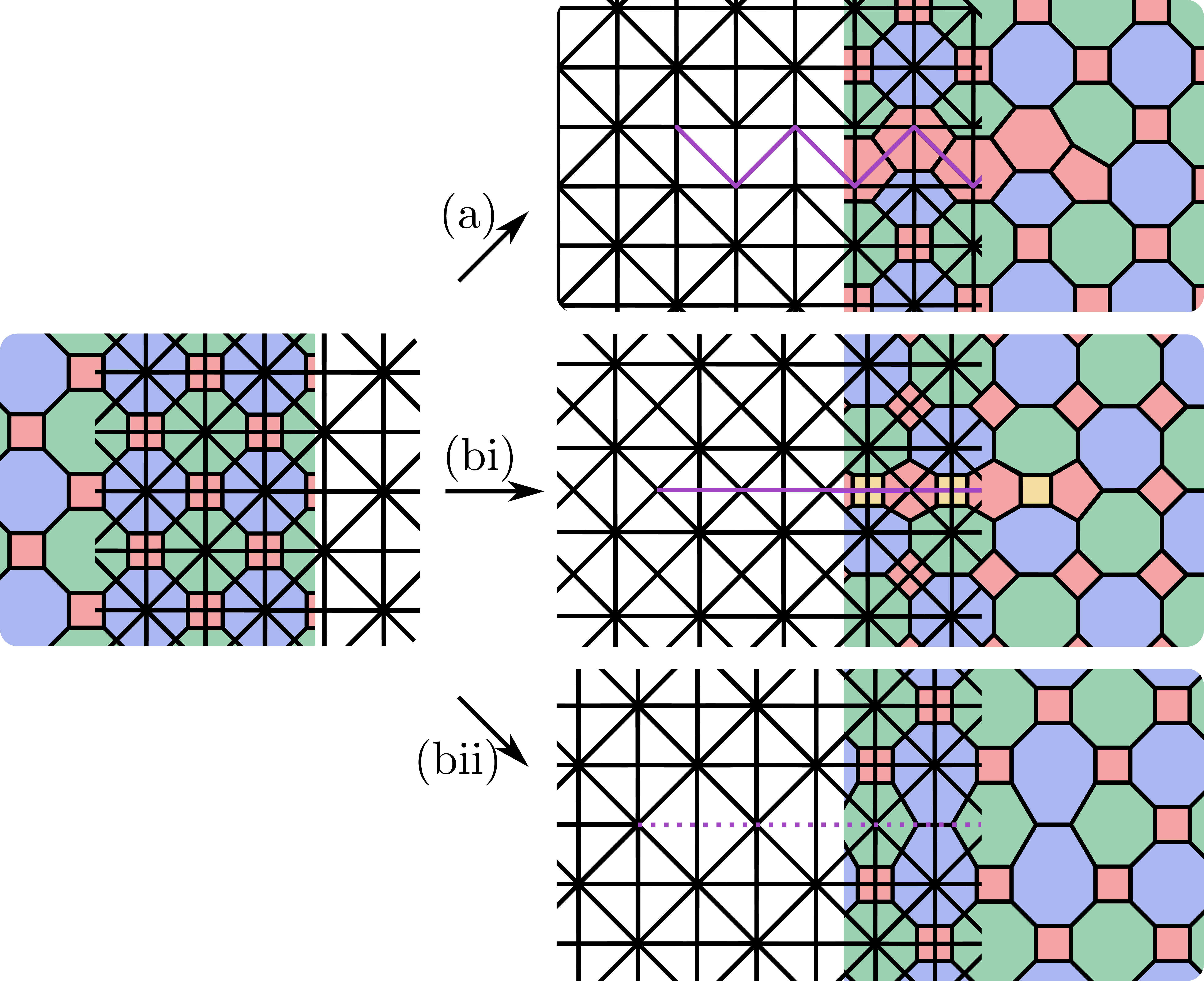}
		\caption{
			The three Pachner moves from Fig.~\ref{fig:PachnerMoves} applied on the 4.8.8 color code.
			The resulting dislocation lines are shown in purple.
			Note, the dashed line in (bii) indicates the removed edges.
			The lattice in the middle on the right side is rotated $45^{\circ}$ compared to the rest.
		}
		\label{fig:PachnerMoveTwists488}
	\end{figure}

	In the dual color code lattice, the triangular faces correspond to physical qubits and the vertices to stabilizers.
	Thus, the (a) Pachner move changes the number of qubits on which the four involved stabilizers have support by $\pm 1$.
	This will turn the usually even plaquettes into odd plaquettes, which has the effect that $x$-type and $z$-type stabilizers no longer commute.
	To go back to to a commuting Hamiltonian, one of the stabilizers has to be removed.
	If, however, a series of (a) moves are applied along a line, as shown in Fig.~\ref{fig:PachnerMoveTwists666}~(a) and Fig.~\ref{fig:PachnerMoveTwists488}~(a), the parity of vertices along the line gets changed twice and is even in the end.
	This means the only odd plaquettes lie at the terminal points and constitute the twists of the created domain wall.

	The (b) Pacher move only changes two surrounding plaquettes, where it either adds (bi) or removes (bii) an edge.
	Additionally, it adds or removes a square stabilizer.
	Again, these moves can be applied along a line such that vertices along the line change parity twice.
	This restores the parity of all plaquettes to be even, except for the ones at the terminal points.
	The single plaquette at each terminal point changes its support to an odd number of qubits and its stabilizers need to be changed.
	See Fig.~\ref{fig:PachnerMoveTwists666}~(bi) and Fig.~\ref{fig:PachnerMoveTwists488}~(bi) and (bii).

\begin{figure}
		\centering
		\includegraphics[width=.75\linewidth]{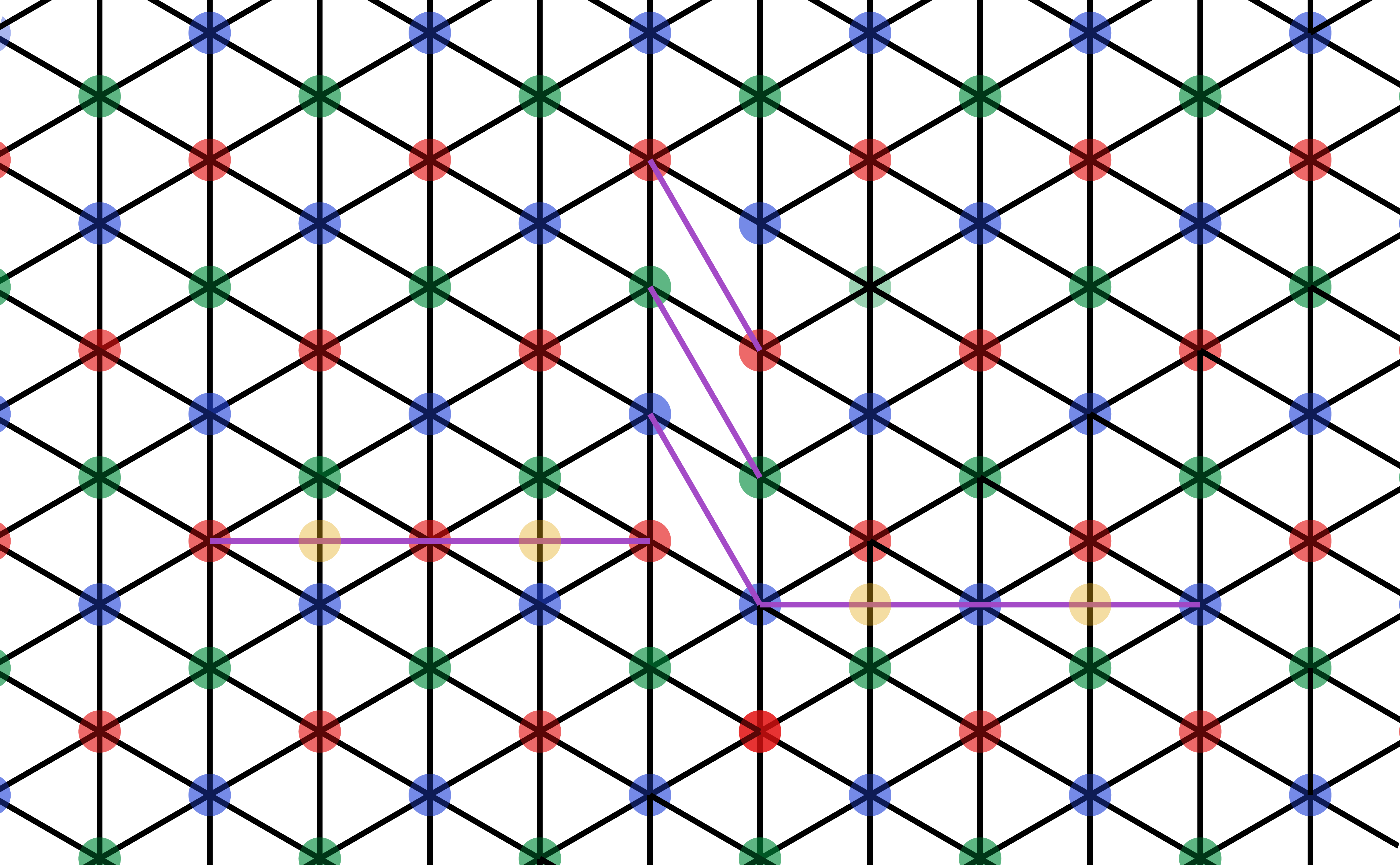}
		\caption{
			A configuration of a branching domain wall terminated by three twists is shown on the dual lattice.
			On the left an $R$ twist, on the right a $B$ twist and in the top an $RB$ twist.
			Alternatively, this picture can be understood as showing the fusion between an $R$ and a $B$ twist to form an $RB$ twist.
			Fig.~\ref{fig:666TwistFusionPrimal} depicts the exact same configuration on the primal lattice.
		}
		\label{fig:666TwistFusionDual}
	\end{figure}

	Extending and shortening the line along which the Pachner moves are applied, allows for twists to be moved and fused together.
	Consider as an example Fig.~\ref{fig:666TwistFusionDual}, where three twists are shown, an $R$ twist on the left, a $B$ twist on the right and an $RB$ twist at the top.
	This configuration can be obtained by fusing an $R$ and a $B$ twist together.
	Notice how at the point where the three domain walls meet, not a single stabilizer has odd weight.
	Hence, there is no non-trivial twist located there.
	The same configuration was shown in the main text in Fig.~\ref{fig:666TwistFusionPrimal}.

	\begin{figure}[b!]
		\centering
		\includegraphics[width=.55\linewidth]{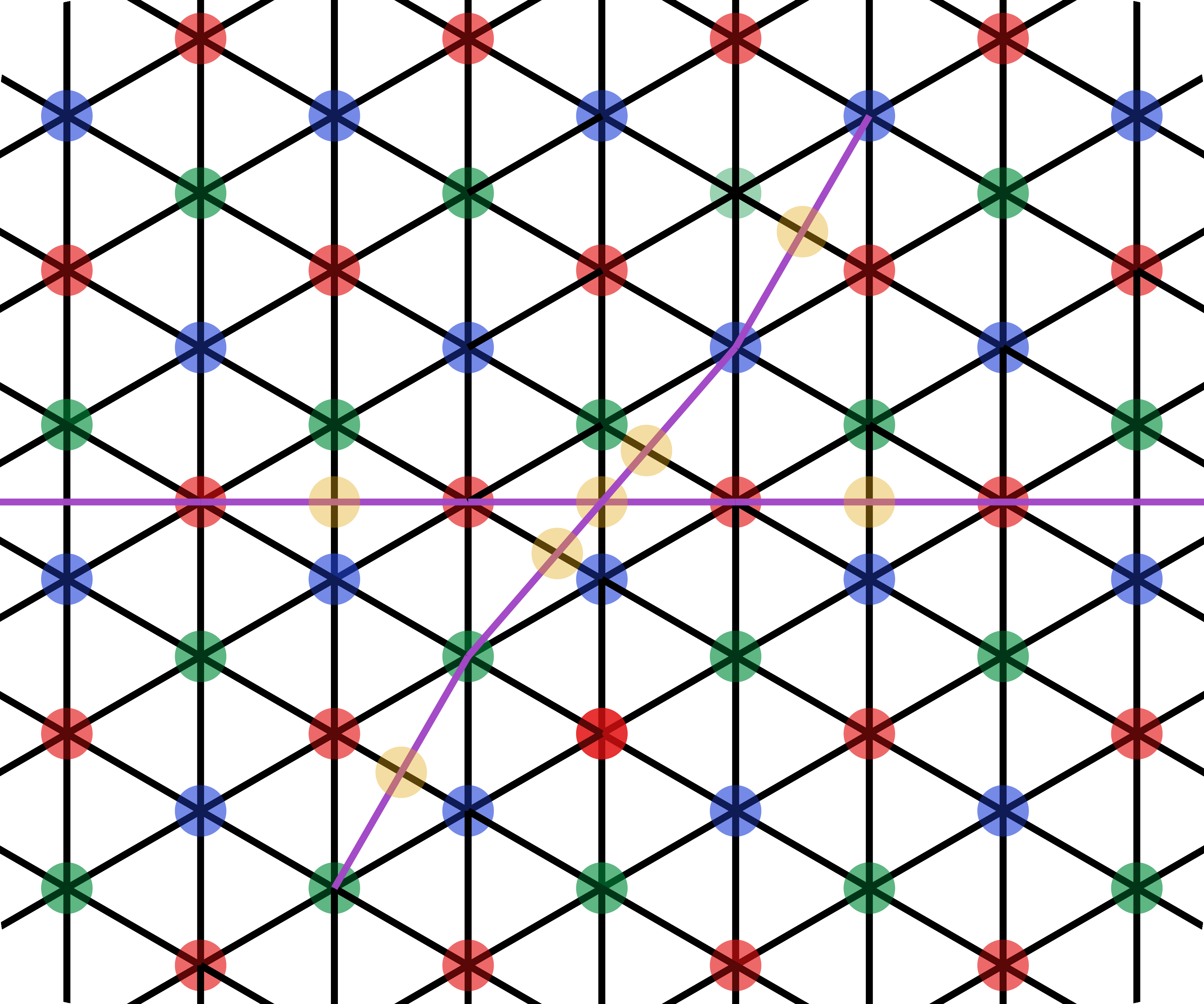}
		\caption{
			A $G$ twist (bottom), when moved over an $R$ domain wall (middle) changes into a $B$ twist (top).
			This process can be understood by starting with an $R$ domain wall and then applying Pachner moves which move the $G$ twist over said $R$ domain wall.
			Since the lattice along the $R$ domain wall hosts yellow square stabilizers, the series of Pachner moves changes such, that after crossing the $R$ domain wall, it is terminated by a $B$ twist on a blue vertex.
			Fig.~\ref{fig:TwistCrossingDWPrimal} depicts the same configuration of twists in the primal lattice.
		}
		\label{fig:TwistCrossingDWGeneralDual}
	\end{figure}
	The way color twists change when moved over color domain walls can also be visualized using Pachner moves.
	In Fig.~\ref{fig:TwistCrossingDWGeneralDual}, two crossing domain walls are depicted.
	We can understand the configuration as a pair of $G$ twists located at the bottom, one of which was moved over an $R$ domain wall in the middle which transformed in to a $B$ twist at the top.
	In doing so, the series of Pachner moves changed from connecting green vertices to connecting blue vertices as soon as the twist was moved over the domain wall.
	This shows, how a $G$ twist changes to a $B$ twist if moved over an $R$ domain wall.
	See Fig.~\ref{fig:TwistCrossingDWPrimal} for the same configuration in the primal color code lattice.

	\begin{figure}
		\centering
		\includegraphics[width=.75\linewidth]{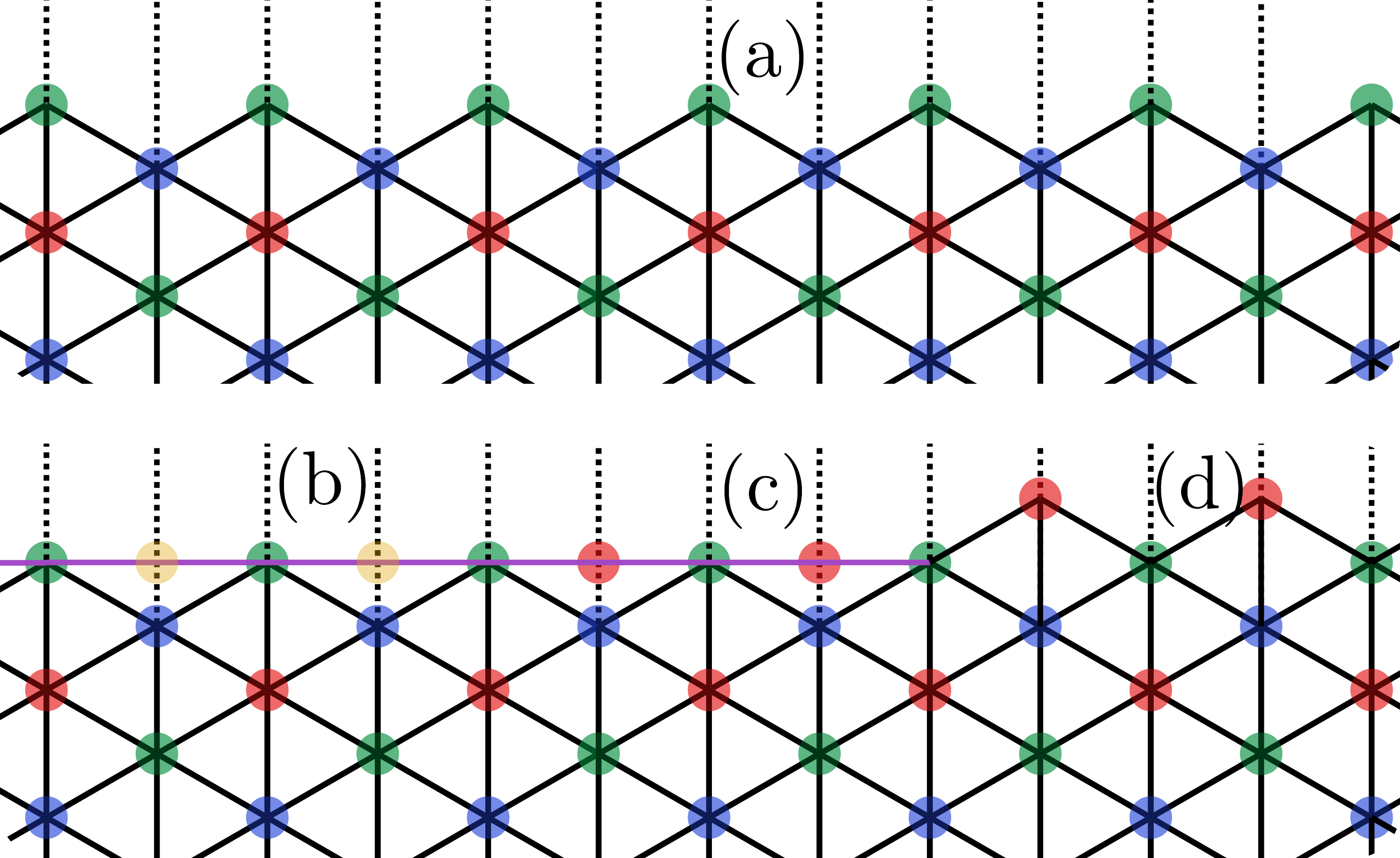}
		\caption{
			Adding a color domain wall changes the color boundaries.
			\textbf{(a)} shows a common red color boundary in the dual lattice.
			The dashed lines going upwards indicate the deleted red vertex (plaquette in the primal lattice) that was removed to create the red color boundary.
			In \textbf{(b)}, a $G$ domain wall was added to the boundary by applying (bi) Pachner moves.
			In \textbf{(c)}, the in (b) introduces yellow plaquettes has been changed to red.
			And finally, in \textbf{(d)} the lattice was a bit deformed to make it apparent that the obtained boundary is a blue boundary.
			We summarize, a $G$ domain wall changes a red to a blue color boundary.
		}
		\label{fig:DWChangeBoundariesDual}
	\end{figure}
	Next, let study how color domain walls change the color boundaries of the color code.
	Consider Fig.~\ref{fig:DWChangeBoundariesDual}, where we show how a red color boundary turns into a blue color boundary when a $G$ domain wall is added.
	The red color boundary corresponds to a single red vertex (plaquette in the primal lattice), whose stabilizer generators were removed from the code Hamiltonian.
	This is indicated in the figure by the dashed lines extending to a red vertex far outside the top of the figure.
	Adding a $G$ domain wall at the boundary by applying (bi) Pachner moves, adds new yellow vertices.
	This transforms the boundary to a blue color boundary, as is made clear in the figure by first recoloring the vertices and then slightly deforming the lattice.

	\begin{figure}[b!]
		\centering
		\includegraphics[width=.5\linewidth]{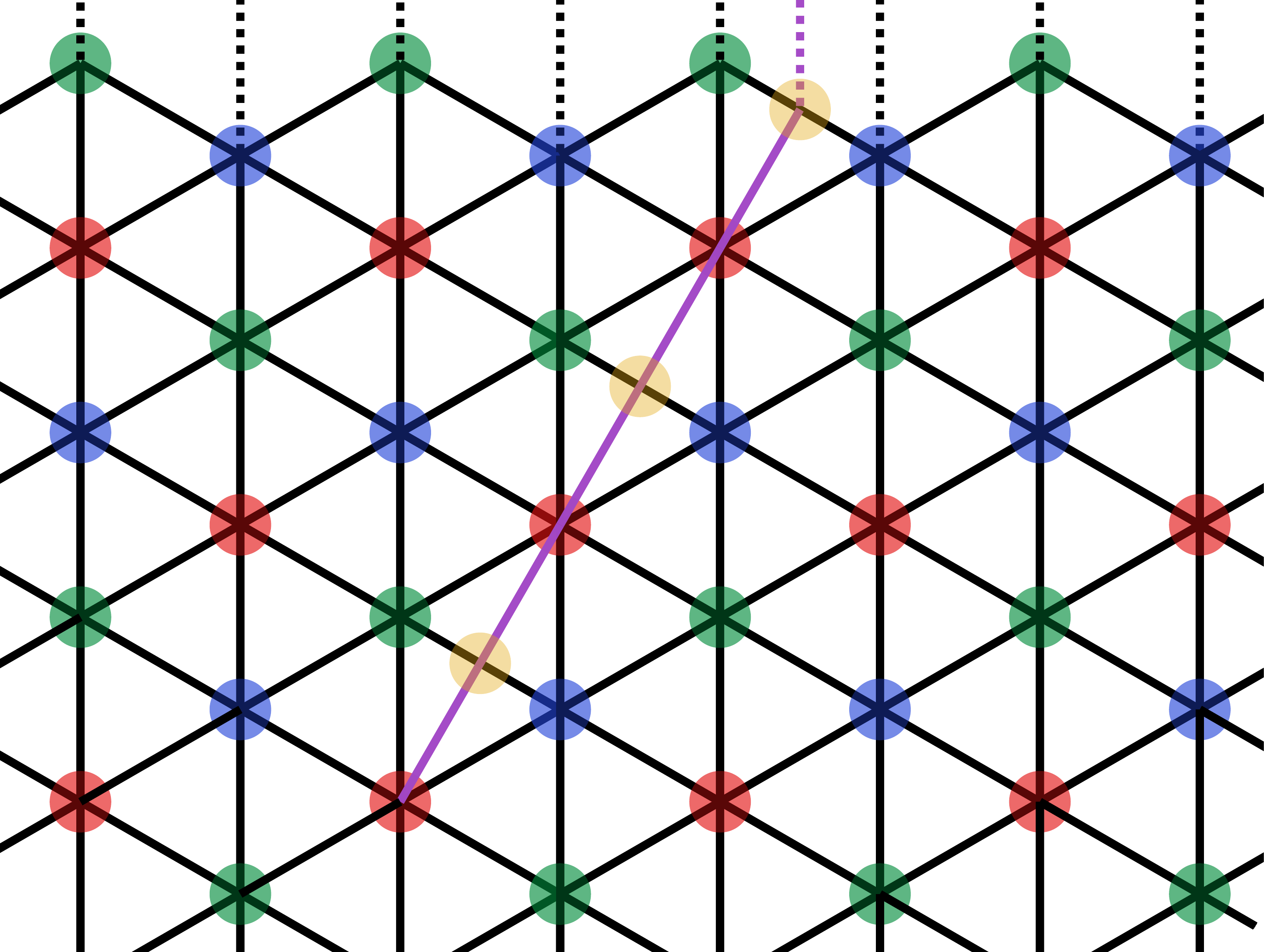}
		\caption{
			An $R$ defect line condensed at a red color boundary as drawn on the dual lattice.
			The color boundary on the top corresponds to a removed red plaquette, indicated by the dashed edges.
			An $R$ domain wall introduced by (bi) Pachner moves applied along a line can terminate at the red boundary.
		}
		\label{fig:TwistCondensationDual}
	\end{figure}
	Lastly, let us turn to the condensation of a color twist at a color boundary.
	Fig.~\ref{fig:TwistCondensationDual} depicts an $R$ domain wall which extends over a red color boundary.
	As discussed, the red boundary corresponds to a removed red vertex, as indicated by the dashed edges.
	An $R$ domain wall created by (bi) Pachner moves can be extended such that it terminates at the removed red vertex.
	This corresponds to the condensation of the terminating $R$ twist at the red boundary.
	Fig.~\ref{fig:TwistCondensationLattices}~(a) depicts the same configuration of twists in the primal lattice.

	We find the connection between color permuting twists and Pachner moves is illuminating as it visualizes the algebraic structure of the color permuting subgroup of the color code symmetry.
	It also implicitly offers a protocol to braid and condense color twists through code deformation techniques.
	Additionally, the Pachner move construction likely generalizes to show extensive twists~\cite{Mesaros13} in higher dimensional color codes~\cite{Bombin07, Bombin07b, Bombin15, Kubica15a, Watson15}, see also the supplemental material of Ref.~\cite{Brown16a}.

\section{Stellated surface codes}
	\label{app:StellatedSurfaceCodes}

\begin{figure}[b!]
		\centering
		\includegraphics[width=1\linewidth]{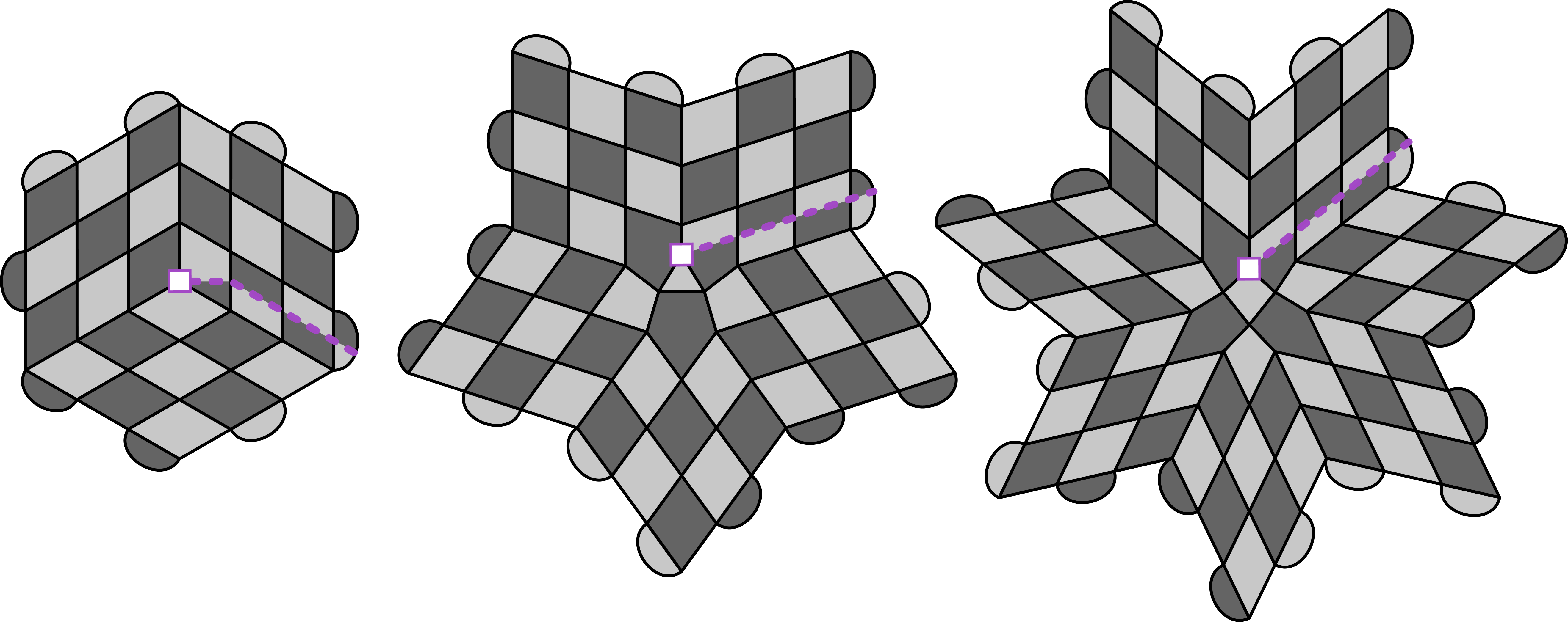}
		\caption{
			The three first stellated surface codes with odd $s$, shown with a code distance $d = 7$.
			Each plaquette hosts one stabilizer which acts in the $z$-($x$-)basis on the qubit if the plaquette is colored light (dark) gray where it meets the qubit.
			Along the domain wall, indicated by the dashed purple line, stabilizers act in both bases on different qubits.
			At the center of the code a stabilizer acts in the $y$-basis on the qubit which it meets with two colors.
			\textbf{Left:} The $s = 3$ stellated surface code is the Yoder-Kim triangular surface code~\cite{Yoder16}.
			The $s = 5$ and the $s = 7$ stellated toric codes are shown in the \textbf{middle} and on the \textbf{right}, respectively.
		}
		\label{fig:StellatedSurfaceCode}
	\end{figure}

	As mentioned in Sec.~\ref{sec:TwistsApplications} where stellated color codes were introduced, there also exist similar codes for the toric code model.
	We call them stellated surface codes.
	The triangular surface code introduced in Ref.~\cite{Yoder16} can be interpreted as the $s = 3$ member of this code family, where $s$ denotes the rotational symmetry of the code.
	Stellated surface codes with odd symmetry $s$ host a toric code twist in their center, which is connected to the boundary via a domain wall, shown in purple in Fig.~\ref{fig:StellatedSurfaceCode}.
	The $c$ value increases with increasing $s$.
	We find $c = 2 - \frac{2}{s}$ for odd $s$ and $c = 2 - \frac{4}{s}$ for even $s$, both approaching $2$ for large values of $s$.

	The stellated surface codes with the three smallest odd $s$ are shown in Fig.~\ref{fig:StellatedSurfaceCode}.
	We use the convention of placing physical qubits at vertices and associating each plaquette with one stabilizer~\cite{Wen03}.
	Light gray plaquettes correspond to a $z$ stabilizer, which checks the parity of qubits adjacent to the plaquette in the $z$-basis.
	Similarly, dark gray plaquettes correspond to $x$ stabilizers.
	Plaquettes along the purple dashed domain wall are colored in both shades, dark and light gray.
	They also host a single stabilizer each which acts in different bases;
	qubits which lie on their light gray side are measured in the $z$-basis, the ones on the dark gray side in the $x$-basis, and the single qubit where both shades of gray meet is measured in the $y$-basis.



\begin{thebibliography}{105}
\providecommand{\natexlab}[1]{#1}
\providecommand{\url}[1]{\texttt{#1}}
\expandafter\ifx\csname urlstyle\endcsname\relax
  \providecommand{\doi}[1]{doi: #1}\else
  \providecommand{\doi}{doi: \begingroup \urlstyle{rm}\Url}\fi

\bibitem[Wen(2004)]{Wen04}
Xiao-Gang Wen.
\newblock \emph{Quantum field theory of many-body systems: from the origin of
  sound to an origin of light and electrons}.
\newblock Oxford University Press on Demand, 2004.
\newblock URL \url{https://books.google.com/books?id=llnlrfdR4YgC}.

\bibitem[Kitaev(2003)]{Kitaev03}
Alexei~Yu. Kitaev.
\newblock Fault-tolerant quantum computation by anyons.
\newblock \emph{Ann. Phys.}, 303:\penalty0 2, 2003.
\newblock \doi{10.1016/S0003-4916(02)00018-0}.

\bibitem[Dennis et~al.(2002)Dennis, Kitaev, Landahl, and Preskill]{Dennis02}
Eric Dennis, Alexei Kitaev, Andrew Landahl, and John Preskill.
\newblock Topological quantum memory.
\newblock \emph{J. Math. Phys.}, 43:\penalty0 4452, 2002.
\newblock \doi{10.1063/1.1499754}.

\bibitem[Preskill(2017)]{Preskill17lectures}
J.~Preskill.
\newblock Topological quantum computation, 2017.
\newblock URL
  \url{http://www.theory.caltech.edu/~preskill/ph219/topological.pdf}.
\newblock (Chapter 9 of Lecture Notes on Quantum Computation).

\bibitem[Nayak et~al.(2008)Nayak, Simon, Stern, Freedman, and Sarma]{Nayak08}
Chetan Nayak, Steven~H. Simon, Ady Stern, Michael Freedman, and Sankar~Das
  Sarma.
\newblock Non-{A}belian anyons and topological quantum computation.
\newblock \emph{Rev. Mod. Phys.}, 80:\penalty0 1083, 2008.
\newblock \doi{10.1103/RevModPhys.80.1083}.

\bibitem[Pachos(2012)]{Pachos12}
Jiannis~K Pachos.
\newblock \emph{Introduction to topological quantum computation}.
\newblock Cambridge University Press, 2012.
\newblock URL \url{https://books.google.com/books?id=XDciVh6bAE0C}.

\bibitem[Terhal(2015)]{Terhal15}
B.~M. Terhal.
\newblock Quantum error correcton for quantum memories.
\newblock \emph{Rev. Mod. Phys.}, 87:\penalty0 307, 2015.
\newblock \doi{10.1103/RevModPhys.87.307}.

\bibitem[Brown et~al.(2016{\natexlab{a}})Brown, Loss, Pachos, Self, and
  Wootton]{Brown16}
B.~J. Brown, D.~Loss, J.~K. Pachos, C.~N. Self, and J.~R. Wootton.
\newblock Quantum memories at finite temperature.
\newblock \emph{Rev. Mod. Phys.}, 88:\penalty0 045005, 2016{\natexlab{a}}.
\newblock \doi{10.1103/RevModPhys.88.045005}.

\bibitem[Campbell et~al.(2017)Campbell, Terhal, and Vuillot]{Campbell17}
Earl~T Campbell, Barbara~M Terhal, and Christophe Vuillot.
\newblock Roads towards fault-tolerant universal quantum computation.
\newblock \emph{Nature}, 549\penalty0 (7671):\penalty0 172, 2017.
\newblock \doi{10.1038/nature23460}.

\bibitem[Kitaev(2006)]{Kitaev06}
Alexei Kitaev.
\newblock Anyons in an exactly solved model and beyond.
\newblock \emph{Ann. Phys.}, 321\penalty0 (1):\penalty0 2--111, 2006.
\newblock \doi{10.1016/j.aop.2005.10.005}.

\bibitem[Barkeshli and Wen(2010)]{Barkeshli10}
Maissam Barkeshli and Xiao-Gang Wen.
\newblock {$U(1)\times U(1) \rtimes \mathbb{Z}_2$} {C}hern-{S}imons theory and
  ${Z}_4$ parafermion fractional quantum {H}all states.
\newblock \emph{Phys. Rev. B}, 81:\penalty0 045323, 2010.
\newblock \doi{10.1103/PhysRevB.81.045323}.

\bibitem[Bomb\'{i}n(2010)]{Bombin10}
Hector Bomb\'{i}n.
\newblock Topological order with a twist: {I}sing anyons from an {A}belian
  model.
\newblock \emph{Phys. Rev. Lett.}, 105:\penalty0 030403, 2010.
\newblock \doi{10.1103/PhysRevLett.105.030403}.

\bibitem[Beigi et~al.(2011)Beigi, Shor, and Whalen]{Beigi11}
Salman Beigi, Peter~W. Shor, and Daniel Whalen.
\newblock The quantum double model with boundary: Condensations and symmetries.
\newblock \emph{Commun. Math. Phys.}, 313:\penalty0 351, 2011.
\newblock \doi{10.1007/s00220-011-1294-x}.

\bibitem[Kitaev and Kong(2012)]{Kitaev12}
Alexei Kitaev and Liang Kong.
\newblock Model for gapped boundaries and domain walls.
\newblock \emph{Commun. Math. Phys.}, 313:\penalty0 351, 2012.
\newblock \doi{10.1007/s00220-012-1500-5}.

\bibitem[Barkeshli et~al.(2014)Barkeshli, Bonderson, Cheng, and
  Wang]{Barkeshli14}
Maissam Barkeshli, Parsa Bonderson, Meng Cheng, and Zhenghan Wang.
\newblock Symmetry, defects, and gauging of topological phases.
\newblock \emph{arXiv:1410.4540}, 2014.
\newblock URL \url{https://arxiv.org/abs/1410.4540}.

\bibitem[Barter et~al.(2018)Barter, Bridgeman, and Jones]{Barter18}
Daniel Barter, Jacob~C. Bridgeman, and Corey Jones.
\newblock Domain walls in topological phases and the {B}rauer-{P}icard ring for
  $\text{Vec}(\mathbb{Z}/p\mathbb{Z})$.
\newblock \emph{arXiv:1806.01279}, 2018.
\newblock URL \url{https://arxiv.org/abs/1806.01279}.

\bibitem[Bomb\'{i}n(2011)]{Bombin11}
Hector Bomb\'{i}n.
\newblock Clifford gates by code deformation.
\newblock \emph{New J. Phys.}, 13:\penalty0 043005, 2011.
\newblock \doi{10.1088/1367-2630/13/4/043005}.

\bibitem[Barkeshli et~al.(2013{\natexlab{a}})Barkeshli, Jian, and
  Qi]{Barkeshli13}
Maissam Barkeshli, Chao-Ming Jian, and Xiao-Liang Qi.
\newblock Twist defects and projective non-{A}belian braiding statistics.
\newblock \emph{Phys. Rev. B}, 87:\penalty0 045130, 2013{\natexlab{a}}.
\newblock \doi{10.1103/PhysRevB.87.045130}.

\bibitem[Brown et~al.(2014)Brown, Al-Shimary, and Pachos]{Brown14}
Benjamin~J. Brown, Abbas Al-Shimary, and Jiannis~K. Pachos.
\newblock Entropic barriers for two-dimensional quantum memories.
\newblock \emph{Phys. Rev. Lett.}, 112:\penalty0 120503, 2014.
\newblock \doi{10.1103/PhysRevLett.112.120503}.

\bibitem[Hastings and Geller(2015)]{Hastings15}
Matthew~B. Hastings and A.~Geller.
\newblock Reduced space-time costs {I}sing dislocation codes and arbitrary
  ancillas.
\newblock \emph{Quant. Inf. Comp.}, 15:\penalty0 0962, 2015.
\newblock URL \url{https://arxiv.org/abs/1408.3379}.

\bibitem[Wootton(2015)]{Wootton15a}
James~R. Wootton.
\newblock A family of stabilizer codes for {$ D({\mathbb Z} _2)$} anyons and
  majorana modes.
\newblock \emph{J. Phys. A: Math. Theor.}, 48:\penalty0 215302, 2015.
\newblock \doi{10.1088/1751-8113/48/21/215302}.

\bibitem[Yoder and Kim(2017)]{Yoder16}
Theodore~J. Yoder and Isaac~H. Kim.
\newblock The surface code with a twist.
\newblock \emph{Quantum}, 1:\penalty0 2, 2017.
\newblock \doi{10.22331/q-2017-04-25-2}.

\bibitem[Brown et~al.(2017)Brown, Laubscher, Kesselring, and Wootton]{Brown17}
B.~J. Brown, K.~Laubscher, M.~S. Kesselring, and J.~R. Wootton.
\newblock Poking holes and cutting corners to achieve {C}lifford gates with the
  surface code.
\newblock \emph{Phys. Rev. X}, 7:\penalty0 021029, 2017.
\newblock \doi{10.1103/PhysRevX.7.021029}.

\bibitem[Yoshida(2017)]{Yoshida17}
Beni Yoshida.
\newblock Gapped boundaries, group cohomology and fault-tolerant logical gates.
\newblock \emph{Ann. Phys.}, 377:\penalty0 387--413, 2017.
\newblock \doi{10.1016/j.aop.2016.12.014}.

\bibitem[Roberts et~al.(2017)Roberts, Yoshida, Kubica, and Bartlett]{Roberts17}
Sam Roberts, Beni Yoshida, Aleksander Kubica, and Stephen~D. Bartlett.
\newblock Symmetry-protected topological order at nonzero temperature.
\newblock \emph{Phys. Rev. A}, 96:\penalty0 022306, 2017.
\newblock \doi{10.1103/PhysRevA.96.022306}.

\bibitem[Zhu et~al.(2017)Zhu, Hafezi, and Barkeshli]{Zhu17}
Guanyu Zhu, Mohammad Hafezi, and Maissam Barkeshli.
\newblock Quantum origami: Applying fault-tolerant transversal gates and
  measuring topological order.
\newblock \emph{arXiv:1711.05752}, 2017.
\newblock URL \url{https://arxiv.org/abs/1711.05752}.

\bibitem[Lavasani and Barkeshli(2018)]{Lavasani18}
Ali Lavasani and Maissam Barkeshli.
\newblock Low overhead {C}lifford gates from joint measurements in surface,
  color, and hyperbolic codes.
\newblock \emph{arXiv:1804.04144}, 2018.
\newblock URL \url{https://arxiv.org/abs/1804.04144}.

\bibitem[Bomb\'{i}n and Martin-Delagado(2006)]{Bombin06}
Hector Bomb\'{i}n and Miguel~A. Martin-Delagado.
\newblock Topological quantum distillation.
\newblock \emph{Phys. Rev. Lett.}, 97:\penalty0 180501, 2006.
\newblock \doi{10.1103/PhysRevLett.97.180501}.

\bibitem[Bomb\'{i}n(2015)]{Bombin15}
H\'{e}ctor Bomb\'{i}n.
\newblock Gauge color codes: optimal transveral gates and gauge fixing in
  topological stabilizer codes.
\newblock \emph{New J. Phys.}, 17:\penalty0 083002, 2015.
\newblock \doi{10.1088/1367-2630/17/8/083002}.

\bibitem[Kubica and Beverland(2015)]{Kubica15a}
Aleksander Kubica and Michael~E. Beverland.
\newblock Universal transversal gates with color codes: A simplified approach.
\newblock \emph{Phys. Rev. A}, 91:\penalty0 032330, 2015.
\newblock \doi{10.1103/PhysRevA.91.032330}.

\bibitem[Watson et~al.(2015)Watson, Campbell, Anwar, and Browne]{Watson15}
Fern H.~E. Watson, Earl~T. Campbell, Hussain Anwar, and Dan~E. Browne.
\newblock Qudit color codes and gauge color codes in all spatial dimensions.
\newblock \emph{Phys. Rev. A}, 92:\penalty0 022312, 2015.
\newblock \doi{10.1103/PhysRevA.92.022312}.

\bibitem[Campbell(2016)]{CampbellBlog}
Earl~T. Campbell.
\newblock The smallest interesting color code, 2016.
\newblock URL
  \url{https://earltcampbell.com/2016/09/26/the-smallest-interesting-colour-code/}.

\bibitem[Kubica et~al.(2015)Kubica, Yoshida, and Pastawski]{Kubica15}
Aleksander Kubica, Beni Yoshida, and Fernando Pastawski.
\newblock Unfolding the color code.
\newblock \emph{New Journal of Physics}, 17\penalty0 (8):\penalty0 083026,
  2015.
\newblock \doi{10.1088/1367-2630/17/8/083026}.

\bibitem[Vasmer and Browne(2018)]{Vasmer18}
Michael Vasmer and Dan~E. Browne.
\newblock Universal quantum computing with 3{D} surface codes.
\newblock \emph{arXiv:1801.04255}, 2018.
\newblock URL \url{https://arxiv.org/abs/1801.04255}.

\bibitem[Bravyi and K\"{o}nig(2013)]{Bravyi13}
Sergey Bravyi and Robert K\"{o}nig.
\newblock Classification of topologically protected gates for local stabilizer
  codes.
\newblock \emph{Phys. Rev. Lett.}, 110:\penalty0 170503, 2013.
\newblock \doi{10.1103/PhysRevLett.110.170503}.

\bibitem[Pastawski and Yoshida(2015)]{Pastawski15}
Fernando Pastawski and Beni Yoshida.
\newblock Fault-tolerant logical gates in quantum error-correcting codes.
\newblock \emph{Phys. Rev. A}, 91:\penalty0 012305, 2015.
\newblock \doi{10.1103/PhysRevA.91.012305}.

\bibitem[Jochym-O’Connor et~al.(2018)Jochym-O’Connor, Kubica, and
  Yoder]{OConnor17}
Tomas Jochym-O’Connor, Aleksander Kubica, and Theodore~J Yoder.
\newblock Disjointness of stabilizer codes and limitations on fault-tolerant
  logical gates.
\newblock \emph{Phys. Rev. X}, 8\penalty0 (2):\penalty0 021047, 2018.
\newblock \doi{10.1103/PhysRevX.8.021047}.

\bibitem[Webster and Bartlett(2018)]{Webster17}
Paul Webster and Stephen~D. Bartlett.
\newblock Locality-preserving logical operators in topological stabilizer
  codes.
\newblock \emph{Phys. Rev. A}, 97:\penalty0 012330, 2018.
\newblock \doi{10.1103/PhysRevA.97.012330}.

\bibitem[Raussendorf et~al.(2006)Raussendorf, Harrington, and
  Goyal]{Raussendorf06}
R.~Raussendorf, J.~Harrington, and K.~Goyal.
\newblock A fault-tolerant one-way quantum computer.
\newblock \emph{Ann. Phys.}, 321:\penalty0 2242, 2006.
\newblock \doi{10.1016/j.aop.2006.01.012}.

\bibitem[Bomb{\'\i}n and Martin-Delgado(2009)]{Bombin09}
H.~Bomb{\'\i}n and M.~A. Martin-Delgado.
\newblock Quantum measurements and gates by code deformation.
\newblock \emph{J. Phys. A}, 42:\penalty0 095302, 2009.
\newblock \doi{10.1088/1751-8113/42/9/095302}.

\bibitem[Fowler(2011)]{Fowler11}
Austin~G Fowler.
\newblock Two-dimensional color-code quantum computation.
\newblock \emph{Phys. Rev. A}, 83\penalty0 (4):\penalty0 042310, 2011.
\newblock \doi{10.1103/PhysRevA.83.042310}.

\bibitem[Horsman et~al.(2012)Horsman, Fowler, Devitt, and Meter]{Horsman12}
Clare Horsman, Austin~G. Fowler, Simon Devitt, and Rodney~Van Meter.
\newblock Surface code quantum computing by lattice surgery.
\newblock \emph{New J. Phys.}, 14:\penalty0 123011, 2012.
\newblock \doi{10.1088/1367-2630/14/12/123011}.

\bibitem[Landahl and Ryan-Anderson(2014)]{Landahl14}
Andrew~J. Landahl and Ciaran Ryan-Anderson.
\newblock Quantum computing by color-code lattice surgery.
\newblock \emph{arXiv:1407.5103}, 2014.
\newblock URL \url{https://arxiv.org/abs/1407.5103}.

\bibitem[Teo et~al.(2014)Teo, Roy, and Chen]{Teo14}
Jeffrey~CY Teo, Abhishek Roy, and Xiao Chen.
\newblock Unconventional fusion and braiding of topological defects in a
  lattice model.
\newblock \emph{Phys. Rev. B}, 90\penalty0 (11):\penalty0 115118, 2014.
\newblock \doi{10.1103/PhysRevB.90.115118}.

\bibitem[Yoshida(2015)]{Yoshida15}
Beni Yoshida.
\newblock Topological color code and symmetry-protected topological phases.
\newblock \emph{Phys. Rev. B}, 91:\penalty0 245131, 2015.
\newblock \doi{10.1103/PhysRevB.91.245131}.

\bibitem[Bridgeman et~al.(2017)Bridgeman, Bartlett, and Doherty]{Bridgeman17}
Jacob~C. Bridgeman, Stephen~D. Bartlett, and Andrew~C. Doherty.
\newblock Tensor networks with a twist: Anyon-permuting domain walls and
  defects in peps.
\newblock \emph{Phys. Rev. B}, 96:\penalty0 245122, 2017.
\newblock \doi{10.1103/PhysRevB.96.245122}.

\bibitem[Williamson et~al.(2017)Williamson, Bultinck, and
  Verstraete]{Williamson17}
Dominic~J Williamson, Nick Bultinck, and Frank Verstraete.
\newblock Symmetry-enriched topological order in tensor networks: Defects,
  gauging and anyon condensation.
\newblock \emph{arXiv:1711.07982}, 2017.
\newblock URL \url{https://arxiv.org/abs/1711.07982}.

\bibitem[Reed et~al.(2012)Reed, DiCarlo, Nigg, Sun, Frunzio, Girvin, and
  Schoelkopf]{Reed12}
M.~D. Reed, L.~DiCarlo, S.~E. Nigg, L.~Sun, L.~Frunzio, S.~M. Girvin, and R.~J.
  Schoelkopf.
\newblock Realization of three-qubit quantum error correction with
  superconducting circuits.
\newblock \emph{Nature}, 482:\penalty0 382, 2012.
\newblock \doi{10.1038/nature10786}.

\bibitem[Barends et~al.(2014)Barends, Kelly, Megrant, Veitia, Sank, Jeffry,
  White, Mutus, Fowler, Campbell, Chen, Chen, Chiaro, Dunsworth, Neill,
  O'Malley, Roushan, Vainsencher, Wenner, Korotkov, Cleland, and
  Martinis]{Barends14}
R.~Barends, J.~Kelly, A.~Megrant, A.~Veitia, D.~Sank, E.~Jeffry, T.~C. White,
  J.~Mutus, A.~G. Fowler, B.~Campbell, Y.~Chen, Z.~Chen, B.~Chiaro,
  A.~Dunsworth, C.~Neill, P.~O'Malley, P.~Roushan, A.~Vainsencher, J.~Wenner,
  A.~N. Korotkov, A.~N. Cleland, and J.~M. Martinis.
\newblock Superconducting quantum circuits at the surface code threshold for
  fault tolerance.
\newblock \emph{Nature}, 508:\penalty0 500, 2014.
\newblock \doi{10.1038/nature13171}.

\bibitem[Nigg et~al.(2014)Nigg, M{\"u}ller, Martinez, Schindler, Hennrich,
  Monz, Martin-Delgado, and Blatt]{Nigg14}
D.~Nigg, M.~M{\"u}ller, E.~A. Martinez, P.~Schindler, M.~Hennrich, T.~Monz,
  M.~A. Martin-Delgado, and R.~Blatt.
\newblock Quantum computations on a topologically encoded qubit.
\newblock \emph{Science}, 345\penalty0 (6194):\penalty0 302--305, 2014.
\newblock \doi{10.1126/science.1253742}.

\bibitem[C{\'o}rcoles et~al.(2015)C{\'o}rcoles, Magesan, Srinivasan, Cross,
  Steffen, Gambetta, and Chow]{Corcoles15}
A.~D. C{\'o}rcoles, Easwar Magesan, Srikanth~J. Srinivasan, Andrew~W. Cross,
  M.~Steffen, Jay~M. Gambetta, and Jerry~M. Chow.
\newblock Demonstration of a quantum error detection code using a square
  lattice of four superconducing qubits.
\newblock \emph{Nat. Comms.}, 6:\penalty0 6979, 2015.
\newblock \doi{10.1038/ncomms7979}.

\bibitem[Albrecht et~al.(2016)Albrecht, Higginbotham, Madsen, Kuemmeth,
  Jespersen, Nyg\r{a}rd, Krogstrup, and Marcus]{Albrecht16}
S.~M. Albrecht, A.~P. Higginbotham, M.~Madsen, F.~Kuemmeth, T.~S. Jespersen,
  J.~Nyg\r{a}rd, P.~Krogstrup, and C.~M. Marcus.
\newblock Exponential protection of zero modes in {M}ajorana islands.
\newblock \emph{Nature}, 531:\penalty0 206, 2016.
\newblock \doi{10.1038/nature17162}.

\bibitem[Takita et~al.(2016)Takita, C\'{o}rcoles, Magesan, Abdo, Brink, Cross,
  Chow, and Gambetta]{Takita16}
Maika Takita, A.~D. C\'{o}rcoles, Easwar Magesan, Baleeg Abdo, Markus Brink,
  Andrew~W. Cross, Jerry~M. Chow, and Jay~M. Gambetta.
\newblock Demonstration of weight-four parity measurements in the surface code
  architecture.
\newblock \emph{Phys. Rev. Lett.}, 117:\penalty0 210505, 2016.
\newblock \doi{10.1103/PhysRevLett.117.210505}.

\bibitem[Linke et~al.(2017)Linke, Gutierrez, Landsman, Figgatt, Debnath, Brown,
  and Monroe]{Linke17}
Norbert~M. Linke, Mauricio Gutierrez, Kevin~A. Landsman, Caroline Figgatt,
  Shantanu Debnath, Kenneth~R. Brown, and Christopher Monroe.
\newblock Fault-tolerant quantum error detection.
\newblock \emph{Sci. Adv.}, 3:\penalty0 e1701074, 2017.
\newblock \doi{10.1126/sciadv.1701074}.

\bibitem[Bomb{\'\i}n et~al.(2012)Bomb{\'\i}n, Andrist, Ohzeki, Katzgraber, and
  Martin-Delgado]{Bombin12a}
H.~Bomb{\'\i}n, Ruben~S. Andrist, Masayuki Ohzeki, Helmut~G. Katzgraber, and
  M.~A. Martin-Delgado.
\newblock Strong resilience of topological codes to depoloarization.
\newblock \emph{Phys. Rev. X}, 2:\penalty0 021004, 2012.
\newblock \doi{10.1103/PhysRevX.2.021004}.

\bibitem[Landahl et~al.(2011)Landahl, Anderson, and Rice]{Landahl11}
Andrew~J. Landahl, Jonas~T. Anderson, and Patrick~R. Rice.
\newblock Fault-tolerant quantum computing with color codes.
\newblock \emph{arXiv:1108.5738}, 2011.
\newblock URL \url{https://arxiv.org/abs/1108.5738}.

\bibitem[Terhal et~al.(2012)Terhal, Hassler, and DiVincenzo]{Terhal12}
Barbara~M. Terhal, Fabian Hassler, and David~P. DiVincenzo.
\newblock From majorana fermions to topological order.
\newblock \emph{Phys. Rev. Lett.}, 108:\penalty0 260504, Jun 2012.
\newblock \doi{10.1103/PhysRevLett.108.260504}.

\bibitem[Aasen et~al.(2016)Aasen, Hell, Mishmash, Higginbotham, Danon, Leijnse,
  Jespersen, Folk, Marcus, Flensberg, and Alicea]{Aasen16}
D.~Aasen, M.~Hell, R.~V. Mishmash, A.~Higginbotham, J.~Danon, M.~Leijnse, T.~S.
  Jespersen, J.~A. Folk, C.~M. Marcus, K.~Flensberg, and J.~Alicea.
\newblock Milestones toward majorana-based quantum computing.
\newblock \emph{Phys. Rev. X}, 6:\penalty0 031016, 2016.
\newblock \doi{10.1103/PhysRevX.6.031016}.

\bibitem[Plugge et~al.(2016)Plugge, Landau, Sela, Altland, Flensberg, and
  Egger]{Plugge16}
S.~Plugge, L.~A. Landau, E.~Sela, A.~Altland, K.~Flensberg, and R.~Egger.
\newblock Roadmap to majorana surface codes.
\newblock \emph{Phys. Rev. B}, 94:\penalty0 174514, Nov 2016.
\newblock \doi{10.1103/PhysRevB.94.174514}.

\bibitem[Landau et~al.(2016)Landau, Plugge, Sela, Altland, Albrecht, and
  Egger]{Landau16}
L.~A. Landau, S.~Plugge, E.~Sela, A.~Altland, S.~M. Albrecht, and R.~Egger.
\newblock Towards realistic implementations of a majorana surface code.
\newblock \emph{Phys. Rev. Lett.}, 116:\penalty0 050501, Feb 2016.
\newblock \doi{10.1103/PhysRevLett.116.050501}.

\bibitem[Litinski et~al.(2017)Litinski, Kesselring, Eisert, and von
  Oppen]{Litinski17}
D.~Litinski, M.~Kesselring, J.~Eisert, and F.~von Oppen.
\newblock Combining topological hardware and topological software: Color code
  quantum computing with topological superconductor networks.
\newblock \emph{Phys. Rev. X}, 7:\penalty0 031048, 2017.
\newblock \doi{10.1103/PhysRevX.7.031048}.

\bibitem[Litinski and von Oppen(2017)]{Litinski17a}
Daniel Litinski and Felix von Oppen.
\newblock Braiding by {M}ajorana tracking and long-range {CNOT} gates with
  color codes.
\newblock \emph{Phys. Rev. B}, 96:\penalty0 205413, 2017.
\newblock \doi{10.1103/PhysRevB.96.205413}.

\bibitem[Bravyi et~al.(2010)Bravyi, Poulin, and Terhal]{Bravyi10}
Sergey Bravyi, David Poulin, and Barbara Terhal.
\newblock Tradeoffs for reliable quantum information storage in 2{D} systems.
\newblock \emph{Phys. Rev. Lett.}, 104\penalty0 (5):\penalty0 050503, 2010.
\newblock \doi{10.1103/PhysRevLett.104.050503}.

\bibitem[Wen(2003)]{Wen03}
Xiao-Gang Wen.
\newblock Quantum orders in an exact soluble model.
\newblock \emph{Phys. Rev. Lett.}, 90\penalty0 (1):\penalty0 016803, 2003.
\newblock \doi{10.1103/PhysRevLett.90.016803}.

\bibitem[Delfosse et~al.(2016{\natexlab{a}})Delfosse, Iyer, and
  Poulin]{Delfosse16b}
Nicolas Delfosse, Pavithran Iyer, and David Poulin.
\newblock Generalized surface codes and packing of logical qubits.
\newblock \emph{arXiv:1606.07116}, 2016{\natexlab{a}}.
\newblock URL \url{https://arxiv.org/abs/1606.07116}.

\bibitem[Bomb\'{i}n et~al.(2012)Bomb\'{i}n, Duclos-Cianci, and
  Poulin]{Bombin12}
Hector Bomb\'{i}n, Guillaume Duclos-Cianci, and David Poulin.
\newblock Universal topological phase of two-dimensional stabilizer codes.
\newblock \emph{New Journal of Physics}, 14\penalty0 (7):\penalty0 073048,
  2012.
\newblock \doi{10.1088/1367-2630/14/7/073048}.

\bibitem[Bhagoji and Sarvepalli(2015)]{Bhagoji15}
Arjun Bhagoji and Pradeep Sarvepalli.
\newblock Equivalence of 2{D} color codes (without translational symmetry) to
  surface codes.
\newblock \emph{arXiv:1503.03009}, 2015.
\newblock URL \url{https://arxiv.org/abs/1503.03009}.

\bibitem[Criger and Terhal(2016)]{Criger16}
Ben Criger and Barbara Terhal.
\newblock Noise thresholds for the [[4, 2, 2]]-concatenated toric code.
\newblock \emph{arXiv:1604.04062}, 2016.
\newblock URL \url{https://arxiv.org/abs/1604.04062}.

\bibitem[Wang(2017)]{Wang17}
Zhenghan Wang.
\newblock private communication, 2017.

\bibitem[Rowell et~al.(2009)Rowell, Stong, and Wang]{Rowell09}
Eric Rowell, Richard Stong, and Zhenghan Wang.
\newblock On classification of modular tensor categories.
\newblock \emph{Commun. Math. Phys.}, 292\penalty0 (2):\penalty0 343--389,
  2009.
\newblock \doi{10.1007/s00220-009-0908-z}.

\bibitem[Moussa(2016)]{Moussa16}
Jonathan~E. Moussa.
\newblock Transversal {C}lifford gates on folded surface codes.
\newblock \emph{Phys. Rev. A}, 94:\penalty0 042316, 2016.
\newblock \doi{10.1103/PhysRevA.94.042316}.

\bibitem[Gottesman(1997)]{Gottesman97}
Daniel Gottesman.
\newblock \emph{Stabilizer Codes and Quantum Error Correction}.
\newblock PhD thesis, California Institute of Technology, 1997.
\newblock URL \url{https://arxiv.org/abs/quant-ph/9705052}.

\bibitem[Levin(2013)]{Levin13}
M.~Levin.
\newblock Protected edge modes without symmetry.
\newblock \emph{Phys. Rev. X}, 3\penalty0 (2):\penalty0 021009, 2013.
\newblock \doi{10.1103/PhysRevX.3.021009}.

\bibitem[Barkeshli et~al.(2013{\natexlab{b}})Barkeshli, Jian, and
  Qi]{Barkeshli13a}
Maissam Barkeshli, Chao-Ming Jian, and Xiao-Liang Qi.
\newblock Classification of topological defects in {A}belian topological
  states.
\newblock \emph{Phys. Rev. B}, 88:\penalty0 241103(R), 2013{\natexlab{b}}.
\newblock \doi{10.1103/PhysRevB.88.241103}.

\bibitem[Cong et~al.(2017{\natexlab{a}})Cong, Cheng, and Wang]{Cong17a}
Iris Cong, Meng Cheng, and Zhenghan Wang.
\newblock Defects between gapped boundaries in two-dimensional topological
  phases of matter.
\newblock \emph{Phys. Rev. B}, 96:\penalty0 195129, 2017{\natexlab{a}}.
\newblock \doi{10.1103/PhysRevB.96.195129}.

\bibitem[Burnell(2017)]{Burnell17}
FJ~Burnell.
\newblock Anyon condensation and its applications.
\newblock \emph{Annu. Rev. Condens. Matter Phys.}, \penalty0 (0), 2017.
\newblock \doi{10.1146/annurev-conmatphys-033117-054154}.

\bibitem[Cong et~al.(2016)Cong, Cheng, and Wang]{Cong16}
Iris Cong, Meng Cheng, and Zhenghan Wang.
\newblock Topological quantum computation with gapped boundaries.
\newblock \emph{arXiv:1609.02037}, 2016.
\newblock URL \url{https://arxiv.org/abs/1609.02037}.

\bibitem[Cong et~al.(2017{\natexlab{b}})Cong, Cheng, and Wang]{Cong17}
Iris Cong, Meng Cheng, and Zhenghan Wang.
\newblock Universal quantum computation with gapped boundaries.
\newblock \emph{Phys. Rev. Lett.}, 119\penalty0 (17):\penalty0 170504,
  2017{\natexlab{b}}.
\newblock \doi{10.1103/PhysRevLett.119.170504}.

\bibitem[Lindner et~al.(2012)Lindner, Berg, Refael, and Stern]{Lindner12}
Netanel~H Lindner, Erez Berg, Gil Refael, and Ady Stern.
\newblock Fractionalizing majorana fermions: Non-abelian statistics on the
  edges of abelian quantum hall states.
\newblock \emph{Phys. Rev. X}, 2\penalty0 (4):\penalty0 041002, 2012.
\newblock \doi{10.1103/PhysRevX.2.041002}.

\bibitem[Tarantino et~al.(2016)Tarantino, Lindner, and Fidkowski]{Tarantino16}
Nicolas Tarantino, Netanel~H Lindner, and Lukasz Fidkowski.
\newblock Symmetry fractionalization and twist defects.
\newblock \emph{New J. Phys.}, 18\penalty0 (3):\penalty0 035006, 2016.
\newblock \doi{10.1088/1367-2630/18/3/035006}.

\bibitem[Teo(2016)]{Teo16}
J.~C.~Y. Teo.
\newblock Globally symmetric topological phase: from anyonic symmetry to twist
  defect.
\newblock \emph{J. Phys.}, 28\penalty0 (14):\penalty0 143001, 2016.
\newblock \doi{10.1088/0953-8984/28/14/143001}.

\bibitem[Beverland et~al.(2016)Beverland, Buerschaper, Koenig, Pastawski,
  Preskill, and Sijher]{Beverland16}
Michael~E Beverland, Oliver Buerschaper, Robert Koenig, Fernando Pastawski,
  John Preskill, and Sumit Sijher.
\newblock Protected gates for topological quantum field theories.
\newblock \emph{J. Math. Phys.}, 57\penalty0 (2):\penalty0 022201, 2016.
\newblock \doi{10.1063/1.4939783}.

\bibitem[Kitaev and Preskill(2006)]{Kitaev06a}
Alexei Kitaev and John Preskill.
\newblock Topological entanglement entropy.
\newblock \emph{Phys. Rev. Lett.}, 96:\penalty0 110404, 2006.
\newblock \doi{10.1103/PhysRevLett.96.110404}.

\bibitem[Dong et~al.(2008)Dong, Fradkin, Leigh, and Nowling]{Dong08}
Shiying Dong, Eduardo Fradkin, Robert~G. Leigh, and Sean Nowling.
\newblock Topological entanglement entropy in {C}hern-{S}imons theories and
  quantum {H}all fluids.
\newblock \emph{JHEP}, 05:\penalty0 016, 2008.
\newblock \doi{10.1088/1126-6708/2008/05/016}.

\bibitem[Brown et~al.(2013)Brown, Bartlett, Doherty, and Barrett]{Brown13}
Benjamin~J. Brown, Stephen~D. Bartlett, Andrew~C. Doherty, and Sean~D. Barrett.
\newblock Topological entanglement entropy with a twist.
\newblock \emph{Phys. Rev. Lett.}, 111:\penalty0 220402, 2013.
\newblock \doi{10.1103/PhysRevLett.111.220402}.

\bibitem[Liu et~al.(2017)Liu, M\"{o}ller, and Bergholtz]{Liu17}
Zhao Liu, Gunnar M\"{o}ller, and Emil~J. Bergholtz.
\newblock Exotic non-{A}belian topological defects in lattice fractional
  quantum {H}all states.
\newblock \emph{Phys. Rev. Lett.}, 119:\penalty0 106801, 2017.
\newblock \doi{10.1103/PhysRevLett.119.106801}.

\bibitem[Bonderson et~al.(2017)Bonderson, Knapp, and Patel]{Bonderson17}
Parsa Bonderson, Christina Knapp, and Kaushal Patel.
\newblock Anyonic entanglement and topological entanglement entropy.
\newblock \emph{Ann. Phys.}, 385:\penalty0 399, 2017.
\newblock \doi{10.1016/j.aop.2017.07.018}.

\bibitem[You and Wen(2012)]{You12}
Y.-Z. You and X.-G. Wen.
\newblock Projective non-abelian statistics of dislocation defects in a
  {${\mathbb Z}_N$} rotor model.
\newblock \emph{Phys. Rev. B}, 86\penalty0 (16):\penalty0 161107, 2012.
\newblock \doi{10.1103/PhysRevB.86.161107}.

\bibitem[Nautrup et~al.(2017)Nautrup, Friis, and Briegel]{Nautrup17}
Hendrik~Poulsen Nautrup, Nicolai Friis, and Hans~J Briegel.
\newblock Fault-tolerant interface between quantum memories and quantum
  processors.
\newblock \emph{Nature Commun.}, 8\penalty0 (1):\penalty0 1321, 2017.
\newblock \doi{10.1038/s41467-017-01418-2}.

\bibitem[Litinski and von Oppen(2018)]{Litinski18}
Daniel Litinski and Felix von Oppen.
\newblock Lattice surgery with a twist: Simplifying clifford gates of surface
  codes.
\newblock \emph{Quantum}, 2:\penalty0 62, 2018.
\newblock \doi{10.22331/q-2018-05-04-62}.

\bibitem[Bomb{\'\i}n and Martin-Delgado(2006)]{Bombin06a}
H~Bomb{\'\i}n and MA~Martin-Delgado.
\newblock Topological quantum error correction with optimal encoding rate.
\newblock \emph{Phys. Rev. A}, 73\penalty0 (6):\penalty0 062303, 2006.
\newblock \doi{10.1103/PhysRevA.73.062303}.

\bibitem[Freedman et~al.(2002)Freedman, Meyer, and Luo]{Freedman02}
Michael~H Freedman, David~A Meyer, and Feng Luo.
\newblock {$Z_2$}-systolic freedom and quantum codes.
\newblock In \emph{Mathematics of quantum computation}, pages 287--320. 2002.
\newblock URL \url{https://books.google.de/books?id=evPKBQAAQBAJ}.

\bibitem[Delfosse(2013)]{Delfosse13}
Nicolas Delfosse.
\newblock Tradeoffs for reliable quantum information storage in surface codes
  and color codes.
\newblock In \emph{2013 IEEE International Symposium on Information Theory},
  pages 917--921. IEEE, 2013.
\newblock \doi{10.1109/ISIT.2013.6620360}.

\bibitem[Breuckmann et~al.(2017)Breuckmann, Vuillot, Campbell, Krishna, and
  Terhal]{Breuckmann17}
N.~P. Breuckmann, C.~Vuillot, E.~T. Campbell, A.~Krishna, and B.~M. Terhal.
\newblock Hyperbolic and semi-hyperbolic surface codes for quantum storage.
\newblock \emph{Quantum Sc. Tech.}, 2:\penalty0 035007, 2017.
\newblock \doi{10.1088/2058-9565/aa7d3b}.

\bibitem[Bomb{\'\i}n and Martin-Delgado(2007)]{Bombin07b}
H~Bomb{\'\i}n and MA~Martin-Delgado.
\newblock Exact topological quantum order in d=3 and beyond: Branyons and
  brane-net condensates.
\newblock \emph{Phys. Rev. B}, 75\penalty0 (7):\penalty0 075103, 2007.
\newblock \doi{10.1103/PhysRevB.75.075103}.

\bibitem[Bravyi et~al.(2006)Bravyi, Hastings, and Verstraete]{Bravyi06}
S.~Bravyi, M.~B. Hastings, and F.~Verstraete.
\newblock {L}ieb-{R}obinson bounds and the generation of correlations and
  topological quantum order.
\newblock \emph{Phys. Rev. Lett.}, 97:\penalty0 050401, 2006.
\newblock \doi{10.1103/PhysRevLett.97.050401}.

\bibitem[Chen et~al.(2010)Chen, Gu, and Wen]{Chen10}
Xie Chen, Zheng-Cheng Gu, and Xiao-Gang Wen.
\newblock Local unitary transformation, long-range quantum entanglement, wave
  function renormalization and topological order.
\newblock \emph{Phys. Rev. B}, 82:\penalty0 155138, 2010.
\newblock \doi{10.1103/PhysRevB.82.155138}.

\bibitem[Bravyi and Kitaev(1998)]{Bravyi98}
Sergey~B Bravyi and A~Yu Kitaev.
\newblock Quantum codes on a lattice with boundary.
\newblock \emph{arXiv:9811052}, 1998.
\newblock URL \url{https://arxiv.org/abs/quant-ph/9811052}.

\bibitem[Delfosse et~al.(2016{\natexlab{b}})Delfosse, Iyer, and
  Poulin]{Delfosse16a}
Nicolas Delfosse, Pavithran Iyer, and David Poulin.
\newblock A linear-time benchmarking tool for generalized surface codes.
\newblock \emph{arXiv:1611.04256}, 2016{\natexlab{b}}.
\newblock URL \url{https://arxiv.org/abs/1611.04256}.

\bibitem[Bomb{\'\i}n(2010)]{Bombin10b}
H.~Bomb{\'\i}n.
\newblock {Topological subsystem codes}.
\newblock \emph{Phys. Rev. A}, 81\penalty0 (3):\penalty0 32301, 2010.
\newblock \doi{10.1103/PhysRevA.81.032301}.

\bibitem[Pachner(1991)]{Pachner91}
Udo Pachner.
\newblock {P}. {L}. homeomorphic manifolds are equivalent by elementary
  shellings.
\newblock \emph{Europ. J. Combinatorics}, 12:\penalty0 129, 1991.
\newblock \doi{10.1016/S0195-6698(13)80080-7}.

\bibitem[Nakahara(2003)]{Nakahara}
Mikio Nakahara.
\newblock \emph{Geometry, Topology and Physics}.
\newblock Institute of Physics, 2003.
\newblock URL \url{https://books.google.com/books?id=cH-XQB0Ex5wC}.

\bibitem[Mesaros et~al.(2013)Mesaros, Kim, and Ran]{Mesaros13}
Andrej Mesaros, Yong~Baek Kim, and Ying Ran.
\newblock Changing topology by topological defects in three-dimensional
  topologically ordered phases.
\newblock \emph{Phys. Rev. B}, 88:\penalty0 035141, 2013.
\newblock \doi{10.1103/PhysRevB.88.035141}.

\bibitem[Bomb\'{i}n and Martin-Delagado(2007)]{Bombin07}
H.~Bomb\'{i}n and M.~A. Martin-Delagado.
\newblock Topological computation without braiding.
\newblock \emph{Phys. Rev. Lett.}, 98:\penalty0 160502, 2007.
\newblock \doi{10.1103/PhysRevLett.98.160502}.

\bibitem[Brown et~al.(2016{\natexlab{b}})Brown, Nickerson, and
  Browne]{Brown16a}
Benjamin~J. Brown, Naomi~H. Nickerson, and Dan~E. Browne.
\newblock Fault-tolerant error correction with the gauge color code.
\newblock \emph{Nat. Commun.}, 7:\penalty0 12302, 2016{\natexlab{b}}.
\newblock \doi{10.1038/ncomms12302}.

\end{thebibliography}

\end{document}